\documentclass{article}
\usepackage[english]{babel}
\usepackage[utf8]{inputenc}
\usepackage{color}

\usepackage{arxiv}

\usepackage{graphicx} 
\usepackage{amsmath} 
\usepackage{amssymb} 
\usepackage{amsthm} 
\usepackage{tabularx} 
\usepackage{siunitx} 
\usepackage{algorithm}
\usepackage{float} 
\usepackage{caption} 
\usepackage{subcaption} 
\usepackage{filecontents} 
\usepackage{epstopdf}
\usepackage[makeroom]{cancel} 
\usepackage[noend]{algpseudocode}
\usepackage{gensymb} 
\usepackage{comment}

\usepackage[T1]{fontenc}    
\usepackage{lmodern}
\usepackage[colorlinks=true, urlcolor=black, linkcolor=blue, citecolor=red]{hyperref}
\usepackage{url}            
\usepackage{booktabs}       
\usepackage{amsfonts}       
\usepackage{nicefrac}       
\usepackage{lipsum}		
\usepackage{doi}

\usepackage{newclude}			

\usepackage[sc,osf]{mathpazo}   
\everymath{\displaystyle}
\graphicspath{{img/}{Figures/}}

\theoremstyle{remark}

\title{Level-Set modeling of grain growth in 316L stainless steel under different assumptions regarding grain boundary properties }

\author{ \href{https://orcid.org/0000-0002-6513-7505}{\includegraphics[scale=0.06]{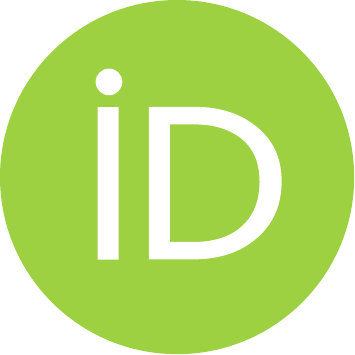}\hspace{1mm}Brayan ~Murgas}$^1$,
	\href{https://orcid.org/0000-0001-6804-1974}{\includegraphics[scale=0.06]{orcid.pdf}\hspace{1mm}Baptiste ~Flipon}$^1$,
	\href{https://orcid.org/0000-0002-8963-977X}{\includegraphics[scale=0.06]{orcid.pdf}\hspace{1mm}Nathalie ~Bozzolo}$^1$,
	\href{https://orcid.org/0000-0002-6677-2850}{\includegraphics[scale=0.06]{orcid.pdf}\hspace{1mm}Marc ~Bernacki}\thanks{Corresponding author: marc.bernacki@mines-paristech.fr} $^1$ \\
	\\
	$^1$ CEMEF – Centre de mise en forme des mat\'{e}riaux, CNRS UMR 7635, \\ Mines-ParisTech, PSL-Research University \\
	CS 10207 rue Claude Daunesse, 06904 Sophia Antipolis Cedex, France \\
}

\renewcommand{\headeright}{}
\renewcommand{\undertitle}{}
\renewcommand{\shorttitle}{Level-Set modeling of grain growth in 316L stainless steel under different assumptions regarding grain boundary properties}

\begin{document}
\maketitle

\begin{abstract}

Two finite element level-set (FE-LS) formulations are compared for the modeling of grain growth of 316L stainless steel in terms of grain size, mean values and histograms. Two kinds of microstructures are considered, some are generated statistically from EBSD maps and the others are generated by immersion of EBSD data in the FE formulation. Grain boundary (GB) mobility is heterogeneously defined as a function of the GB disorientation. On the other hand, GB energy is considered as heterogeneous or anisotropic, respectively defined as a function of the disorientation and both the GB misorientation and the GB inclination. In terms of mean grain size value and grain size distribution (GSD), both formulations provide similar responses. However, the anisotropic formulation better respects the experimental disorientation distribution function (DDF) and predicts more realistic grain morphologies. It was also found that the heterogeneous GB mobility described with a sigmoidal function only affects the DDF and the morphology of grains. Thus, a slower evolution of twin boundaries (TBs) is perceived.

\end{abstract}


\section{Introduction}

As most metallic materials exist in the form of polycrystals, determining the kinetics of metallurgical mechanisms such as recovery, grain growth (GG) and recrystallization is crucial since they determine the final microstructure and properties \cite{rollett2017recrystallization}. Grain boundary (GB) engineering refers to the control of GBs with the aim of obtaining high performance materials. Thus, numerical models have emerged to help us in predicting the evolution of microstructures submitted to different thermomechanical loads and the microstructure-property relationship.

The migration of GBs is classically described at the polycrystalline scale by the well-known equation $v=\mu P$, where $v$ is the GB velocity, $\mu$ the GB mobility and $P$ the driving pressure. During GG, the evolution of GBs is driven by the reduction of interfacial energy and the driving pressure is classically defined as a curvature flow driving pressure $P=-\gamma \kappa$, where $\gamma$ is the GB energy and $\kappa$ the mean curvature (i.e. the trace of the curvature tensor in 3D). At the polycrystalline scale, this kinematic equation is widely accepted while largely questioned \cite{doi:10.1126/science.abj3210, FLOREZ2022117459}. Moreover, a definition of the reduced mobility ($\mu \gamma$) within the misorientation and inclination 5D space is not straightforward \cite{OlmstedI2009, OlmstedII2009, bulatov2014grain, RUNNELS2016174}. 

GG has been widely studied at the polycrystalline scale with various numerical approaches like Phase-Field \cite{garcke1999multiphase, miyoshi2017multi, moelans2009comparative}, Monte Carlo \cite{gao1996real, upmanyu2002boundary}, Molecular Dynamics \cite{hoffrogge2017grain}, Orientated Tessellation Updating Method \cite{SAKOUT2020261}, Vertex \cite{BarralesMora2010}, Front-tracking Lagrangian or Eulerian formulations in a Finite Element (FE) context \cite{wakai2000,Florez2020,Florez2020b}, Level-Set (LS) \cite{bernacki2011level, miessen2015advanced, Fausty2020, ma14143883} and Kobayashi--Warren--Carter models \cite{kim2021crystal}, to cite some examples. GB energy and mobility have been widely studied since they were reported as being anisotropic by Smith \cite{Smith1948introduction} and Kohara \cite{kohara1958anisotropy}. The simplest models use constant values for the GB energy $\gamma$ and a temperature dependent mobility, $\mu(T)$, referred to as isotropic models \cite{anderson1984computer, gao1996real, lazar2011more, bernacki2011level, garcke1999multiphase}. Heterogeneous models were also proposed, in which each boundary has its own energy and mobility \cite{rollett1989simulation, hwang1998simulation, upmanyu2002boundary, Fausty2018, zollner2019texture, miyoshi2016validation, chang2019effect, miyoshi2019accuracy, miessen2015advanced, Fausty2020, holm2001misorientation} trying to reproduce more complex microstructures with local heterogeneity, such as twin boundaries. Each grain has its own crystal orientation, and GB energy and mobility depend on the disorientation angle between two grains \cite{miyoshi2017multi, Fausty2020}, but the effect of the misorientation axis and GB inclination is frequently omitted. General frameworks were thus proposed including the GB properties dependence on misorientation and inclination \cite{kazaryan2002grain, FAUSTY202128, hallberg2019modeling}, categorized as anisotropic models. It must be highlighted that the difference between 3-parameter (heterogeneous) and 5-parameter (anisotropic) full-field formulations is often unclear in the literature, heterogeneous GB properties being often categorized as anisotropic.  

The main reason why most of the studies are carried out using heterogeneous GB properties is the lack of data of GB properties. The early measurements of GB properties (mainly GB reduced mobility) were carried out on bicrystals \cite{viswanathan1973kinetics, demianczuk1975effect, maksimova1988transformation, gottstein1992true, winning2002mechanisms, ivanov2006kinetics} leading to the well known Sigmoidal model \cite{rollett2017recrystallization}. As experimental and computational technologies are improved, new experimental and computational 3D techniques allow to study GG and recrystallization using, for instance, X-ray \cite{ZHANG2017229, zhang2018determination, ZHANG2020211, JUULJENSEN2020100821, Fang:fc5052} or molecular dynamics \cite{janssens2006computing, olmsted2009survey, olmsted2009surveyii}. Hence, at the mesoscopic scale, few studies have been carried out in 2D using anisotropic GB properties designed by mathematical models \cite{kazaryan2002grain, FAUSTY202128} or by fitting data from molecular dynamics \cite{hallberg2019modeling}. Nevertheless, these 2D models neglect a part of the 3D space, i.e. the GB inclination is measured in the sample plane, and GB properties are simplified. Finally, regarding the study of GG in 3D, one frequently finds heterogeneous GB properties based on mathematical descriptors of GB properties \cite{FJELDBERG2010267, CHANG20191262, song2020effect, MIYOSHI2021109992} or based on databases of GB energy values \cite{KIM20111152, kim2014phase}. Based on this, two open questions arise: can GB properties be described in 2D using the classical Read-Shockley \cite{ReadShockley} and Sigmoidal \cite{humphreys1997unified} model? Is the effect of anisotropy stronger in 3D ? The latter implying to carry out 3D simulations instead of 2D, thus using a better description of GB properties in the 5D GB space.

In a preceding paper \cite{ma14143883}, four different formulations using a FE-LS approach have been compared. The first is an isotropic formulation used to model different annealing phenomena, such as GG, recrystallization and GG in presence of second phase particles \cite{Bernacki2008, Bernacki2009, bernacki2011level, scholtes2015new, Maire2017}. The second is an extension of the isotropic formulation considering heterogeneous values of GB energy and mobility \cite{ma14143883}. The third formulation was proposed for triple junctions in \cite{Fausty2018} and extended to model GG using heterogeneous GB energy in \cite{Fausty2020} and both heterogeneous GB energy and mobility in \cite{ma14143883}. The last one is an anisotropic formulation based on thermodynamics and differential geometry, it was first applied to bicrystals \cite{FAUSTY202128} and extended to polycrystals with heterogeneous GB energy and mobility in \cite{ma14143883}. In \cite{ma14143883}, academic cases of triple junctions and polycrystalline microstructures were presented. The main conclusion was that the isotropic formulation can reproduce grain size, mean values and distributions when the anisotropy level is moderated. However, when the anisotropy level increases, the anisotropic formulation leads to more physical predictions in terms of grain morphology, global surface energy evolution and multiple junction equilibrium. 

The goal of this work is to criticize the capacity of the isotropic and anisotropic formulations to model GG in a real material, here a 316L austenitic stainless steel, in terms of mean grain size, grain size distributions and mean GB properties. We compare the effect of the initial microstructure using statistically representative Laguerre-Voronoï tessellation \cite{Hitti2012} and digital twin microstructures from EBSD data. The effect of the GB energy definition is illustrated with two different frameworks: a 1-parameter, well-known as the Read-Shockley formulation \cite{ReadShockley}; and a 5-parameter one using the the GB5DOF code proposed in \cite{bulatov2014grain}. The effect of the GB mobility description using an isotropic and a Sigmoidal model \cite{humphreys1997unified} is also discussed. The paper is organized as follows. First, in section~\ref{sec:FrameworkDevelopment}, crystallographic definitions, LS treatments and FE-LS formulations are presented briefly. The methodology to estimate the GB reduced mobility from experimental data is presented in section~\ref{sec:Material}. Finally, the results using the isotropic and anisotropic formulations are compared using statistically representative Laguerre-Voronoï tessellations (Section~\ref{sec:StatCase}), immersed microstructures with heterogeneous GB properties (Section~\ref{sec:PX}) and immersed microstructures with anisotropic GB energy using the GB5DOF code \cite{bulatov2014grain} (Section~\ref{sec:PXGB5DOF}).

\section{The numerical formulation}
\label{sec:FrameworkDevelopment}

The LS method was firstly proposed in \cite{Osher1988} to describe curvature flow problems, enhanced later for evolving multiple junctions \cite{Merriman1994, Zhao1996} and applied to recrystallization and grain growth in \cite{Bernacki2008, Bernacki2011}. The principle for modeling polycrystals is the following: grains are defined by LS functions $\phi$ in the space $\Omega$  

\begin{align}
  \left\{
  \begin{array}{l}
    \phi(X) = \pm d(X,\Gamma), \quad X \in \Omega, \quad \Gamma=\partial G \\
    \phi(X \in \Omega) = 0 \quad \rightleftharpoons \quad X \in \Gamma,
\end{array}
\label{eqn:LS}
\right .
\end{align}
more precisely the grain interface $\Gamma$ is described by the zero-isovalue of the corresponding $\phi$ function. In Eq.~\ref{eqn:LS}, $d$ is the signed Euclidean distance to $\Gamma$ and $\phi$ is classically chosen as positive inside the grain and negative outside. The dynamics of the interface is studied by following the evolution of the LS field. When the interface evolution is characterized by a velocity field $\vec{v}$, its movement can be obtained through the resolution of the following transport equation \cite{Osher1988}:     

\begin{align}
    \dfrac{\partial \phi}{\partial t} + \vec{v} \cdot \vec{\nabla} \phi = 0.
    \label{eqn:transport}
\end{align}

Classically, one LS function is used to describe one grain and Eq.~\ref{eqn:transport} is solved for each grain to describe the grain boundary network evolution. However, when the number of grains, $N_G$, increases, one may use a graph coloring/recoloring strategy \cite{scholtes2015new} in order to limit drastically the number of involved LS functions $\Phi = \{ \phi_i, i=1, \dotsc, N \}$ with $N \ll N_G$. Two more treatments are necessary. Firstly, the LS functions are generally reinitialized at each time step in order to keep their initial metric property when they are initially built as distance functions to the grain interface as proposed in Eq.\ref{eqn:LS}:  

\begin{align}
    \| \nabla \phi \| = 1.
    \label{eqn:normLS}
\end{align}
In the proposed numerical framework, the algorithm developed in \cite{Shakoor2015amm} is used.
Secondly, the LS evolutions may not preserve the impenetrability/overlapping constraints leading to potential overlaps/voids between grain interfaces at multiple junctions. The solution proposed in \cite{Merriman1994} and largely used in LS context \cite{Bernacki2011} is adopted.

The main interest of this global numerical front-capturing framework lies in its ability to define different physical phenomena when they are encapsulated in the velocity field and to deal easily with topological events such as grain disappearance. In the next section different formulations of the GB velocity and the subsequent FE resolution are presented.

\subsection{GB velocity formulation}

The isotropic formulation uses a homogeneous GB energy and mobility \cite{bernacki2011level}, the velocity field is thus defined as 

\begin{align}
    \vec{v} = - \mu \gamma \kappa \vec{n},
    \label{eqn:VelFieldHomo}
\end{align}
where $\kappa$ is the mean curvature of the boundary in $2D$ and the trace of the curvature tensor in $3D$, and $n$ the outward unit normal to the boundary. By verifying Eq.\ref{eqn:normLS} and assuming the LS function to be positive inside the corresponding grains and negative outside, the unitary normal and so the mean curvature may be defined as $\vec{n} = - \vec{\nabla} \phi$ and $\kappa = \vec{\nabla}\cdot\vec{n}=- \Delta \phi$ and, then the velocity in Eq.~\ref{eqn:VelFieldHomo} may be written:
\begin{align}
    \vec{v} = - \mu \gamma \Delta \phi \vec{\nabla} \phi.
    \label{eqn:VelFieldHomoLS}
\end{align} 

\begin{figure}[h!]
  \centering
  \includegraphics[scale=0.15]{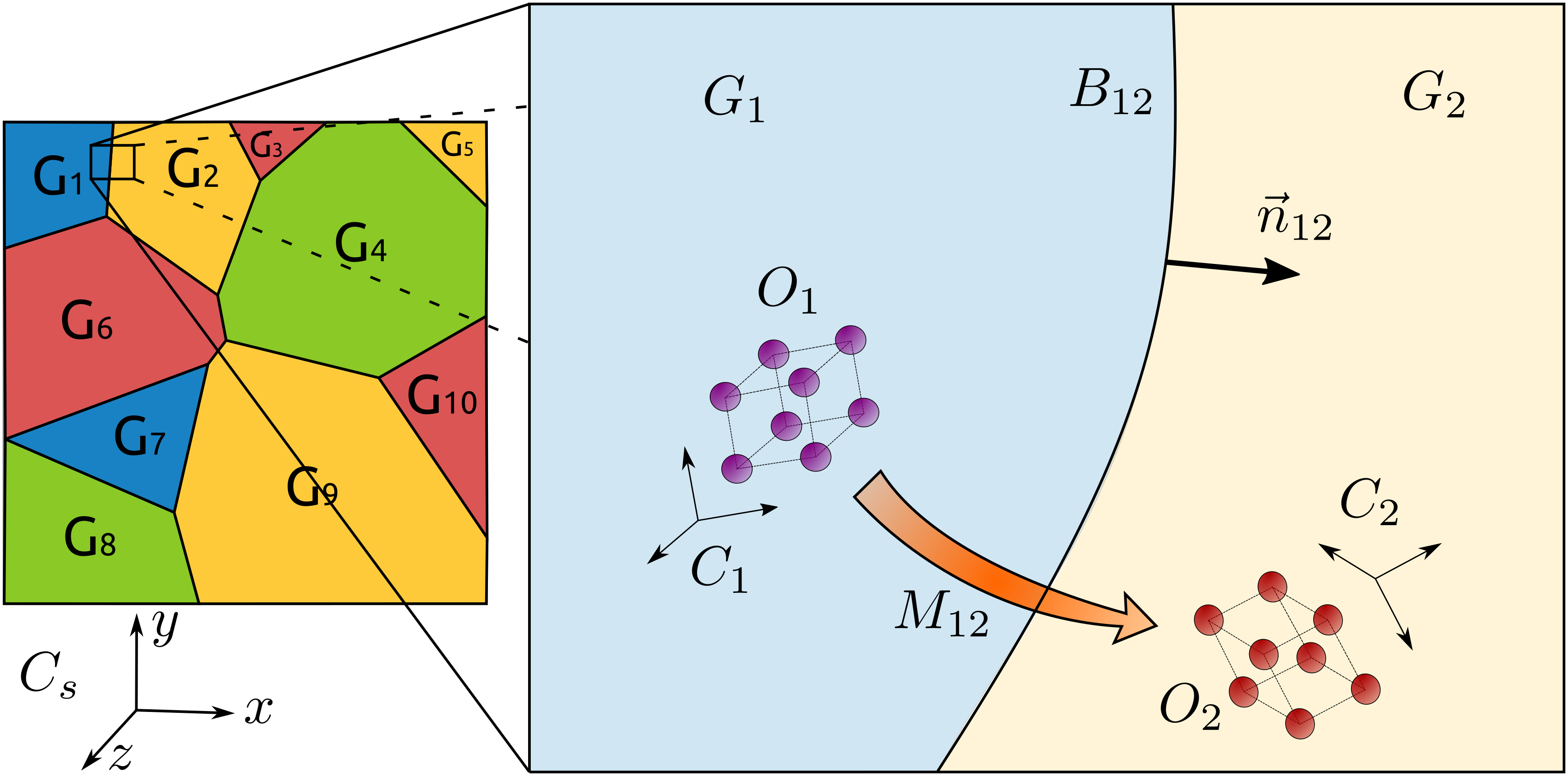}
  \caption{Scheme depicting one GB and its parameters. Image available online at {Flickr} (\url{https://flic.kr/p/2m5JQkz}, Uploaded on June 15, 2021
) licensed under {CC BY 2.0} (\url{https://creativecommons.org/licenses/by/2.0/}, Uploaded on June 15, 2021). Title: 10GGBParam. Author: Brayan Murgas.}\label{fig:GBparameters}
\end{figure}

At the mesoscopic scale, a GB, $B_{ij}$, between grains $G_i$ and $G_j$, is characterized by its morphology and its crystallographic properties which may be summarized by a tuple $B_{ij} = (M_{ij},n_{ij})$ with two shape parameters describing the interfaces through the unitary-outward normal direction $n_{ij}$, and three crystallographic parameters describing the orientation relationship between the two adjacent grains known as the misorientation tensor $M_{ij}$ (see Figure~\ref{fig:GBparameters}). The misorientation is frequently defined with the axis-angle parameterization, i.e. $M_{ij}(a_i,\theta)$, where $a_i$ is the misorientation axis and $\theta$ the disorientation \cite{morawiec2003orientations}. The two quantities of interest, the GB energy $\gamma$ and GB mobility $\mu$, must then be seen as functions from the GB space $\mathcal{B}$ to $\mathbb{R}^{+}$. 

The anisotropic formulation was initially developed using thermodynamics and differential geometry in \cite{FAUSTY202128} and was improved in \cite{ma14143883} in order to consider heterogeneous GB mobility. Both the GB normal and misorientation are taken into account and an intrinsic torque term is present:
\begin{align}
    v = \mu (M) \left( \mathbb{P} \vec{\nabla} \gamma (M,n) \cdot \vec{\nabla} \phi - \left( \vec{\nabla}_{\vec{n}}\vec{\nabla}_{\vec{n}}\gamma (M,n) + \gamma (M,n) \mathbb{I} \right) : \mathbb{K} \right) \vec{\nabla} \phi 
    \label{eqn:VelFieldAniso5LS}
\end{align}
where $\mathbb{I}$ is the unitary matrix, $\mathbb{P}=\mathbb{I}-\vec{n}\otimes\vec{n}$ is the tangential projection tensor, $\vec{\nabla}_{\vec{n}}$ the surface gradient, and $\mathbb{K}=\vec{\nabla}\vec{n}=\vec{\nabla}\vec{\nabla}\phi$ is the curvature tensor. The term $\Gamma(M,n) = \vec{\nabla}_{\vec{n}}\vec{\nabla}_{\vec{n}}\gamma + \gamma \mathbb{I}$ is a tensorial diffusion term, known as GB stiffness tensor \cite{ABDELJAWAD2018440, DU2007467}. The term $ \mathbb{P} \vec{\nabla} \gamma \cdot \vec{\nabla} \phi$ in Eq.~\ref{eqn:VelFieldAniso5LS} should be null in the grain interfaces. However, the front-capturing nature of the LS approach which consists to solve Eq.\ref{eqn:transport} at the GB network and in its vicinity, requires to consider this term which could be non-null around the interfaces. This stabilization term is then totally correlated to the front-capturing nature of the LS approach. The term $\Gamma(M,n)$ is subject of recent studies for twin boundaries (TB) $\Sigma 3$, $\Sigma 5$, $\Sigma 7$, $\Sigma 9$ and $\Sigma 11$ \cite{ABDELJAWAD2018440, MOORE2021117220}. In the present work, the torque term $\vec{\nabla}_{\vec{n}}\vec{\nabla}_{\vec{n}}\gamma$ is neglected but the GB energy still depends on the GB misorientation and inclination, the kinetic equation could be simplified as:
\begin{align}
    v = \mu (M) ( \mathbb{P} \vec{\nabla} \gamma (M,n) \cdot \vec{\nabla} \phi - \gamma (M,n) \Delta \phi ) \vec{\nabla} \phi.
    \label{eqn:VelFieldAnisoLS}
\end{align}

Inserting the kinetic equations \ref{eqn:VelFieldHomoLS}, and \ref{eqn:VelFieldAnisoLS} into Eq.~\ref{eqn:transport}, leads to the weak formulation of the isotropic (Iso) and anisotropic (Aniso) formulation \cite{ma14143883}
\begin{align}
    \int_{\Omega} \dfrac{\partial \phi}{\partial t} \varphi d \Omega + \int_{\Omega} \mu \gamma \vec{\nabla}  \varphi \cdot \vec{\nabla} \phi d \Omega - \int_{\partial \Omega} \mu \gamma \varphi \vec{\nabla} \phi \cdot \vec{n} d (\partial \Omega) = 0,
\label{eqn:WeakFormClassic}
\end{align}
and 
\begin{align}
\begin{split}
    \int_{\Omega} \dfrac{\partial \phi}{\partial t} \varphi d \Omega + \int_{\Omega} \mu (M) \gamma (M) \vec{\nabla}  \varphi \cdot \vec{\nabla} \phi d \Omega - \int_{\partial \Omega} \mu (M) \gamma (M) \varphi \vec{\nabla} \phi \cdot \vec{n} d (\partial \Omega) + \\
    \int_{\Omega} \mu (M) ( \mathbb{P} \cdot \vec{\nabla} \gamma (M) + \vec{\nabla} \gamma (M) ) \varphi \vec{\nabla} \phi d \Omega + \int_{\Omega} \gamma (M) \vec{\nabla} \mu (M) \cdot \vec{\nabla} \phi \varphi d \Omega = 0,
\end{split}
\label{eqn:WeakFormAnisoSimp}
\end{align}
respectively.
\medbreak

If the properties are homogeneous both formulations are then equivalent. The main question is the capability of these two numerical models to reproduce experimental evolution assessed through EBSD data. The next section is dedicated to the parameters identification.  

\clearpage

\section{Parameters Identification}
\label{sec:Material}

\subsection{Material characterisation}

The chemical composition of the 316L stainless steel is reported in Table~\ref{tab:316LCompo}. The samples were machined in the form of rectangular parallelepipeds of $8.5mmx8.5mmx12mm$. The samples were then annealed at $1050\degree C$ during $30min$, $1h$ and $2h$. Afterwards, the samples were prepared for EBSD characterization. The preparation consisted of mechanical polishing, followed by fine polishing and finally electrolitic polishing, the details of the polishing are listed in Table~\ref{tab:316Lpolish}.

\begin{table}[H]
\centering
\begin{tabular}{l  *{9}{c}}
      \toprule
Elem. Wgt\%              &	Fe		&	Si		&	P		&	S		&	Cr	&		Mn	&	Ni		&	Mo	&	N	\\
      \midrule
Min                &	bal.	&	-		&	-		&	-		&	16.0	&	-		&	10.0	&	2.0		&	-	\\

Real            &	65.85	&	0.65	&	0.01	&	0.14	&	18.02	&	1.13	&	11.65	&	2.55	&		\\

Max                &	bal.	&	0.75	&	0.045	&	0.03	&	18.0	&	2.0		&	14.0	&	3.0		&	0.1	\\
      \bottomrule
\end{tabular}
\caption{ Chemical composition of the 316L stainless steel (Weight percent). }
\label{tab:316LCompo} 
\end{table}

\begin{table}[H]
\centering
\small

\caption{ Polishing procedure applied to the 316L stainless steel samples. Plate and tower rate are the parameter of the used automatic polisher. }
\label{tab:316Lpolish} 
\begin{tabular}{ccccc}
\toprule
\textbf{Abrasive}              & \textbf{time} &\textbf{Plate} & \textbf{Tower} & \textbf{Force}  \\
                               & \textbf{[s]} &\textbf{[rpm]} & \textbf{[rpm]} & \textbf{[dN]}  \\
\midrule
320 SiC paper       &    60    &    250    &    150    &    2.5    \\
600 SiC paper       &    60    &    250    &    150    &    2.5    \\
1200 SiC paper      &    60    &    250    &    150    &    2.5    \\
2400 SiC paper      &    60    &    150    &    100    &     1     \\ \midrule
HSV - 3µm Diamond            &    120    &    150    &    100    &     2     \\
solution 0.12mL/8s           &           &           &           &           \\ \midrule
electrolytic polishing       &    30s    &    30V    &    Electrolyte A2 (Struers)    &           \\
\bottomrule
\end{tabular}

\end{table}

Microstructures were analyzed at the center of the sample using a TESCAN FERA 3 Field Emission Gun Scanning Electron Microscope (FEGSEM), equipped with a Symmetry EBSD detector from the Oxford company. The EBSD map at $t=0h$ has a size of $1.138 \ mm \times 0.856 \ mm$ and was acquired with a constant step size of $1.5 \ \mu m$. The other three EBSD maps at $t=30 \ min, \ 1 \ h, \ 2 \ h$ have a size of $1.518 \ mm \times 1.142 \ mm$ and were acquired with a constant step size of $2 \ \mu m$. Grain boundaries have a disorientation above 5 degrees ($\theta>5 \degree$) and $\Sigma 3$ twin boundaries have a misorientation axis $<$$111$$> \pm 5 \degree$ and $\theta=60\pm 5 \degree$.

\begin{figure}[H]
  \centering
  \includegraphics[scale=0.75]{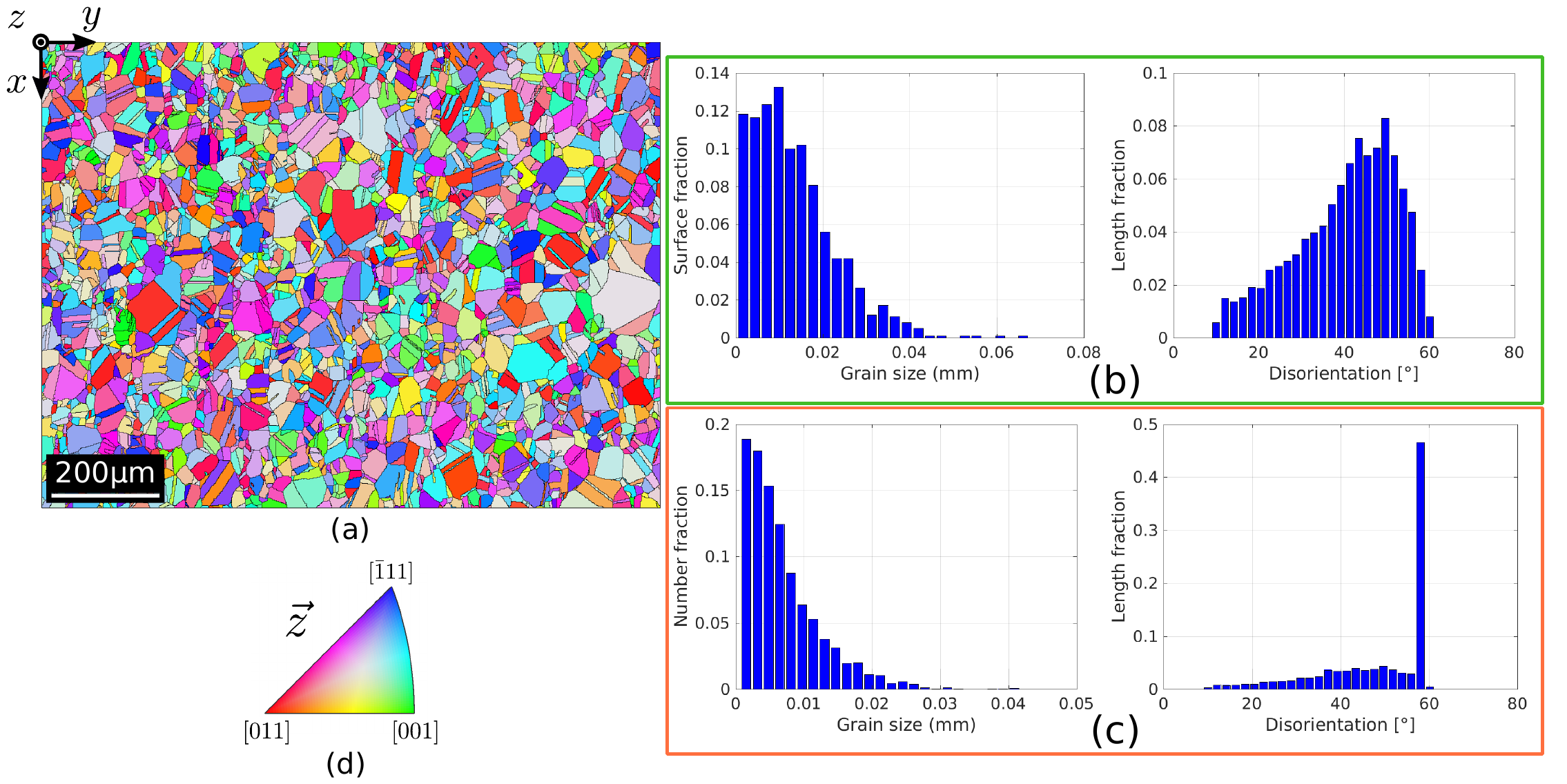}
  \caption{Initial microstructure properties determined by EBSD measurements. (a) IPF-z map. (b) Grain size distribution and DDF of grain boundaries excluding $\Sigma 3$ TBs. (c) Grain size distribution as measured in 2D sections and DDF of grain boundaries including $\Sigma 3$ TBs; the sharp peak on the DDF at $60 \degree$ corresponds to $\Sigma 3$ TBs. (d) Standar triangle used to color the orientation maps IPF-Z (indicating which crystallographic direction is lying parallel to the direction perpendicular to the scanned section). }
  \label{fig:316LInit}
\end{figure}

The main properties of the initial microstructure are reported in Figure~\ref{fig:316LInit}. Figure~\ref{fig:316LInit}(b) illustrates the grain size and disorientation distribution ignoring $\Sigma 3$ twin boundaries (TB). The grain size is defined as an equivalent radius, $R = \sqrt{S/\pi}$, where $S$ is the grain area. The microstructure consists in equiaxed grains with an arithmetic mean radius of $15\ \mu m$, and few bigger grains with a radius around $60\ \mu m$. Additionally, the microstructure presents a Mackenzie-like disorientation distribution function (DDF) typical of random grain orientations. On the other hand, if $\Sigma 3$ TBs are considered, the DDF presents an additional sharp peak at a disorientation angle $\theta = 60^\circ$, which comes from the TBs and then constitute a strong source of anisotropy with regards to GB properties (see Figure~\ref{fig:316LInit}(c)).

\begin{figure}[h!]
  \centering
  \includegraphics[scale=1.3]{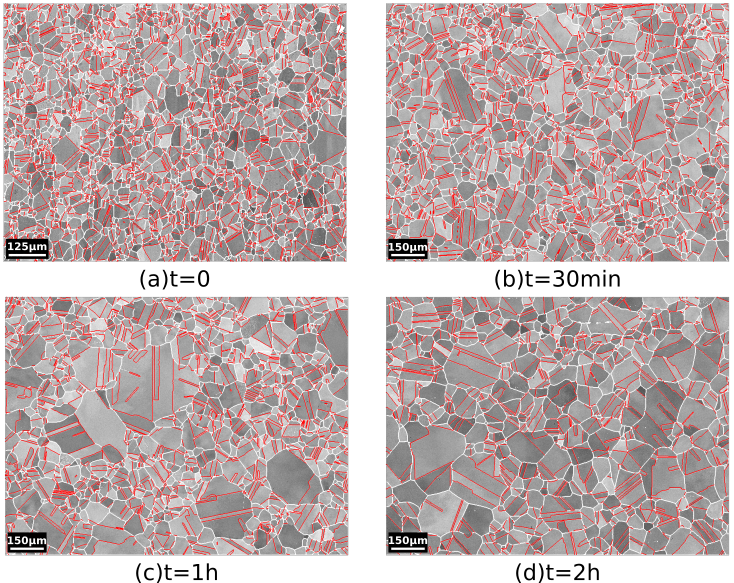}
  \caption{Annealing at $1050\degree C$: band contrast map of the microstructure of 316L steel at (a)t=0s, (b)t=30min, (c)t= 1h, and (d)t= 2h. Grain boundaries are depicted in white and $\Sigma 3$ TBs in red.}
  \label{fig:BCmaps}
\end{figure}

\begin{figure}[H]
  \centering
  \includegraphics[scale=0.8]{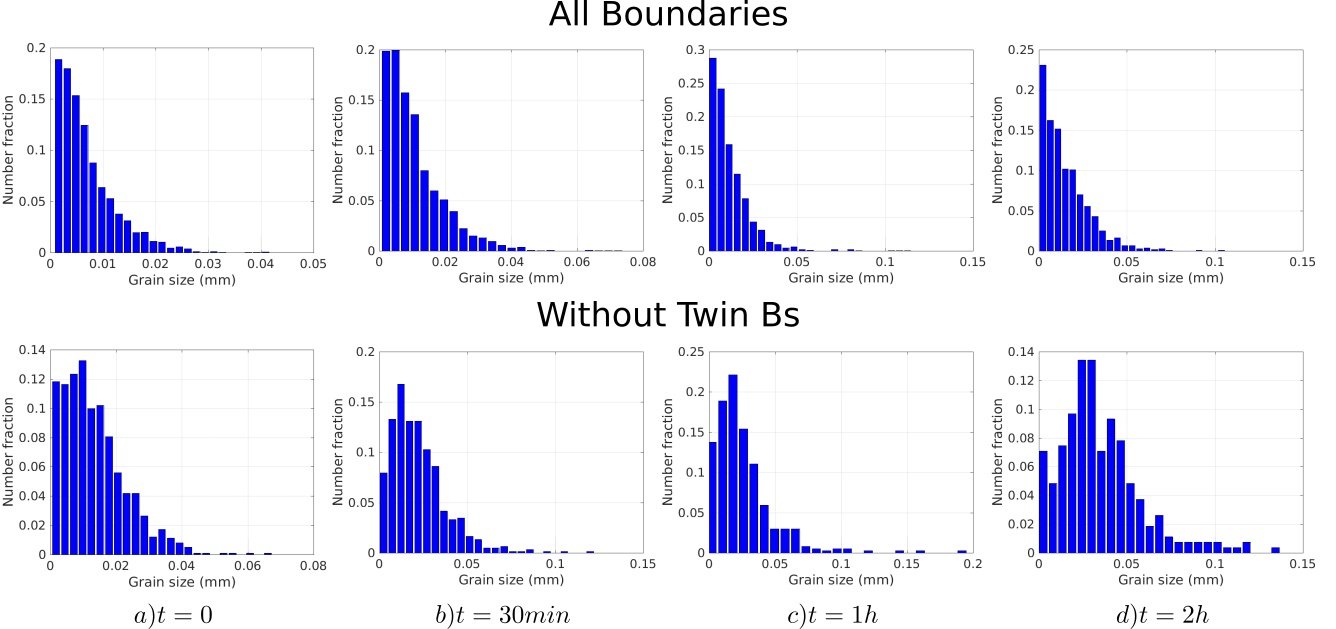}
  \caption{From left to right: evolution of the grain size histograms (in number) at t=0s, t=30min, t=1h and t=2h. Top: All boundaries are considered. Bottom: $\Sigma 3$ twin boundaries are excluded.}
  \label{fig:GSDist}
\end{figure}

\begin{figure}[H]
  \centering
  \includegraphics[scale=0.8]{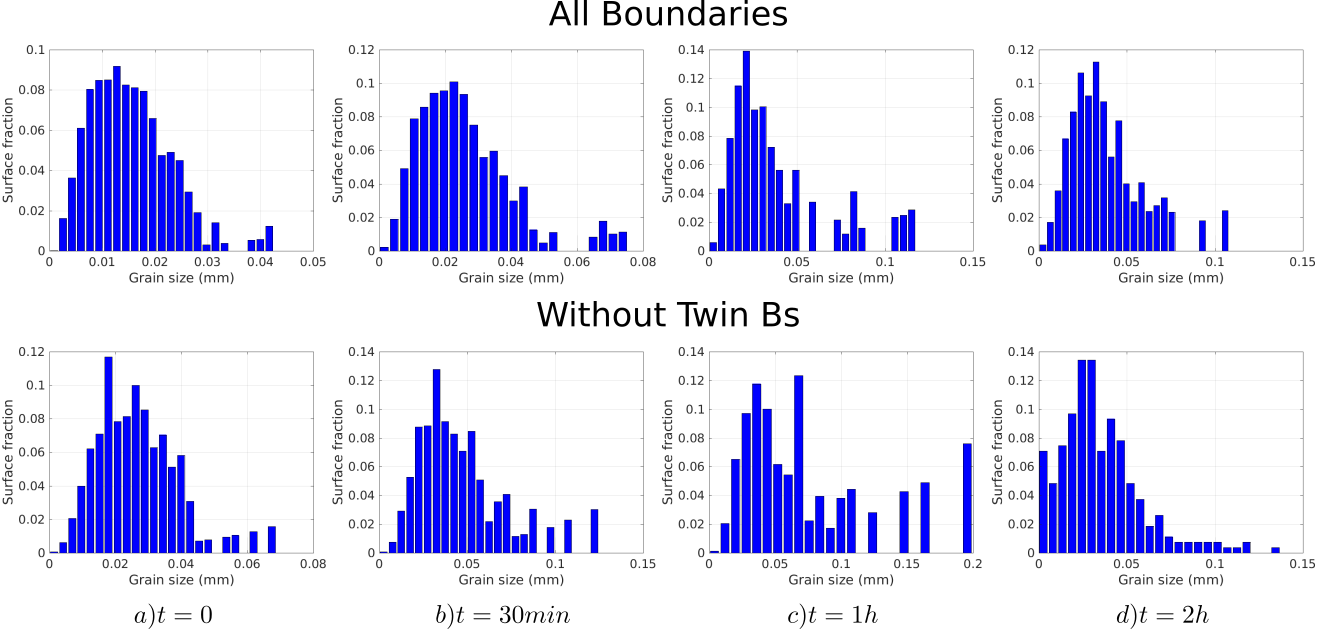}
  \caption{From left to right: evolution of the grain size histograms (in surface) at t=0s, t=30min, t=1h and t=2h. Top: All boundaries are considered. Bottom: $\Sigma 3$ twin boundaries are excluded.}
  \label{fig:GSDistS}
\end{figure}

Figures~\ref{fig:BCmaps}, \ref{fig:GSDist} and \ref{fig:GSDistS} show the band contrast maps and the grain size distributions at $t=0s, \ 30 \ min, \ 1 \ h, \ 2 \ h$.  Based on Figures~\ref{fig:BCmaps} and \ref{fig:GSDist}, the evolution of the microstructure seems to mostly proceed by normal grain growth (NGG) but the surface grain size distribution shows that the microstructure has a bimodal population of grains (see Figure~\ref{fig:GSDistS}). However, some of the grains can reach an equivalent diameter above $0.1\ mm$, much larger than the average grain size.

\clearpage

\clearpage

\subsection{Estimation of the average grain boundary mobility based on the Burke and Turnbull GG method \cite{BURKE1952220}} 
\label{ssec:BurTurn}

\begin{figure}[H]
  \centering
  \includegraphics[scale=0.35]{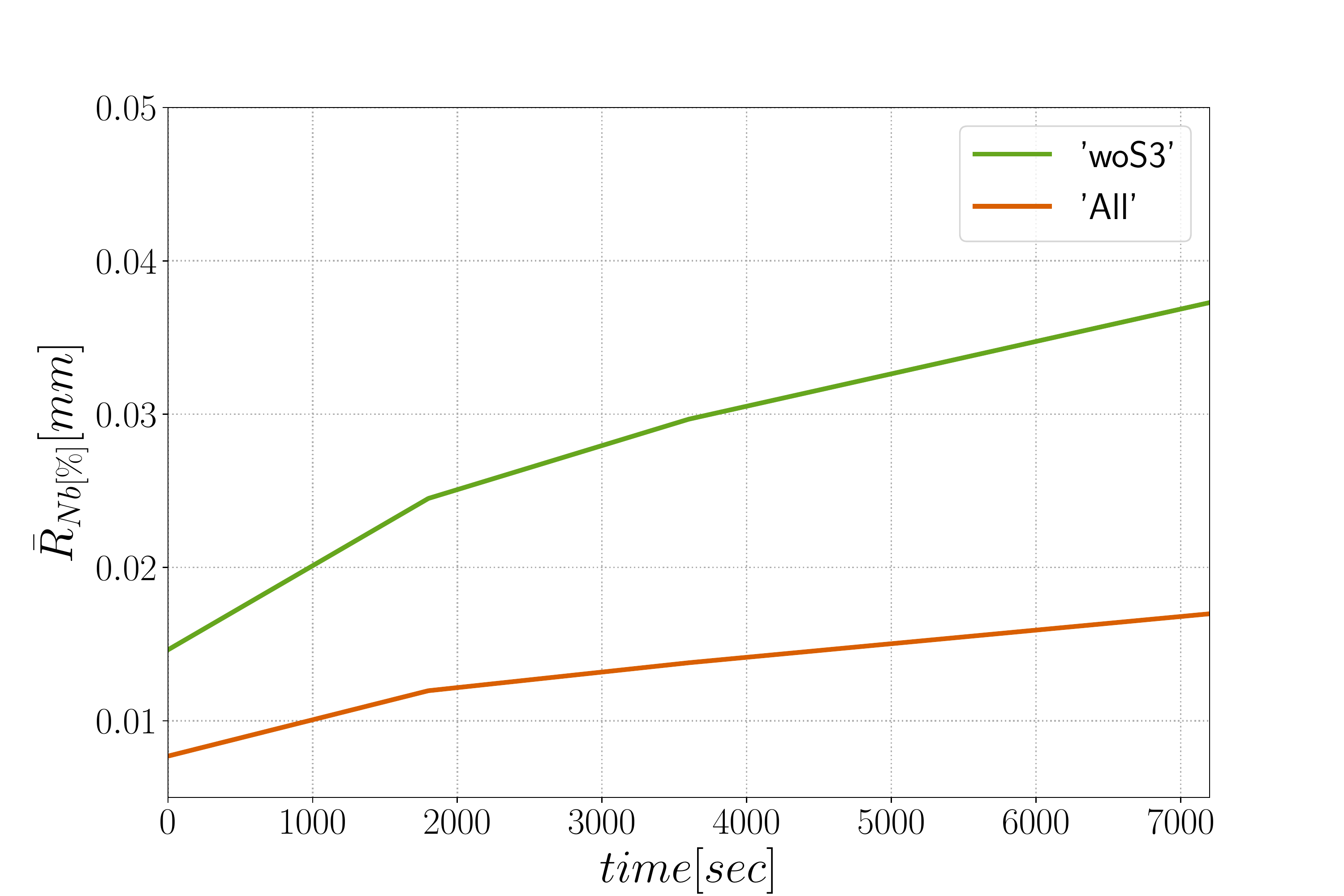}
  \caption{Mean grain radius evolution at $1050\degree C$ from experimental data measured in 2D sections by taking into account all boundaries (in orange) and without TBs (in green). The outer grains which share a boundary with the image borders are not taken into account in the analysis.}
  \label{fig:MeanR}
\end{figure}

In order to compute the average mobility necessary to run full-field simulations, the evolution of the arithmetic mean grain radius $\bar{R}_{Nb}$ must be known. Figure~\ref{fig:MeanR} shows the evolution of $\bar{R}_{Nb}$ as a function of the annealing time. Using the methodology discussed in \cite{CRUZFABIANO2014305,alvarado2021dissolution}, one can obtain an average reduced mobility $\mu \gamma$ using the Burke and Turnbull model \cite{BURKE1952220}. This model, where topological and neighboring effects are neglected, is based on five main assumptions: the driving pressure is proportional to the mean curvature, grains are equixaed, the GB mobility and energy are isotropic, the annealing temperature is constant and no second phase particles are present in the material. In this context, one can obtain a simplified equation describing the mean radius evolution: 

\begin{align}
    \bar{R}_{Nb}(t)^2 - \bar{R}_{Nb}(t=0)^2 = \dfrac{1}{2} \mu \gamma t.
    \label{eqn:RBurkrTurnbull}
\end{align}
This methodology has been used in \cite{CRUZFABIANO2014305, agnoli2014development, maire2016, alvarado2021level,alvarado2021dissolution} assuming general grain boundaries with homogeneous GB energy and mobility. From the evolution of $\bar{R}_{Nb}$ in Figure~\ref{fig:MeanR} (excluding the  $\Sigma 3$ TBs), one can then obtain a first approximation of the product $\mu \gamma$ for the general boundaries at $1050\degree C$. This approximation will be used for the $\mu\gamma$ definition in isotropic simulations. Nevertheless, as illustrated by the second orange curve in Figure~\ref{fig:MeanR}, when $\Sigma 3$ TBs are considered in the analysis, grains are of course smaller, but they also grow much slower, with a direct impact on the apparent reduced mobility. This slow evolution can be produced by the strong anisotropy brought by special GBs in the global grain boundary network migration. Different ways to improve the description of the reduced mobility and their impacts in the results are discussed in the following.   

In the following sections, \textit{general boundaries} make reference to the case without $\Sigma 3$ TBs and the case with $\Sigma 3$ is referred to as \textit{all boundaries}.

\section{Statistical cases}
\label{sec:StatCase}

\begin{figure}[H]
  
  \centering
  \begin{subfigure}[b]{0.48\textwidth}
    \centering
    \includegraphics[width=0.8\textwidth]{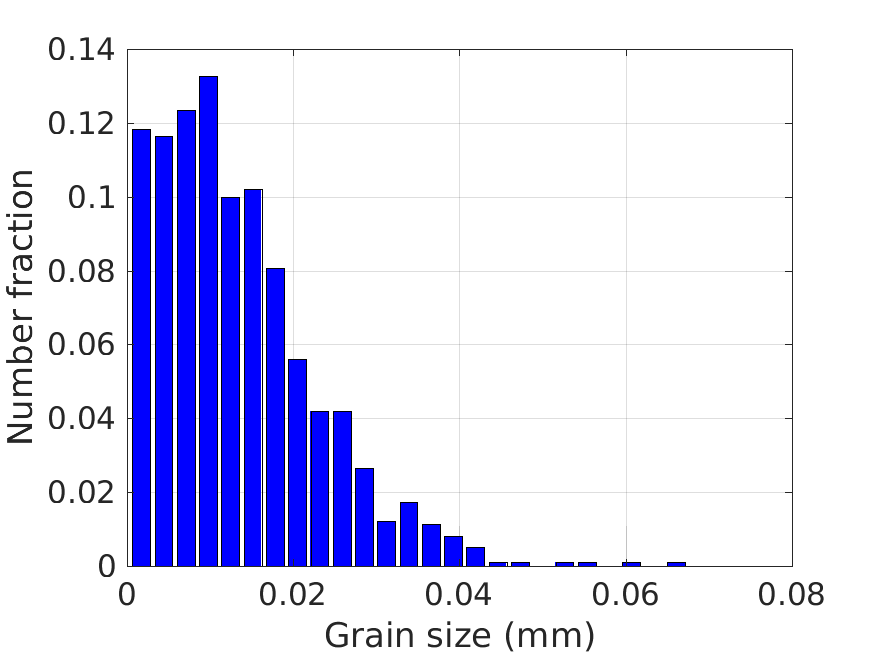}
    \caption{Excluding TBs}
  \end{subfigure}
  \begin{subfigure}[b]{0.48\textwidth}
    \centering
    \includegraphics[width=0.8\textwidth]{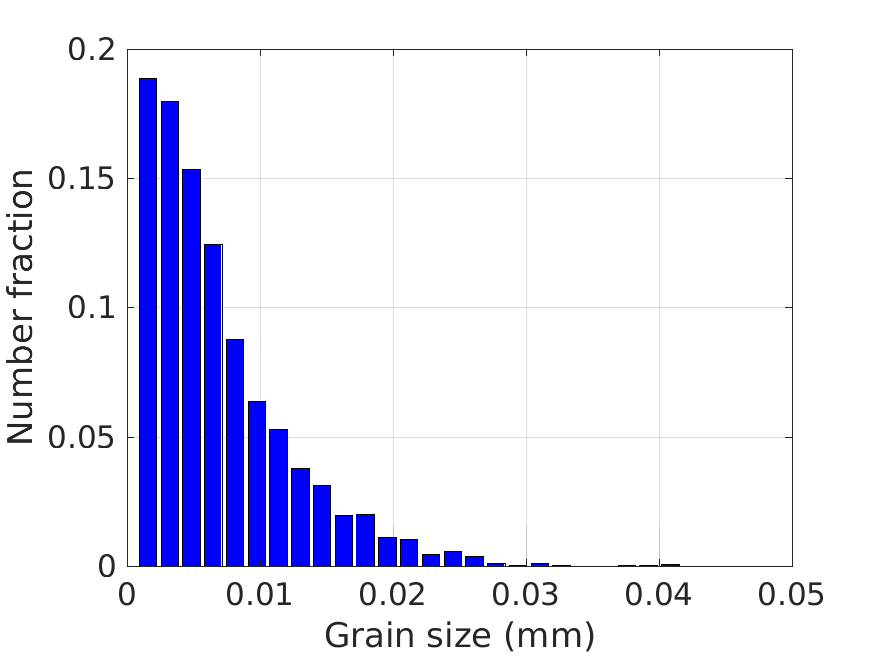}
    \caption{Including TBs}
    \label{fig:InitGSDwTB}
  \end{subfigure}
  \caption{Initial grain size distributions (a) excluding TBs and (b) all grain boundaries obtained from the initial EBSD map shown in Figure~\ref{fig:316LInit}.}
  \label{fig:InitGSD}
\end{figure}

In this section, the 2D GB network is created from the initial experimental grain size distribution shown in Figure~\ref{fig:InitGSD}. The square domain has a length $L=2.0 mm$ and grains are generated using a Laguerre-Voronoi tessellation \cite{Hitti2012} based on an optimized sphere packing algorithm \cite{hitti2013optimized}. Anisotropic remeshing is used with a refinement close to the interfaces, the mesh size in the tangential direction (and far from the interface) is set to $h_{max}=5\ \mu m$ and in the normal direction $h_{min}=1\ \mu m$, with transition distances set to $\phi_{min}=1.2\ \mu m$ and $\phi_{max}=5\ \mu m$ (see \cite{Bernacki2009, ROUX201332, ma14143883} for more details concerning the remeshing procedure and parameters). The time step is set to $\Delta t=10\ s$. The orientation field was generated randomly from the grain orientations measured by EBSD in the initial microstructure (Figure~\ref{fig:316LInit}a). The first part studies grain boundaries without $\Sigma 3$ TB, (Section~\ref{ssec:StatCaseGeneral}). In the second part, $\Sigma 3$ TBs are included in the analysis, (Section~\ref{ssec:StatCaseAll}).

\clearpage

The interfacial energy and average GB properties are computed as 
\begin{equation}
	E_{ \Gamma } = \dfrac{1}{2} \sum_i \sum_{e \in \mathcal{T}} l_e (\phi_i) \gamma_e \text{ and } \bar{x} = \dfrac{1}{2 L_{\Gamma}} \sum_i \sum_{e \in \mathcal{T}} l_e (\phi_i) x_e,
\end{equation}
where $\mathcal{T}$ is the set of elements in the FE mesh, $l_e$ the length of the LS zero iso-values existing in the element $e$ and $i$ refers to the number of LS functions, $L_{\Gamma}$ the total length of the GB network $\Gamma$, and $x_e$ the GB property of the element $e$. 

\subsection{Statistical case with general boundaries}
\label{ssec:StatCaseGeneral}

\begin{figure}[H]
  
  \centering
  \begin{subfigure}[b]{0.48\textwidth}
    \centering
    \includegraphics[scale=0.8]{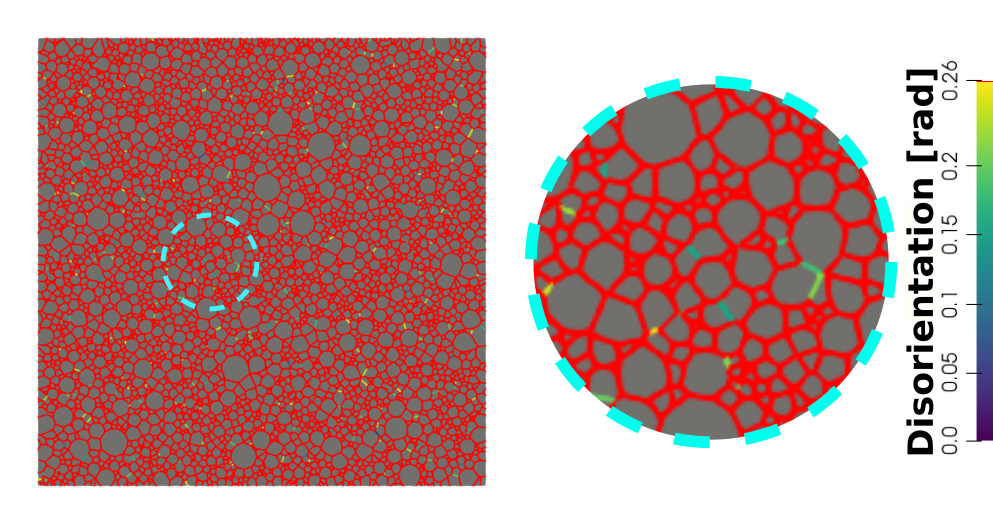}
    \caption{Microstructure disorientation}
    \label{fig:StatMetho}
  \end{subfigure}
  \begin{subfigure}[b]{0.48\textwidth}
    \centering
    \includegraphics[scale=0.25]{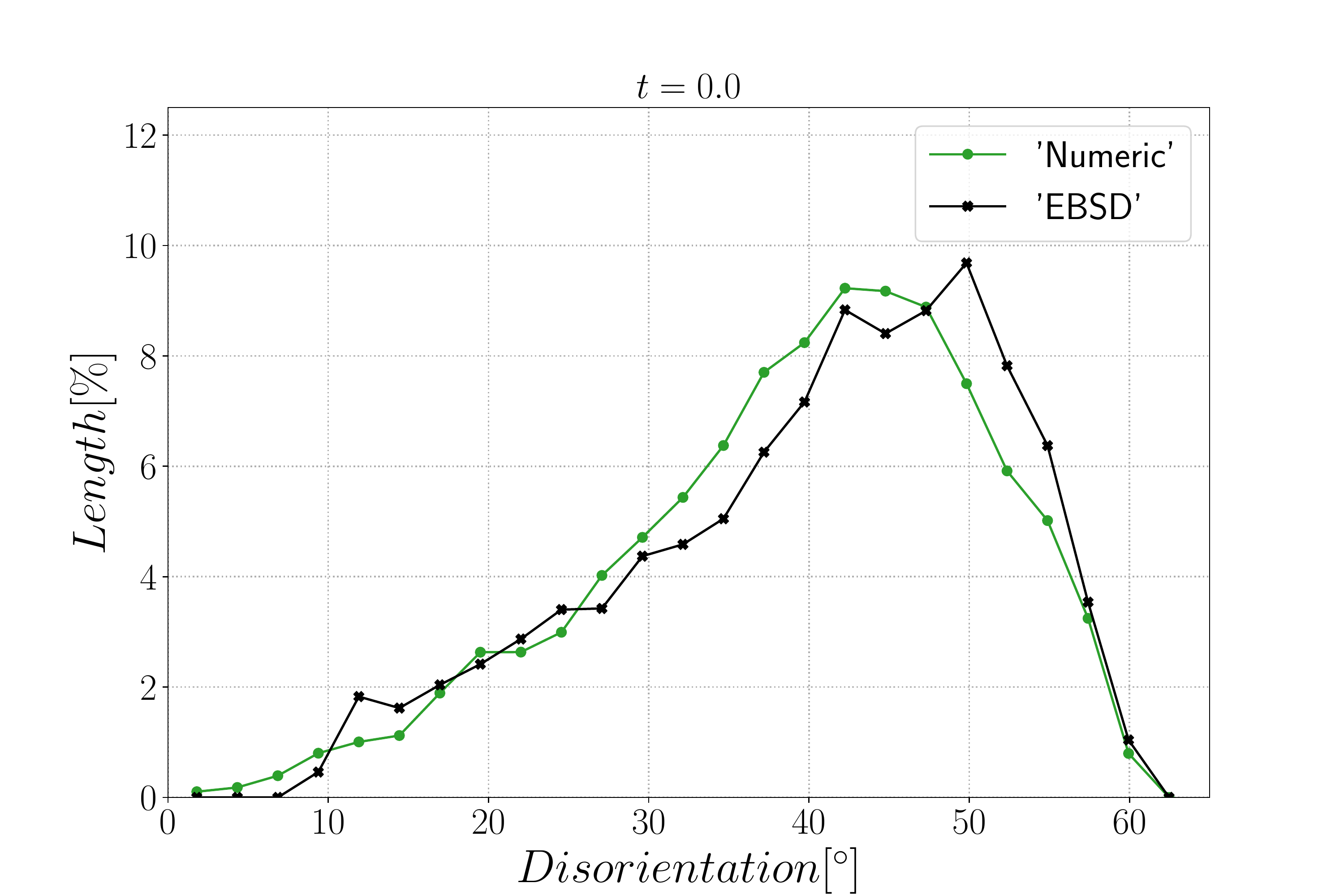}
    \caption{DDF}
    \label{fig:StatwoS3Dist}
  \end{subfigure}
  \caption{Initial (a) microstructure disorientation with a cyan circle which represents the zone shown on the right and (b) the disorientation distribution.}
\end{figure}

The first case with general boundaries is composed of $N_G = 4397$ initial grains. Figures~\ref{fig:StatMetho} and \ref{fig:StatwoS3Dist} show the initial GB disorientation and the initial DDF distribution. Most of the interfaces have a disorientation higher than $15 \degree$ due to the random generation of orientations that leads to a Mackenzie-like DDF, see Figure~\ref{fig:StatwoS3Dist}.

The GB energy and mobility are then defined as being disorientation dependent, using a Read-Shockley (RS) \cite{ReadShockley} and a Sigmoidal (S) function \cite{humphreys1997unified}, respectively: 
\begin{align}
  \left\{
  \begin{array}{l}
    \gamma(\theta) = \gamma_{max} \frac{\theta}{\theta_0} \left( 1 - ln \left( \frac{\theta}{\theta_0} \right) \right), \ \theta < \theta_0 \\
    \gamma_{max}, \ \theta \ge \theta_0
\end{array}
\label{eqn:Gamma}
\right .
\end{align}
and
\begin{align}
  \mu(\theta)=\mu_{max} \left( 1 - exp \left( -5 \left(\frac{ \theta}{\theta_0} \right)^4 \right) \right),  \label{eqn:Mob}
\end{align}
where $\theta$ is the disorientation, $\mu_{max}$ and $\gamma_{max}$ are the GB mobility and energy of general high angle grain boundaries (HAGBs). $\theta_0=15 \degree $ is the disorientation defining the transition from a low angle grain boundary (LAGB) to a HAGB. The maximal value of GB energy is set to $\gamma_{max}=6 \times 10^{-7} J\cdot mm^{-2} $ which is typical for stainless steel \cite{CRUZFABIANO2014305, ratanaphan2019atomistic}. The value of general HAGB mobility was computed using the methodology presented in section~\ref{ssec:BurTurn} and is fixed at $\mu_{max}=0.476$ $mm^4\cdot J^{-1}\cdot s^{-1} $ for both Isotropic and Anisotropic formulations. 

\begin{figure}[H]
  
  \centering
  \begin{subfigure}[c]{0.495\textwidth}
    \centering
    \includegraphics[scale=0.295]{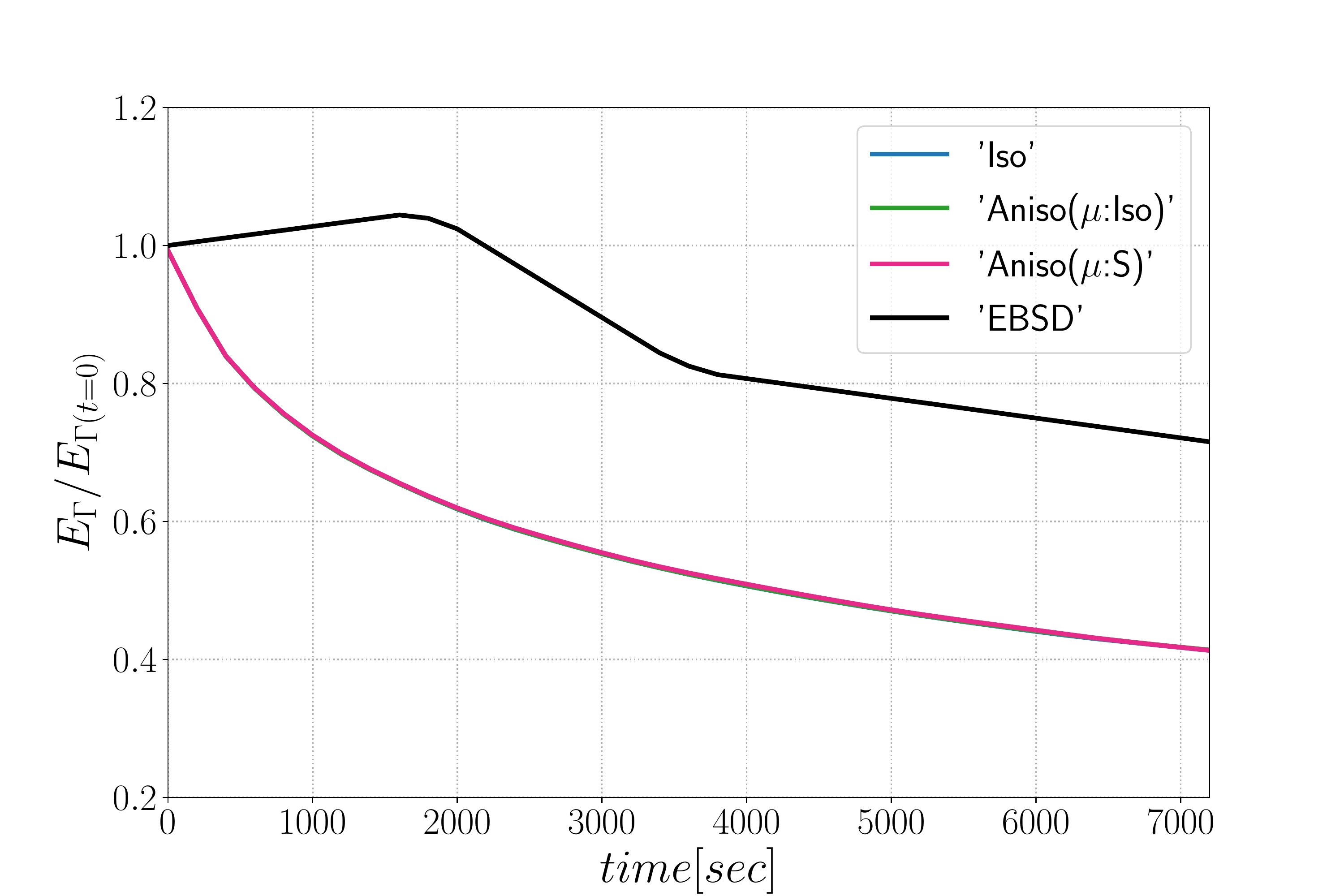}
    \caption{$E_{\Gamma}/E_{\Gamma}(t=0)=f(t)$}
  \end{subfigure} 
  \begin{subfigure}[c]{0.495\textwidth}
    \centering
    \includegraphics[scale=0.295]{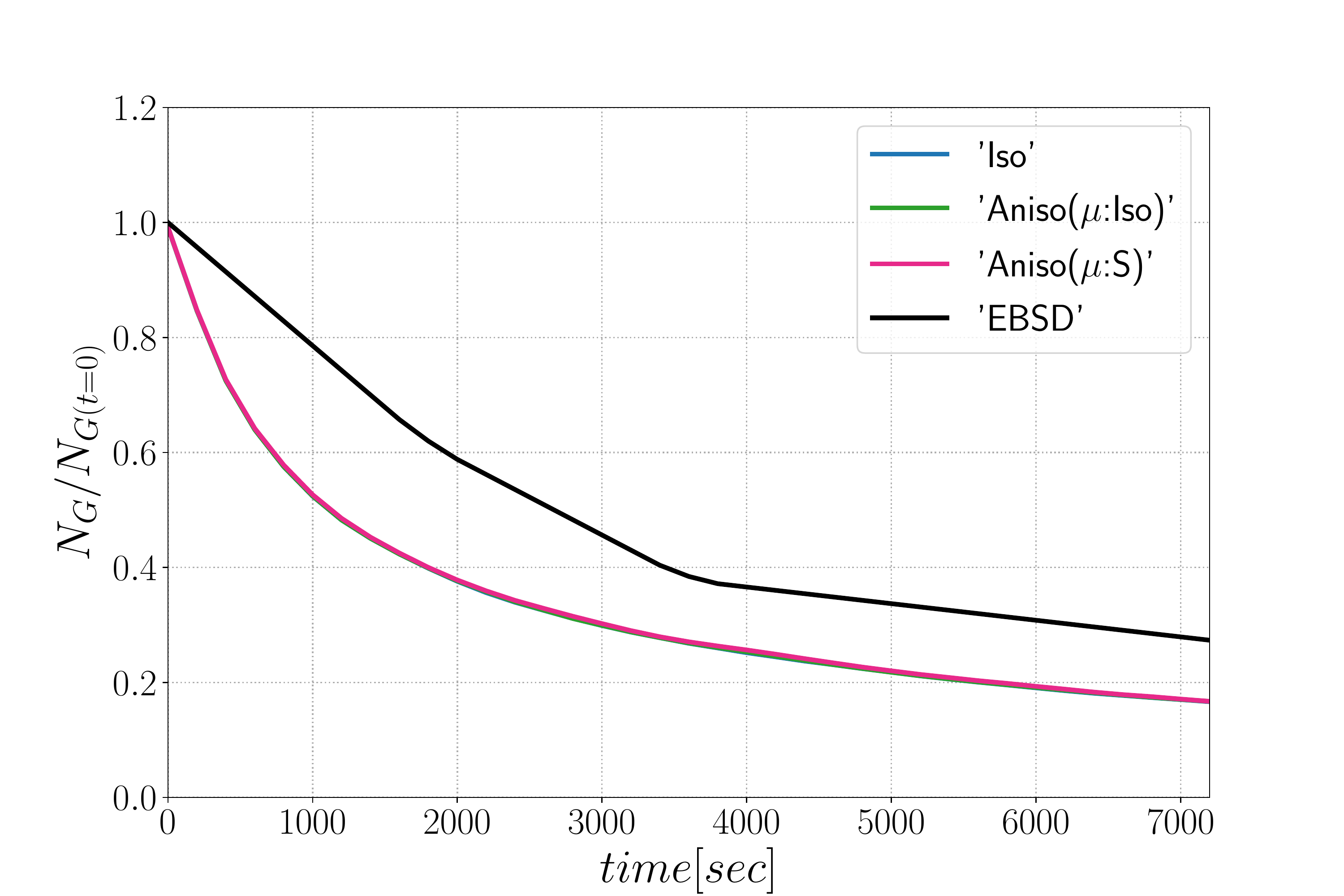}
    \caption{$N_g/N_g(t=0)=f(t)$}
  \end{subfigure} \\
  \begin{subfigure}[c]{0.495\textwidth}
    \centering
    \includegraphics[scale=0.295]{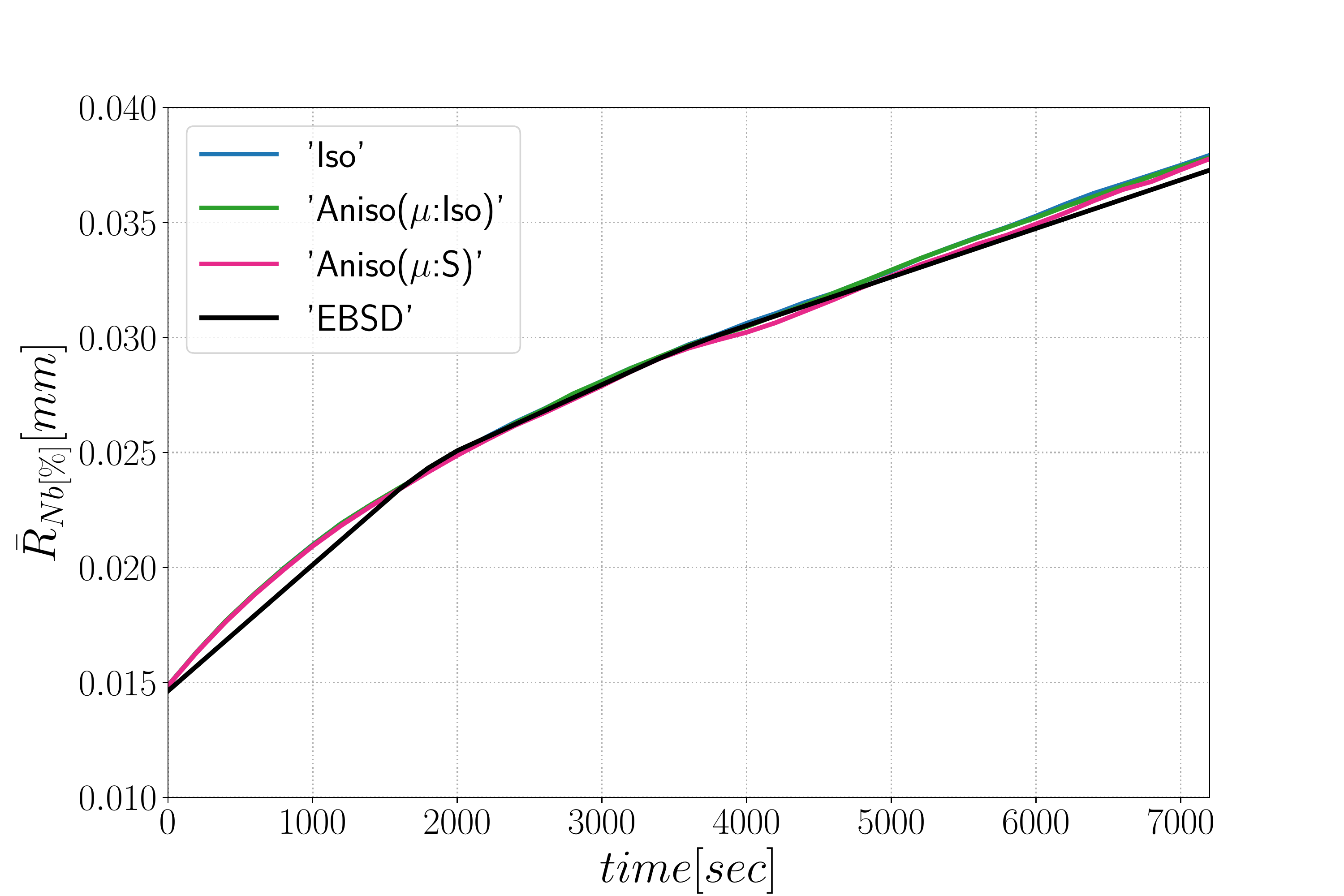}
    \caption{$\bar{R}_{Nb[\%]}=f(t)$}
  \end{subfigure}
  \begin{subfigure}[c]{0.495\textwidth}
    \centering
    \includegraphics[scale=0.295]{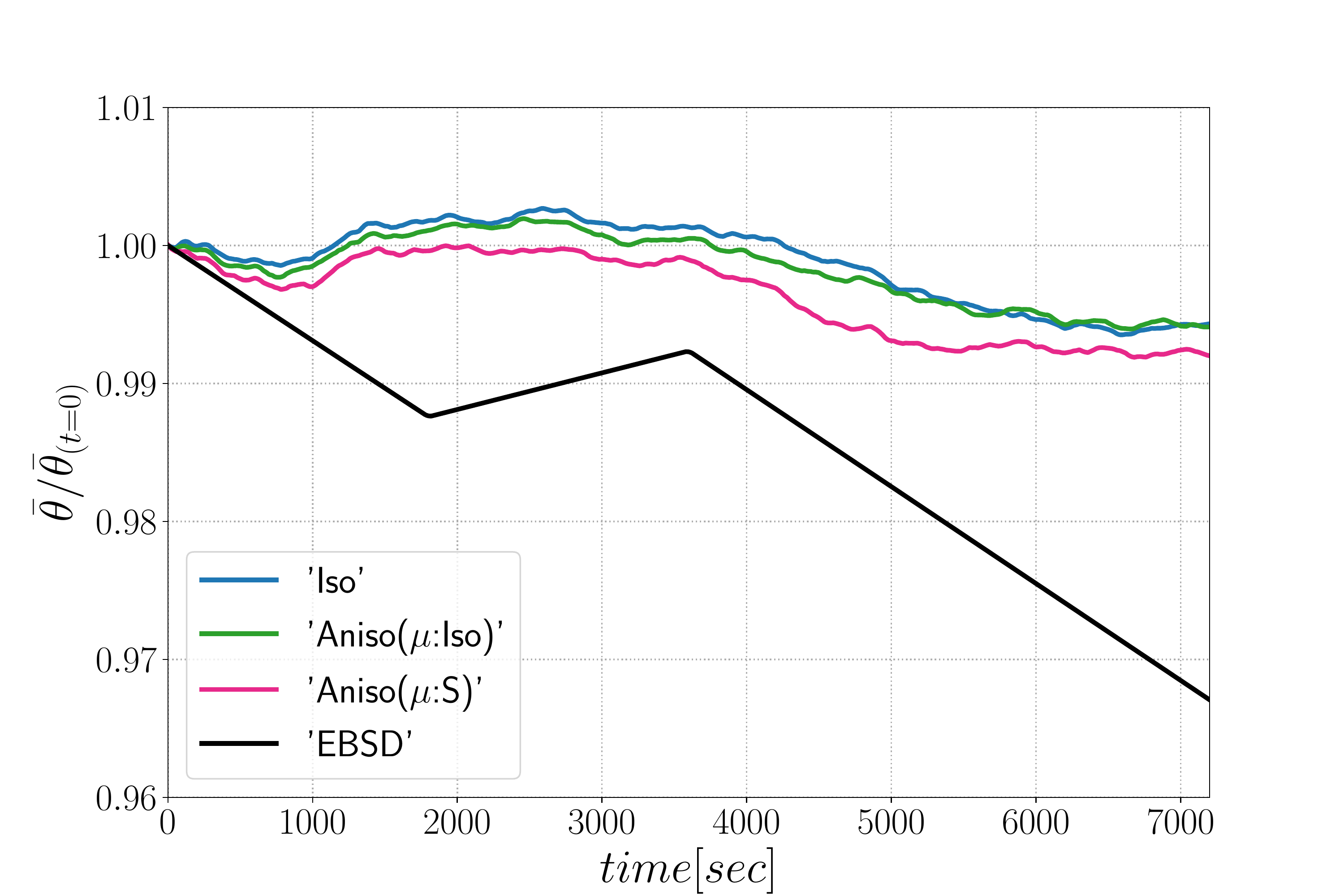}
    \caption{$\bar{\theta}/\bar{\theta}(t=0)=f(t)$}
  \end{subfigure}
  \vspace{5pt}
  \caption{Mean values time evolution for the isotropic (Iso) formulation, anisotropic formulations with isotropic GB mobility (Aniso($\mu$:Iso)) and heterogeneous GB mobility (Aniso($\mu$:S)) and the experimental data (EBSD). Numerical results obtained from the initial microstructure shown in Figure~\ref{fig:StatMetho}.}\label{fig:StatwoS3MuIsoMeanV}
\end{figure}

The simulations carried out using the Anisotropic formulation consider heterogeneous GB energy defined by Equation~\ref{eqn:Gamma} and two descriptions of the mobility. If GB mobility is isotropic, the formulation is referred as ``Aniso($\mu$:Iso)'' and in the cases where GB mobility is heterogeneous (i.e. defined by Equation~\ref{eqn:Mob}), the formulation is referred as ``Aniso($\mu$:S)''. Figure~\ref{fig:StatwoS3MuIsoMeanV} shows the evolution of average quantities: normalized total GB energy $E_{\Gamma}/E_{\Gamma}(t=0)$, normalized number of grains $N_G/N_G(t=0)$, arithmetic mean grain radius $\bar{R}_{Nb}$, and normalized average GB disorientation $\bar{\theta}/\bar{\theta}(t=0)$. One can see that the mean grain radius evolutions agree with the experimental data and that the evolution of the other mean values are close to each other when using the different formulations and present reasonable variations from the experimental data. As stated in \cite{ma14143883}, the effect of an heterogeneous GB mobility does not affect the evolution of the mean values and distributions when orientations are generated randomly and the DDF is similar to a Mackenzie distribution. On can see that the only mean value affected by the GB mobility is the average GB disorientation $\bar{\theta}$, being the ``Aniso($\mu$:S)'' formulation the one that is closer to the experimental evolution.

\begin{figure}[H]
  
  \centering
  \begin{subfigure}[c]{0.495\textwidth}
    \centering
    \includegraphics[scale=0.295]{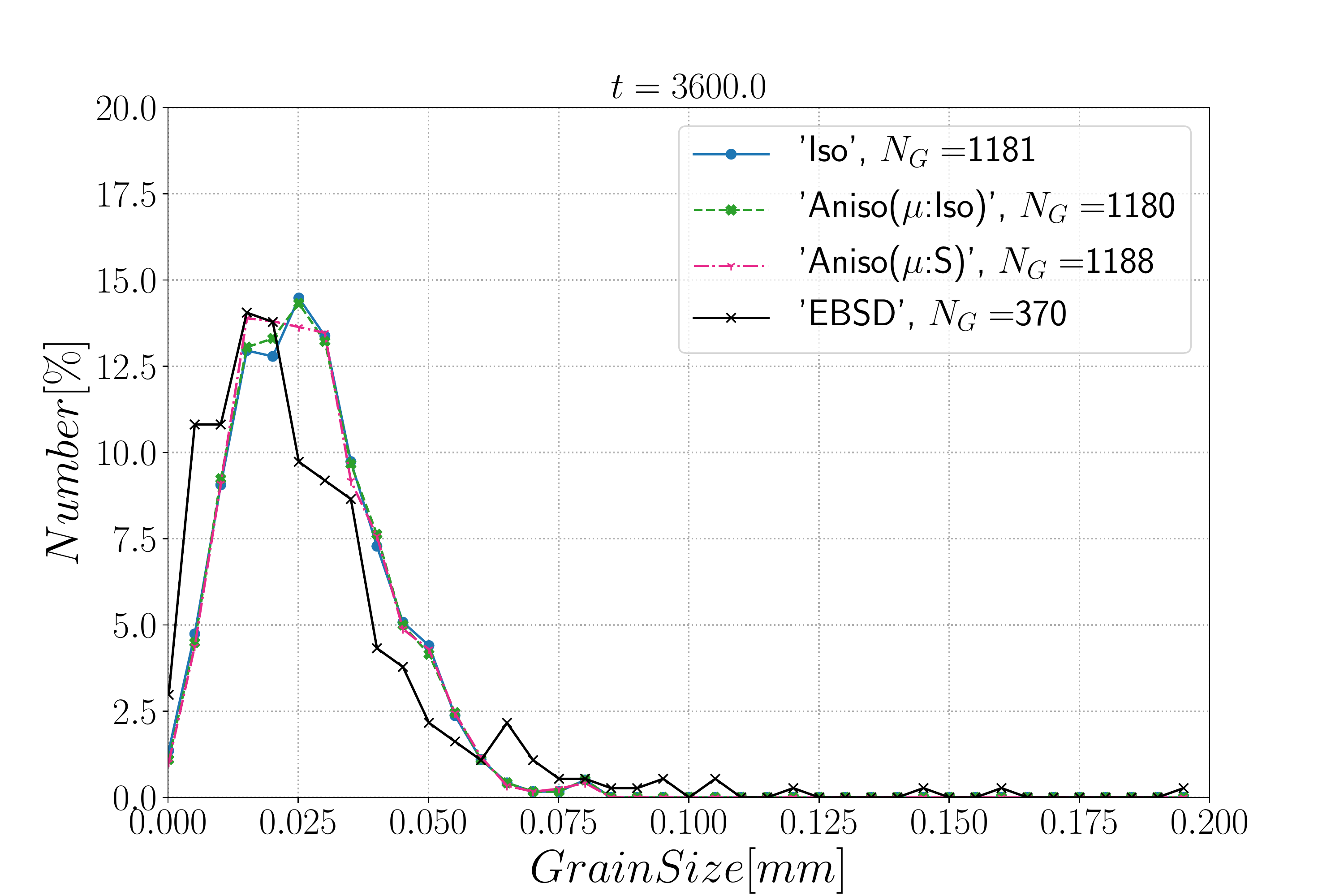}
    \caption{$ t=1\ h $}
  \end{subfigure}
  \begin{subfigure}[c]{0.495\textwidth}
    \centering
    \includegraphics[scale=0.295]{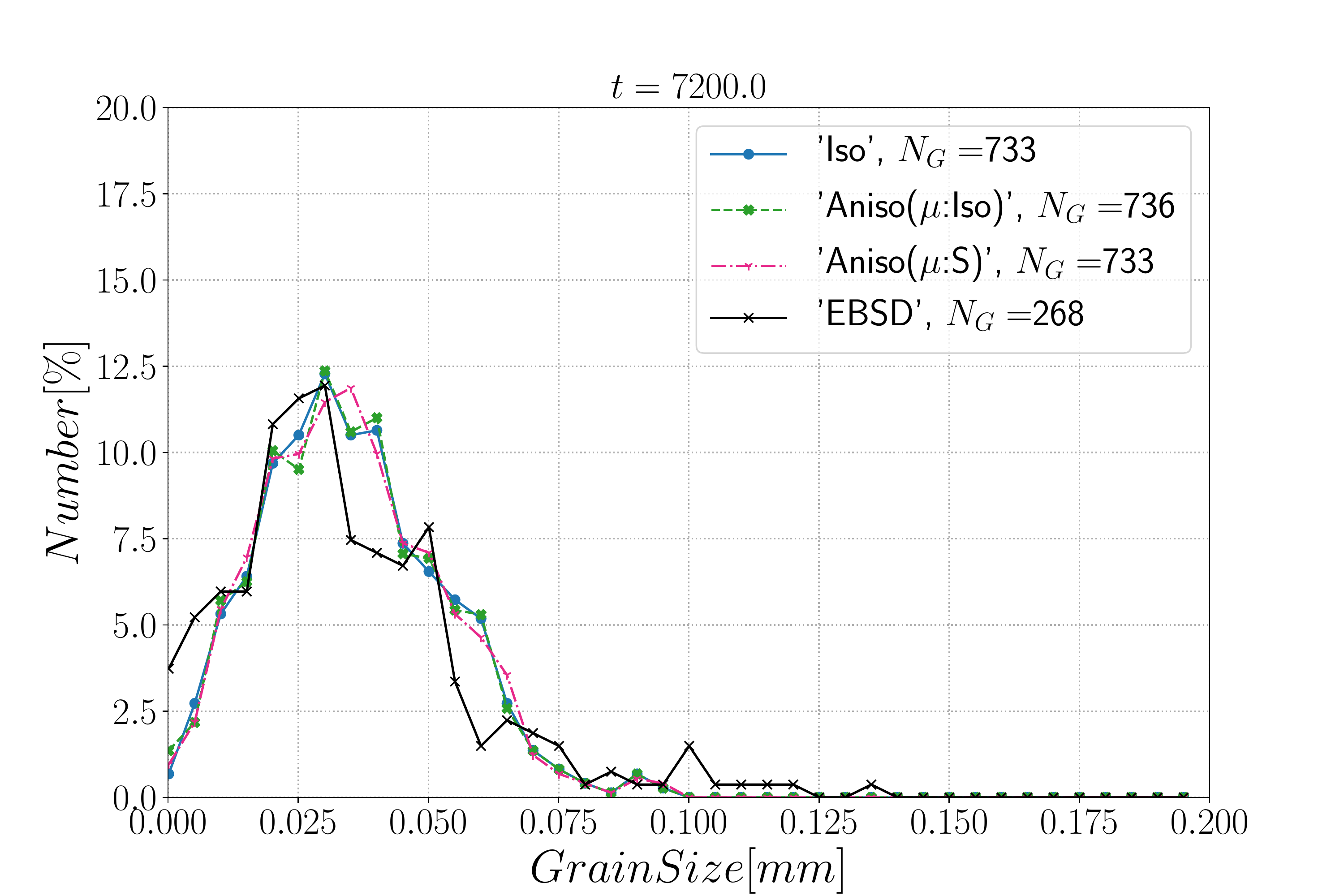}
    \caption{$ t=2\ h $}
  \end{subfigure}
  \caption{Grain Size Distributions obtained excluding TBs at (a) t=1h and (b) t=2h for the isotropic (Iso) formulation, anisotropic formulations with isotropic GB mobility (Aniso($\mu$:Iso)) and heterogeneous GB mobility (Aniso($\mu$:S)) and the experimental data (EBSD). $N_G$ refers to the number.}
  \label{fig:StatwoS3GSDMuIso}
\end{figure}

\begin{figure}[h]
  
  \centering
  \begin{subfigure}[c]{0.495\textwidth}
    \centering
    \includegraphics[scale=0.295]{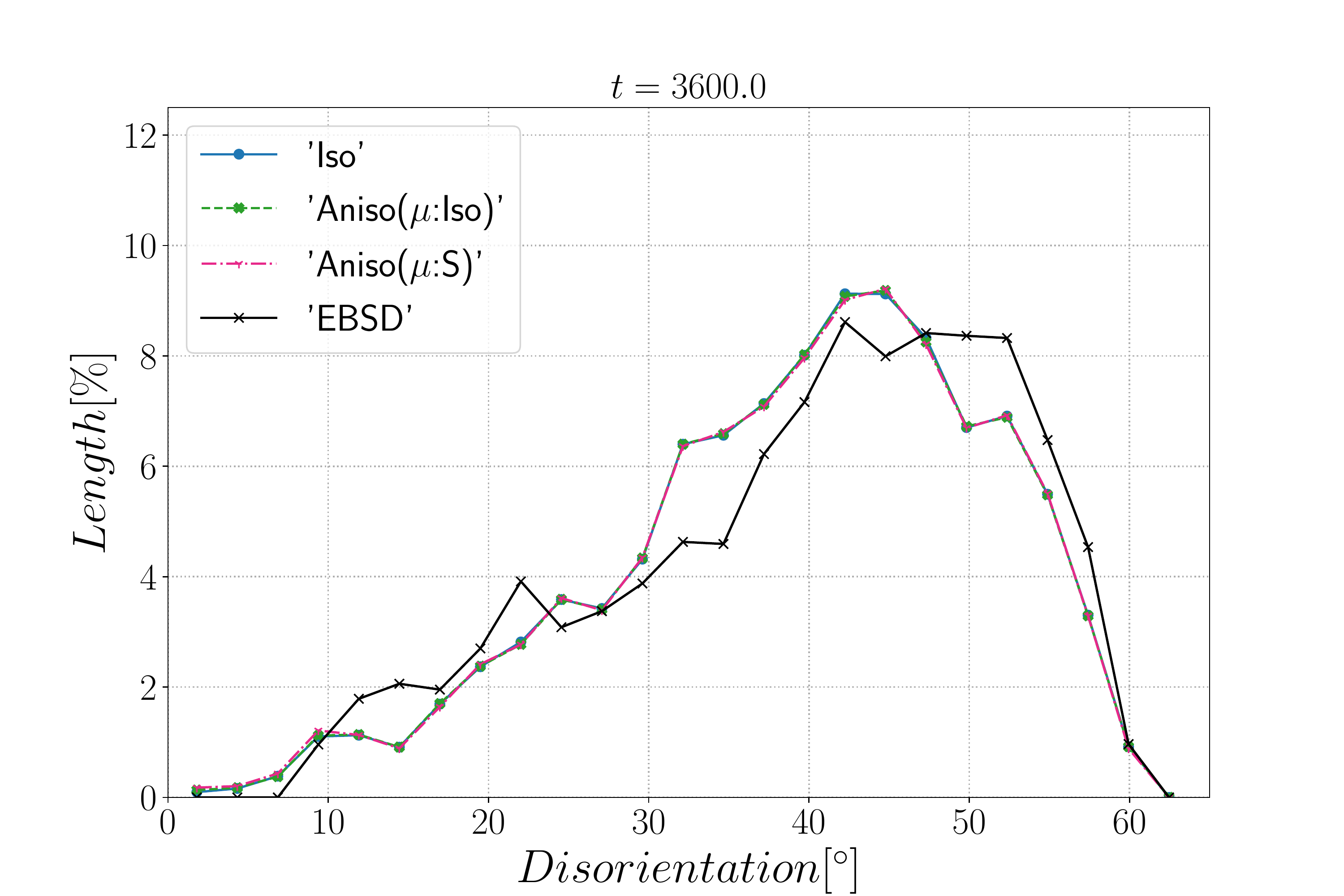}
    \caption{$ t=1\ h $}
  \end{subfigure}
  \begin{subfigure}[c]{0.495\textwidth}
    \centering
    \includegraphics[scale=0.295]{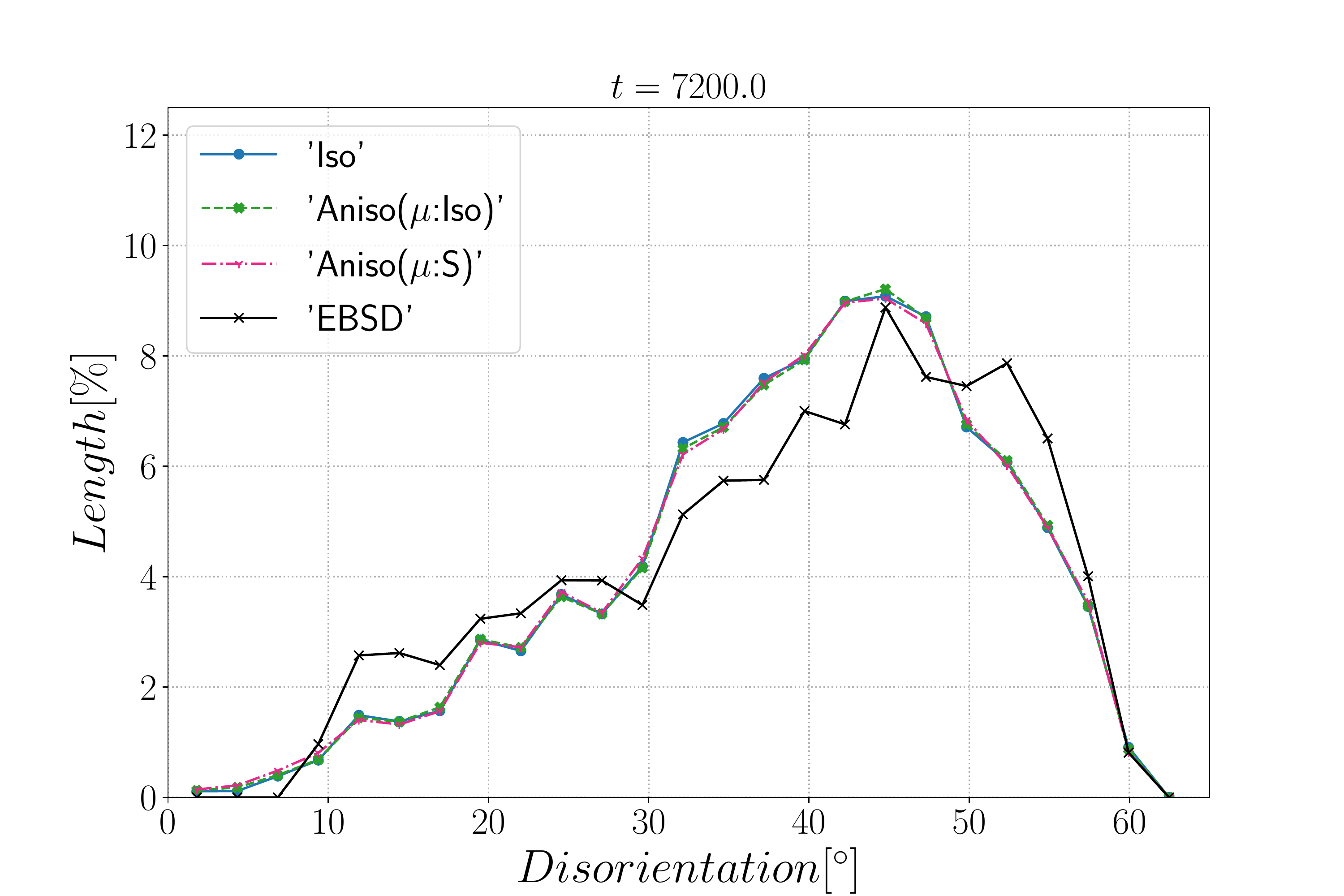}
    \caption{$ t=2\ h $}
  \end{subfigure}
  \caption{Disorientation Distribution obtained excluding TBs at (a) t=1h and (b) t=2h for the isotropic (Iso) formulation, anisotropic formulations with isotropic GB mobility (Aniso($\mu$:Iso)) and heterogeneous GB mobility (Aniso($\mu$:S)) and the experimental data (EBSD). The y-axis represents the GB length percentage.}
  \label{fig:StatwoS3DDFMuIso}
\end{figure}

\begin{figure}[H]
  \centering
  \includegraphics[scale=0.3]{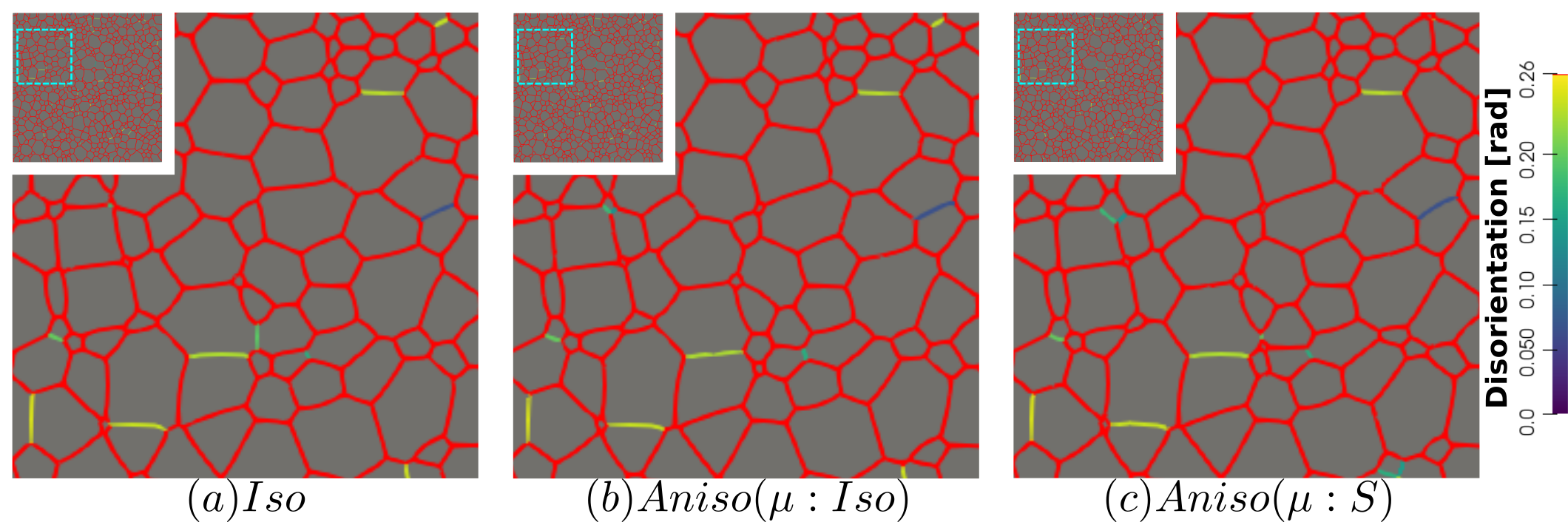}
  \caption{Detail of the GB disorientation at $t=2\ h$ in radians, GBs with a disorientation higher than 0.26 radians ($15\degree$) are colored in red: (a) Isotropic framework, (b) Anisotropic framework with $\gamma$ function of $\theta$ (Equation~\ref{eqn:Gamma}) and $\mu$ constant, and (c) Anisotropic framework with $\gamma$ and $\mu$ functions of $\theta$ through Equation~\ref{eqn:Gamma} and Equation~\ref{eqn:Mob}, respectively. Due to the few GBs with $\theta<0.26$ just a square section at the top-left of the hole microstructure is shown. }\label{fig:PXStatwoS3MuIsoIntMiso}
\end{figure}

Figures~\ref{fig:StatwoS3GSDMuIso} and \ref{fig:StatwoS3DDFMuIso}, show a good match of GSD and DDF between simulation results and experimental data after one and two hours of annealing at 1050°C. One can see that the three cases are alike. The initial Mackenzie-like DDF evolves slowly for cases with random orientations and low anisotropy (as in \cite{holm2001misorientation, chang2014effect, gruber2009misorientation, elsey2013simulations, ma14143883}). Finally, in Figure~\ref{fig:PXStatwoS3MuIsoIntMiso}, one can see the similarity between the microstructures obtained in the different simulations with most of the grains being equiaxed and few LAGBs.        

The results presented here show that the evolution of an untextured polycrystal with an initial Mackenzie-like DDF could be simulated using an Isotropic formulation or an Anisotropic formulation with heterogeneous GB energy or both heterogeneous GB mobility and energy. This methodology has been used in different contexts under different annealing process \cite{CRUZFABIANO2014305, agnoli2014development, maire2016, alvarado2021level} and with academic microstructures in \cite{ma14143883}. In the next section, the same analysis is performed by considering the same domain but by introducing special grain boundaries through an update of the $\mu$ and $\gamma$ fields. 

\clearpage

\subsection{Statistical case with an improved description of the $\gamma$ and $\mu$ fields}
\label{ssec:StatCaseAll}

The microstructure used in this section was generated using the same domain with $L=2.0mm$ and the GSD shown in Figure~\ref{fig:InitGSDwTB}. The initial number of grains is $N_G=14956$ and their orientation is also generated randomly from the EBSD orientations. The initial number of grains is more important comparatively to the previous test case as the GSD described in Figure~\ref{fig:316LInit}(c - left side) where $\Sigma 3$ TBs are taken into account is used to generate the Laguerre-Voronoï polycrystal.

\begin{figure}[H]
  
  \centering
  \begin{subfigure}{0.48\textwidth}
    \centering
    \includegraphics[scale=0.8]{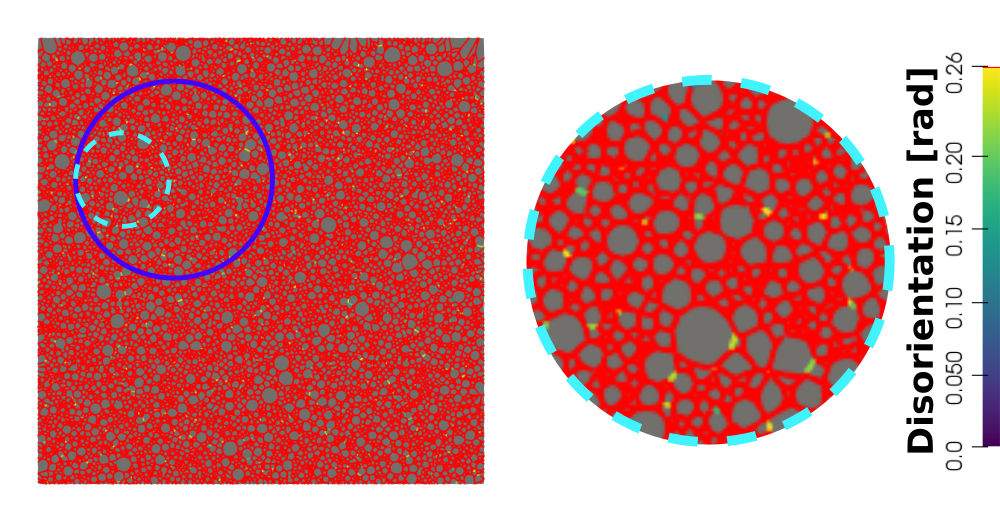}
    \caption{GB energy}
    \label{fig:StatMicro}
  \end{subfigure}
  \begin{subfigure}{0.48\textwidth}
    \centering
    \includegraphics[scale=0.25]{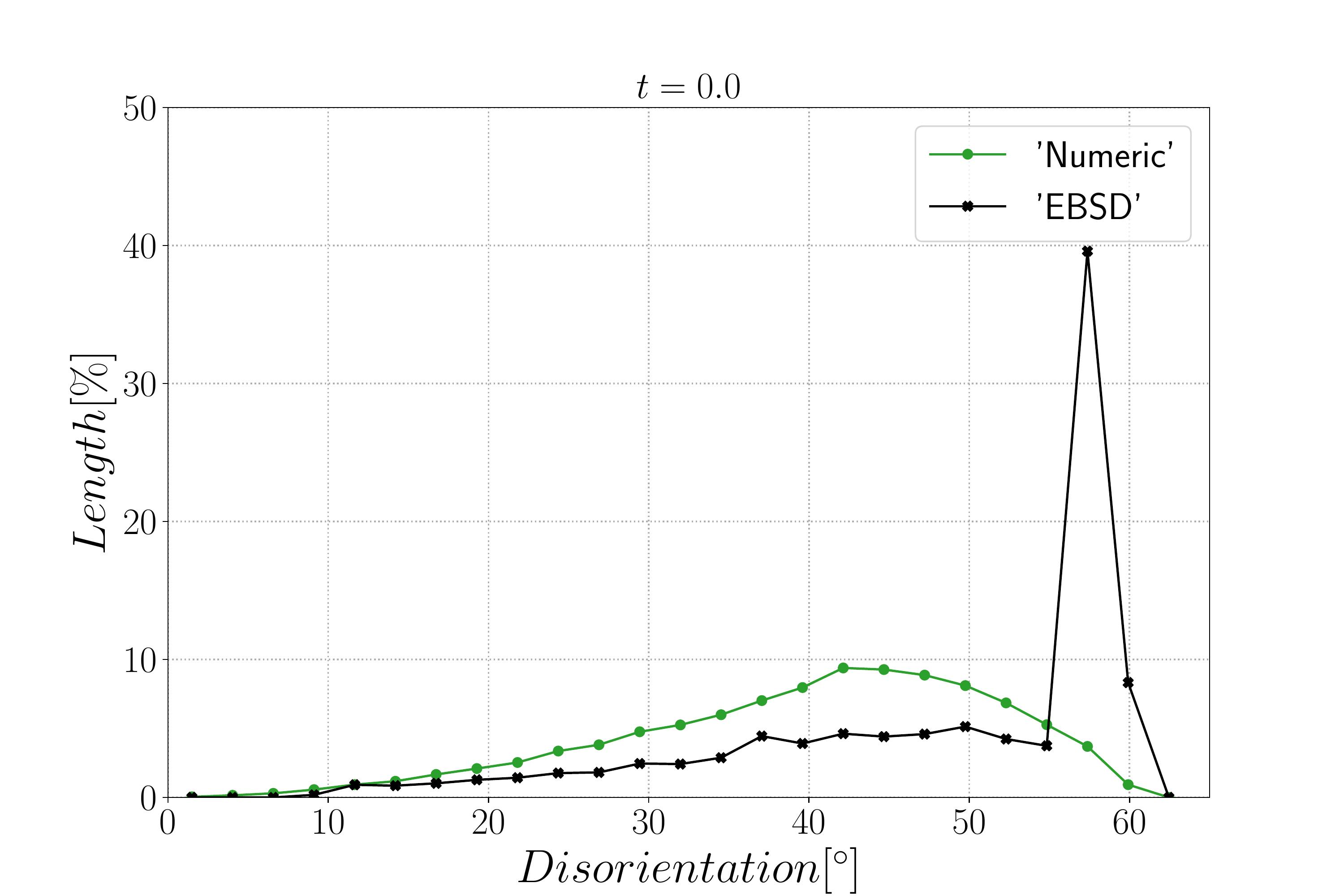}
    \caption{DDF}
    \label{fig:StatDistDDF}
  \end{subfigure}
  \caption{Initial (a) microstructure disorientation with a cyan circle which represents the zone shown on the right and (b) the disorientation distribution.}
  \label{fig:StatMicDist}
\end{figure}

The same mesh and time step than the previous simulations are used in order to be able to fairly compare the obtained results. In order to define the behavior of special GBs with properties close to $\Sigma 3$ TBs, the $\mu$ and $\gamma$ fields are updated as follow: 
\begin{align}
  \left\{
  \begin{array}{ll}
    \gamma(\theta) = \gamma_{max} \frac{\theta}{\theta_0} \left( 1 - ln \left( \frac{\theta}{\theta_0} \right) \right), & \theta < \theta_0 \\
    \\
    \gamma(\theta) = \gamma_{max}, & \theta_0 \le \theta  < \theta_{\Sigma 3} \\
    \\
    \gamma(\theta) = \gamma_{max}*0.1, & \theta \ge \theta_{\Sigma 3} \\
\end{array}
\label{eqn:GammaP}
\right .
\end{align}
\begin{align}
  \left\{
  \begin{array}{ll}
  \mu(\theta)=\mu_{max} \left( 1 - exp \left( -5 \left(\frac{ \theta}{\theta_0} \right)^4 \right) \right), & \theta  < \theta_{\Sigma 3}   \\
  \\
  \mu(\theta) = \mu_{max}*0.1, & \theta \ge \theta_{\Sigma 3} \\
\end{array}
\label{eqn:MobP}
\right .
\end{align}
with $\theta_0 = 15 \degree$ and $\theta_{\Sigma 3} = 60 \degree$ and a value of GB energy and mobility set to $\gamma_{max}*0.1$ and $\mu_{max}*0.1$ for GBs with $\theta \ge \theta_{\Sigma 3}$. The value of $\mu_{max}$ is estimated using the evolution of $\bar{R}_{Nb[\%]}$ considering all GBs (see Figure~\ref{fig:MeanR}). The estimated GB mobility is $\mu_{max} = 0.069 \ mm^4 \cdot J^{-1} \cdot s^{-1}$, and is one order of magnitude lower from the GB mobility estimated without $\Sigma 3$ TBs. The decrease of the GB mobility is proportional to the decrease of grain size (see Figure~\ref{fig:MeanR}) due to the high number of TBs. 

\begin{figure}[H]
  
  \centering
  \begin{subfigure}{0.495\textwidth}
    \centering
    \includegraphics[scale=0.295]{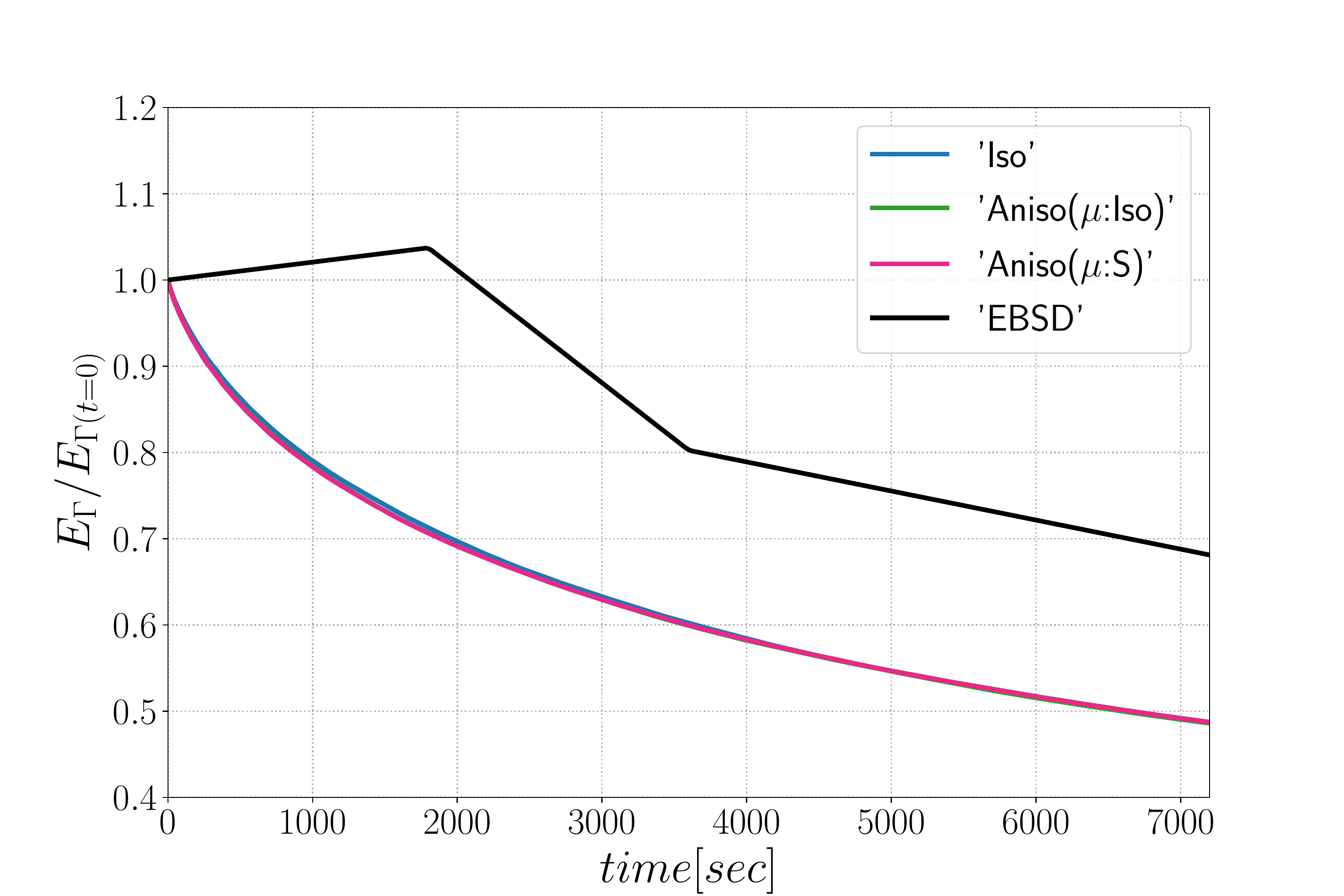}
    \caption{$E_{\Gamma}/E_{\Gamma}(t=0)=f(t)$} 
    \label{fig:StatMuIsoMeanVEtot}
  \end{subfigure} 
  \begin{subfigure}{0.495\textwidth}
    \centering
    \includegraphics[scale=0.295]{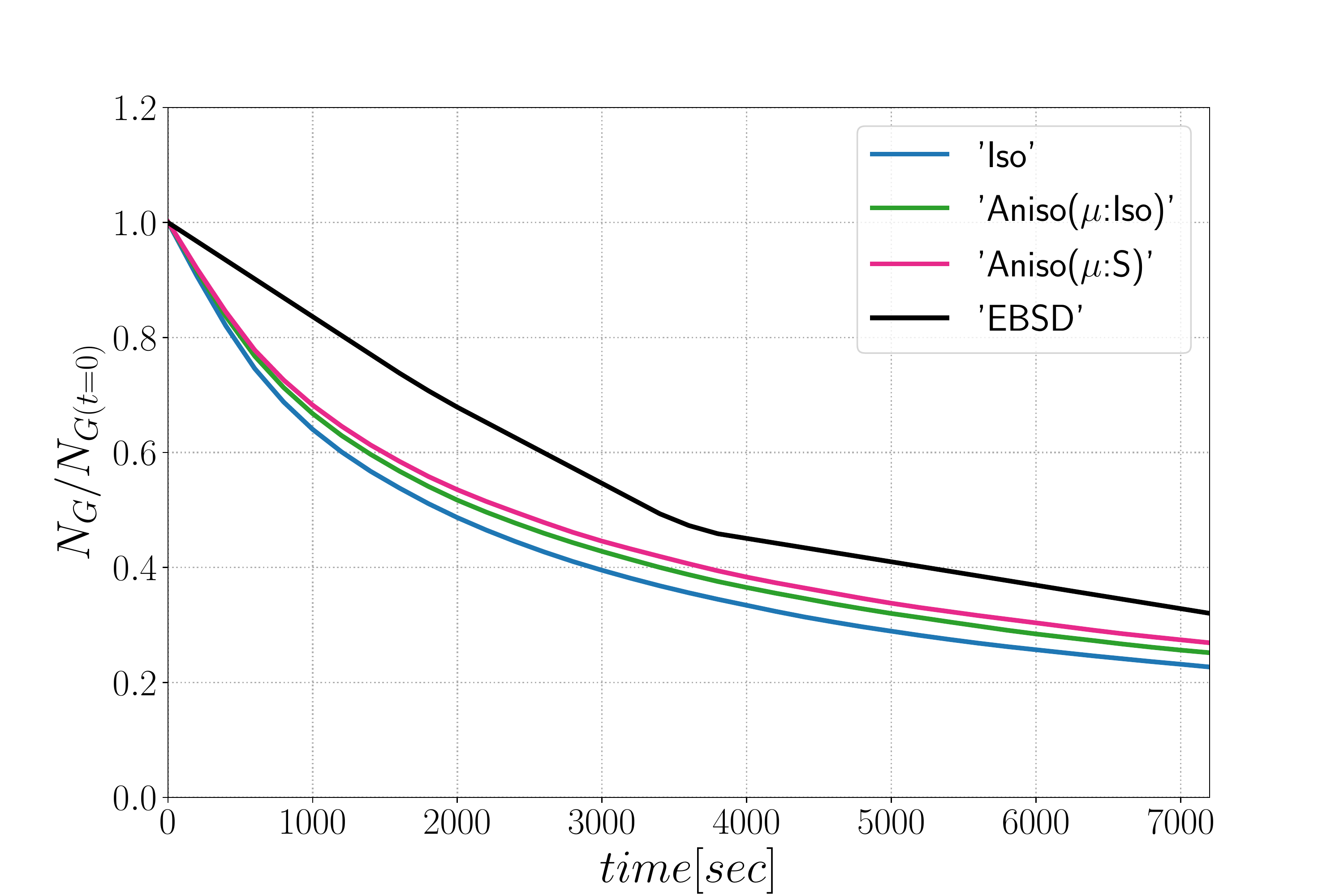}
    \caption{$N_g/N_g(t=0)=f(t)$}
    \label{fig:StatMuIsoMeanVNg} 
  \end{subfigure} \\
  \begin{subfigure}{0.495\textwidth}
    \centering
    \includegraphics[scale=0.295]{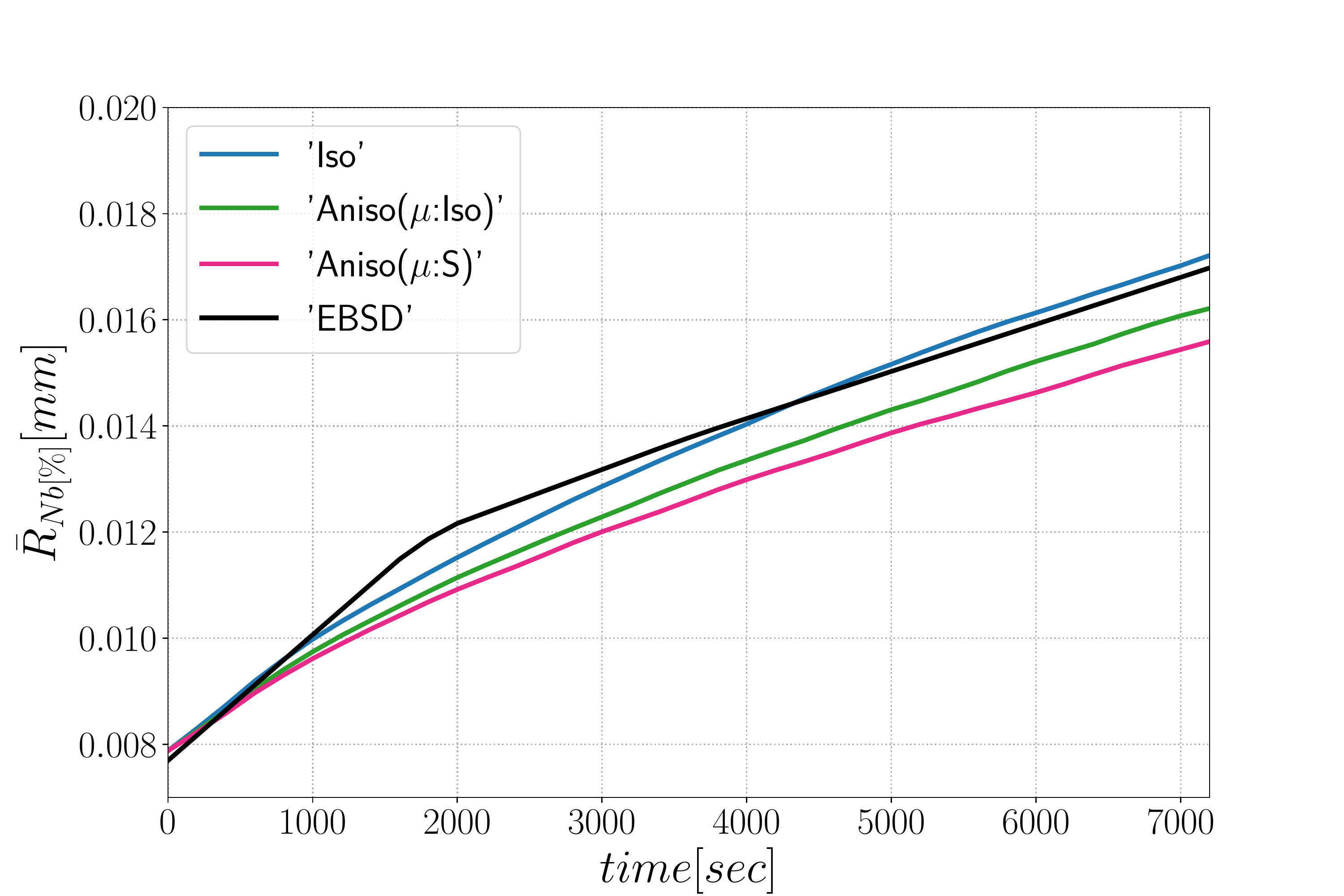}
    \caption{$\bar{R}_{Nb[\%]}=f(t)$}
    \label{fig:StatMuIsoMeanVRn}
  \end{subfigure}
  \begin{subfigure}{0.495\textwidth}
    \centering
    \includegraphics[scale=0.295]{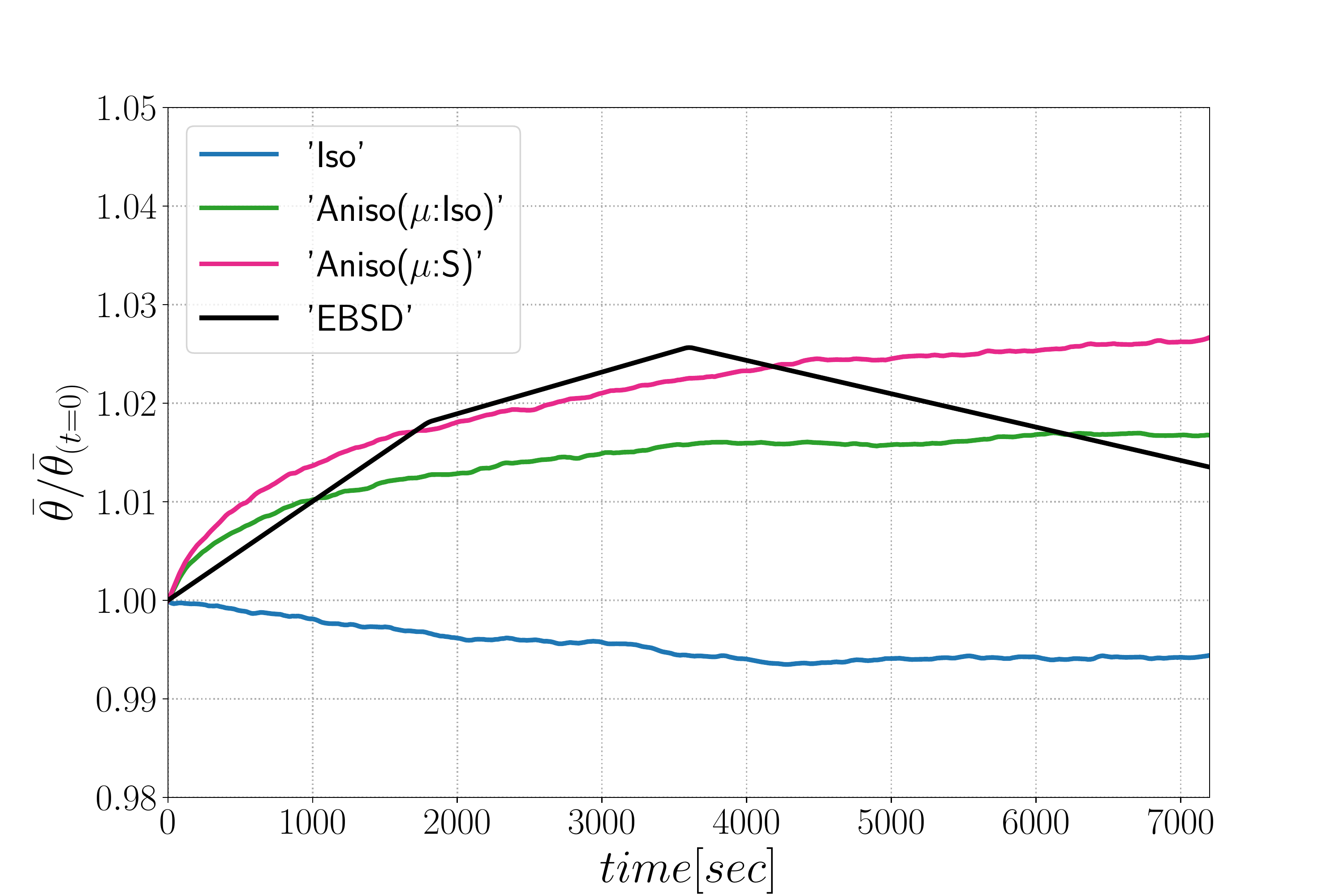}
    \caption{$\bar{\theta}/\bar{\theta}(t=0)=f(t)$}
    \label{fig:StatMuIsoMeanVDis}
  \end{subfigure}
  \vspace{5pt}
  \caption{Mean values time evolution for the isotropic (Iso) formulation, anisotropic formulations with isotropic GB mobility (Aniso($\mu$:Iso)) and heterogeneous GB mobility (Aniso($\mu$:S)) and the experimental data (EBSD). Numerical results obtained from the initial microstructure shown in Figure~\ref{fig:StatMicro}.}\label{fig:StatMuIsoMeanV}
\end{figure}

From the results shown in Figure~\ref{fig:StatMuIsoMeanV}, one can see that all formulations minimize the energy with the same efficiency (Figure~\ref{fig:StatMuIsoMeanVEtot}) and the microstructure evolve at a same rate leading to a good fit of mean grain size and number of grain evolutions comparatively to the experimental data. On the other hand, the Anisotropic formulation shows a better agreement in terms of mean disorientation evolution.

\begin{figure}[h]
  \centering
  \begin{subfigure}{0.48\textwidth}
    \centering
    \includegraphics[scale=0.25]{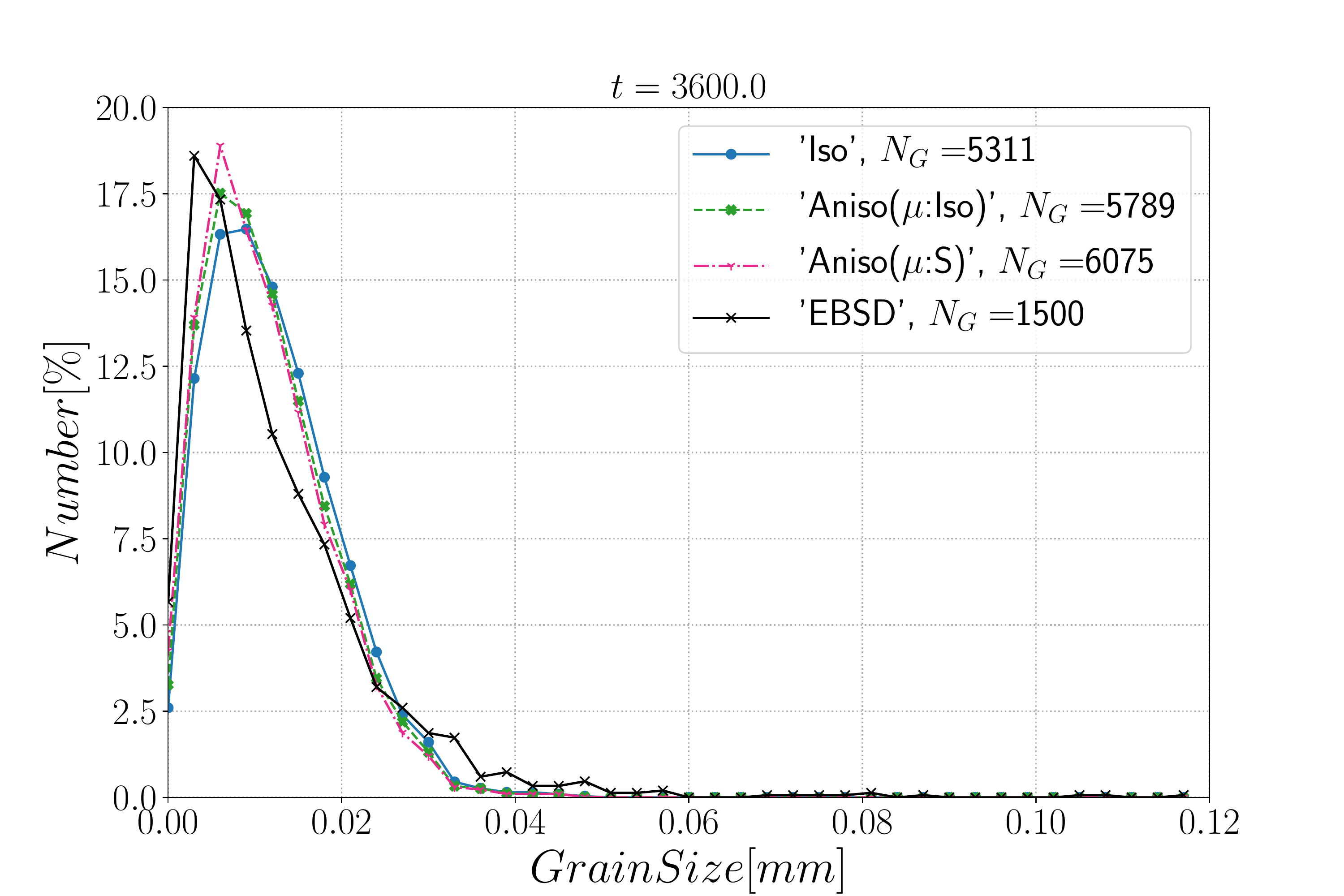}
    \caption{$ t=1h $}
  \end{subfigure}
  \begin{subfigure}{0.48\textwidth}
    \centering
    \includegraphics[scale=0.25]{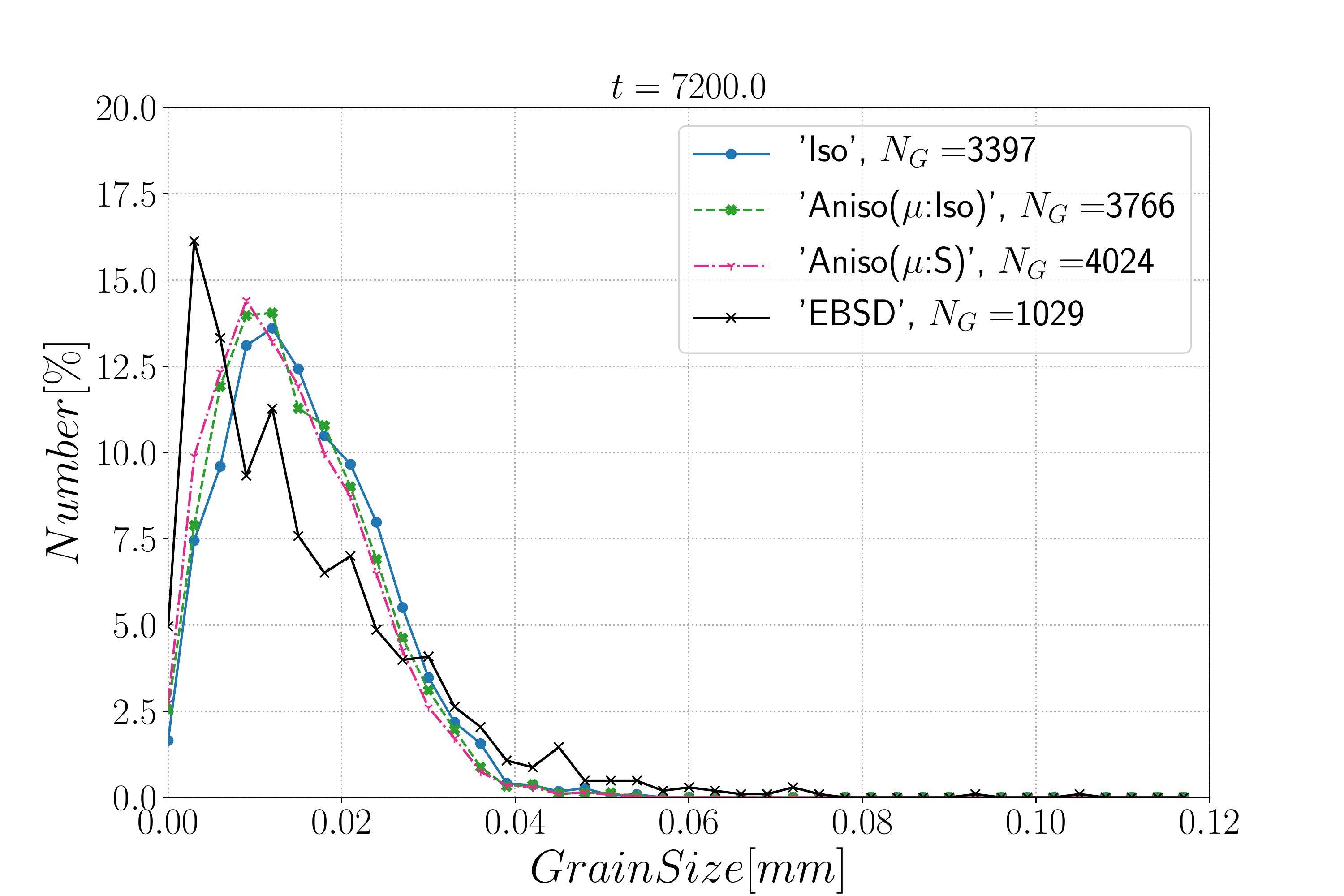}
    \caption{$ t=2h $}
  \end{subfigure}
  \caption{Grain Size Distributions obtained including TBs at (a) $t$=1h and (b) $t$=2h for the isotropic (Iso) formulation, anisotropic formulations with isotropic GB mobility (Aniso($\mu$:Iso)) and heterogeneous GB mobility (Aniso($\mu$:S)) and the experimental data (EBSD). $N_G$ refers to the number.}
  \label{fig:StatGSDMuIso}
\end{figure}

\begin{figure}[h]
  \centering
  \begin{subfigure}{0.48\textwidth}
    \centering
    \includegraphics[scale=0.25]{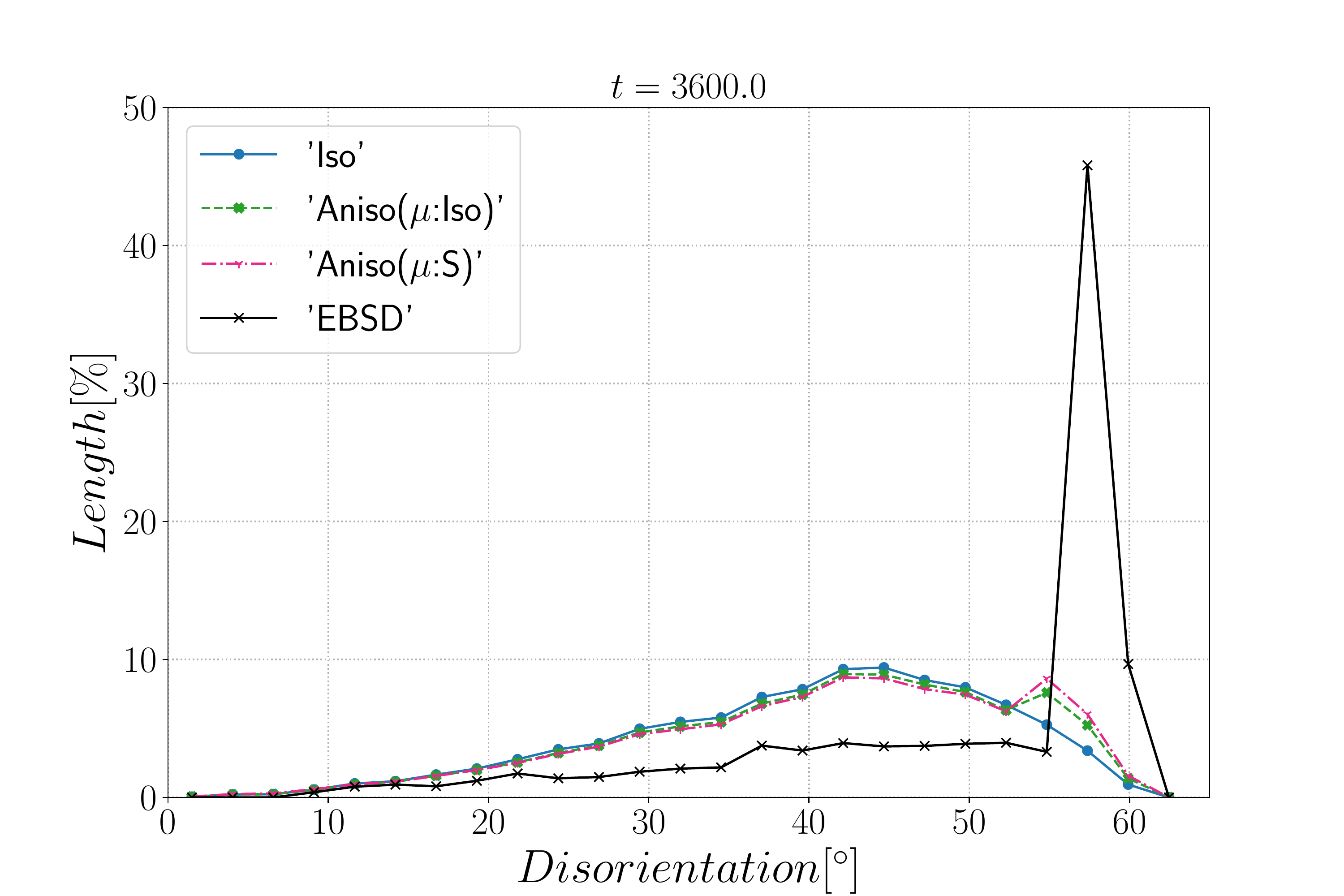}
    \caption{$ t=1h $}
  \end{subfigure}
  \begin{subfigure}{0.48\textwidth}
    \centering
    \includegraphics[scale=0.25]{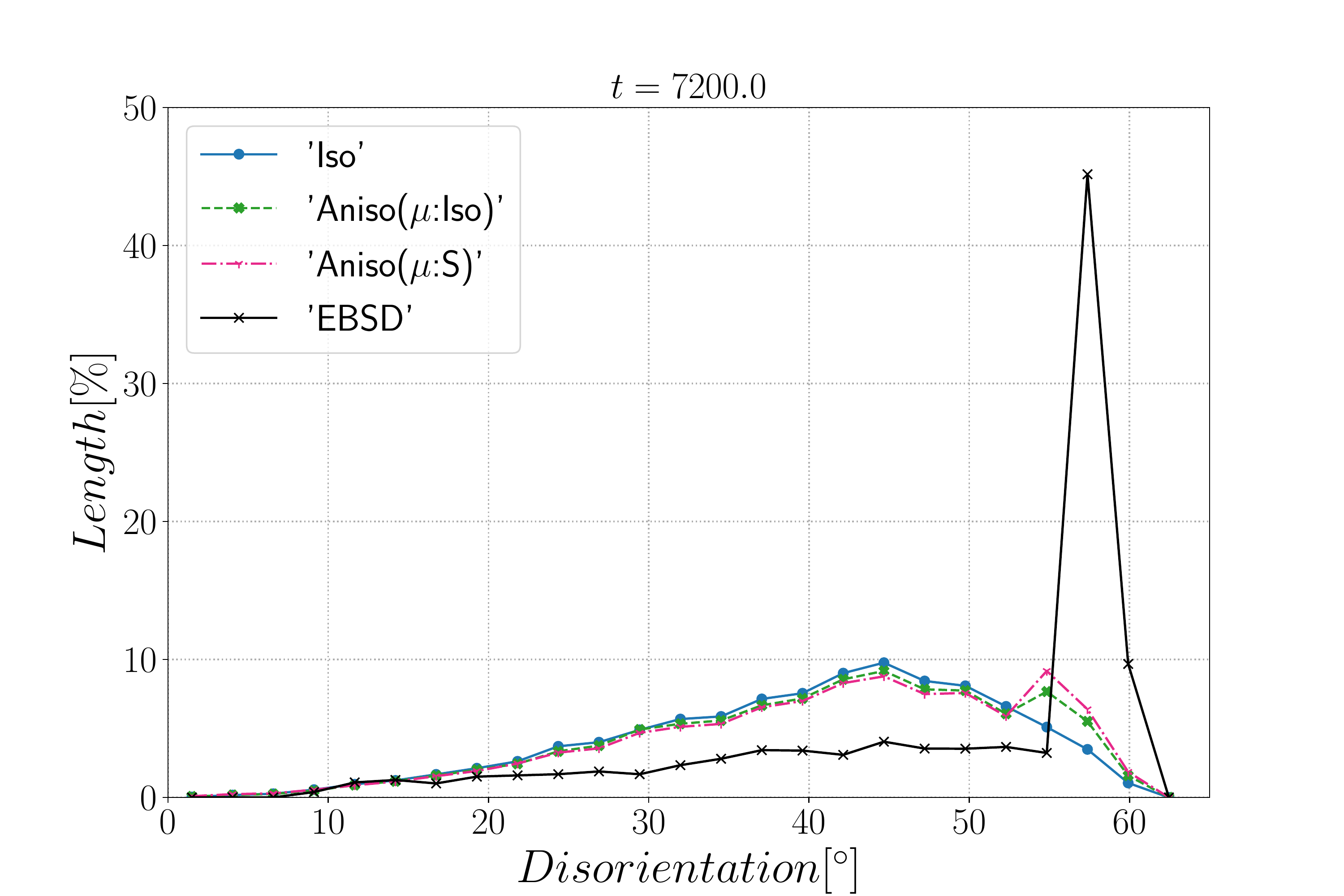}
    \caption{$ t=2h $}
  \end{subfigure}
  \caption{Disorientation Distribution obtained including TBs at (a) $t$=1h and (b) $t$=2h for the isotropic (Iso) formulation, anisotropic formulations with isotropic GB mobility (Aniso($\mu$:Iso)) and heterogeneous GB mobility (Aniso($\mu$:S)) and the experimental data (EBSD). The y-axis represents the GB length percentage.}
  \label{fig:StatDDFMuIso}
\end{figure}

Figure~\ref{fig:StatGSDMuIso} shows the evolution of the grain size distribution at $t$=1h and 2h. GSDs present a good match (Figure~\ref{fig:StatGSDMuIso}) with small differences between the Iso and Anisotropic formulations. However the DDF predictions (Figure~\ref{fig:StatDDFMuIso}) are quite bad for all formulations even if the Anisotropic calculations tends to be better. This result can easily be explained by the use of statistics (GSD and orientations) from EBSD data which are not sufficient to accurately describe the real microstructure. First, the initial topology is simplified but above all, even if the orientation data come from the EBSD measurements, the resulting intial DDF is not accurate as a Mackenzie-like distribution is obtained as illustrated in Figure~\ref{fig:StatMicDist}(b). Hence, the effect of heterogeneous GB energy and mobility anisotropy of the real microstructure are underestimated. A way of improvement of the proposed statistical generation methodology will be to modify the algorithm dedicated to the assignment of the orientation of each virtual grain by imposing also a better respect of the experimental DDF \cite{TU2019268}.

\begin{figure}[H]
  \centering
  \includegraphics[scale=1.0]{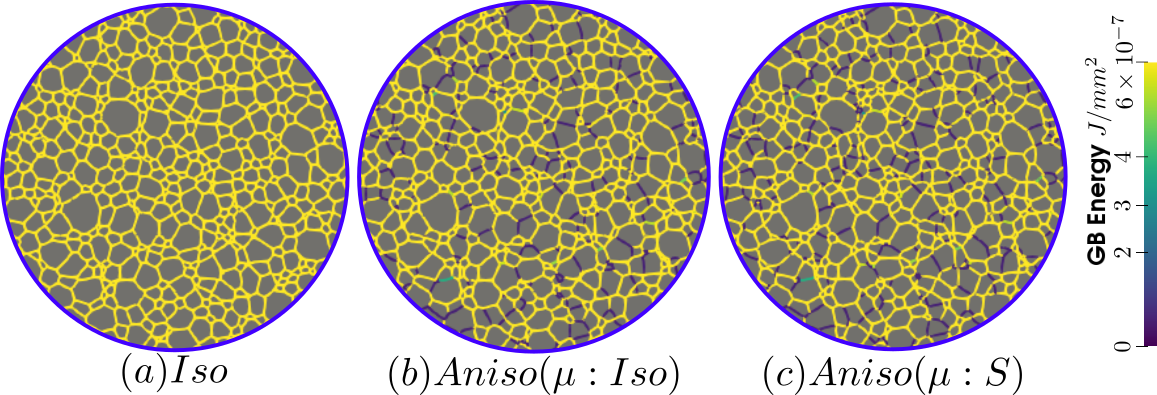}
  \caption{GB energy of the microstructure obtained with the (a) isotropic and anisotropic formulations using (b) isotropic GB mobility and (c) heterogeneous GB mobility at $t$=2h in the same zone shown in Figure~\ref{fig:StatMicro}.}
  \label{fig:PXStatMuIsoIntGBE}
\end{figure}

The GB energy of the microstructure at $t$=2h is shown in Figure~\ref{fig:PXStatMuIsoIntGBE}, a higher number of blue GBs, which correspond to TBs (low value of $\gamma$), is obtained using the Anisotropic formulations and the effect of heterogeneous GB mobility seems negligible.

The next simulations are carried out by immersing the EBSD data in order to overcome the limits discussed above.

\section{Immersion of EBSD data}
\label{sec:PX}

\begin{figure}[H]
  \centering
  \includegraphics[scale=0.85]{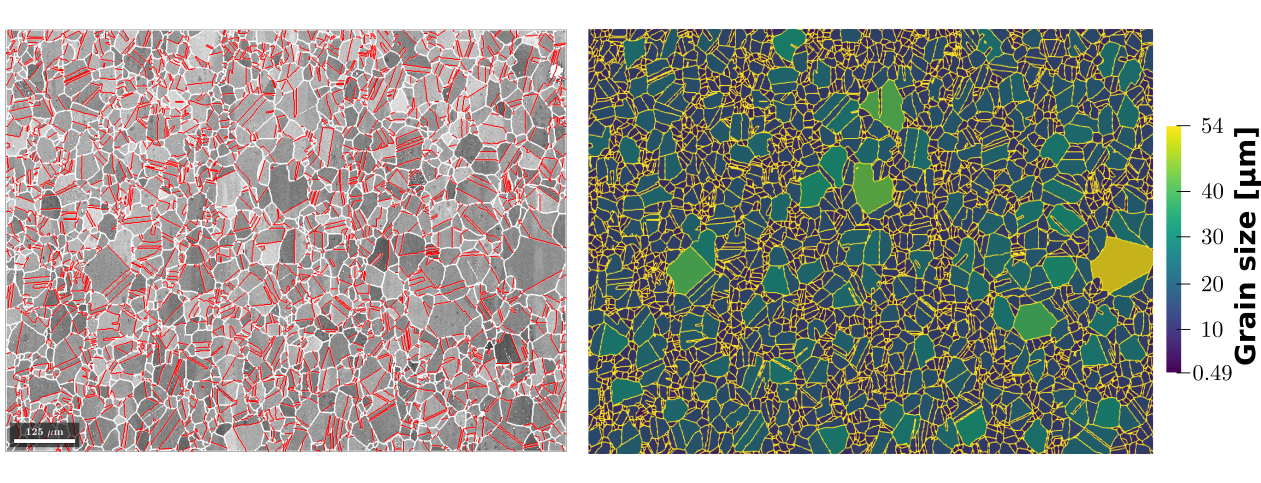}
  \caption{(left) EBSD band contrast map with GBs depicted in white and $\Sigma 3$ TBs colored in red, and (right) its numerical microstructure displayed with a color code related to the grain size and GBs are colored in yellow. Here TBs are considered to calculate grain size, i.e., crystallite size. }
  \label{fig:ImmPX}
\end{figure}

In this section, a digital twin microstructure obtained by immersion of the EBSD map acquired on the initial microstructure (Figure~\ref{fig:316LInit}) is discussed. Figure~\ref{fig:ImmPX} shows the Band Contrast (BC) map of the microstructure and its numerical twin. The dimensions of the domain are $L_x=0.856$ $mm$ and $L_y=1.138$ $mm$ and contains 3472 crystallites. The time step is fixed at $\Delta t = 10$ $s$ and the domain is discretized here using an unstructured static triangular mesh with a mesh size of $h=1$ $\mu m$. This microstructure is more appropriate to compare simulations and experimental data. The evolution of the numerical microstructure is compared to EBSD maps obtained at three different times: $t=30min$, $1h$ and $2h$ (see Figure~\ref{fig:BCmaps}).

\begin{figure}[H]
  \centering
  \includegraphics[scale=0.75]{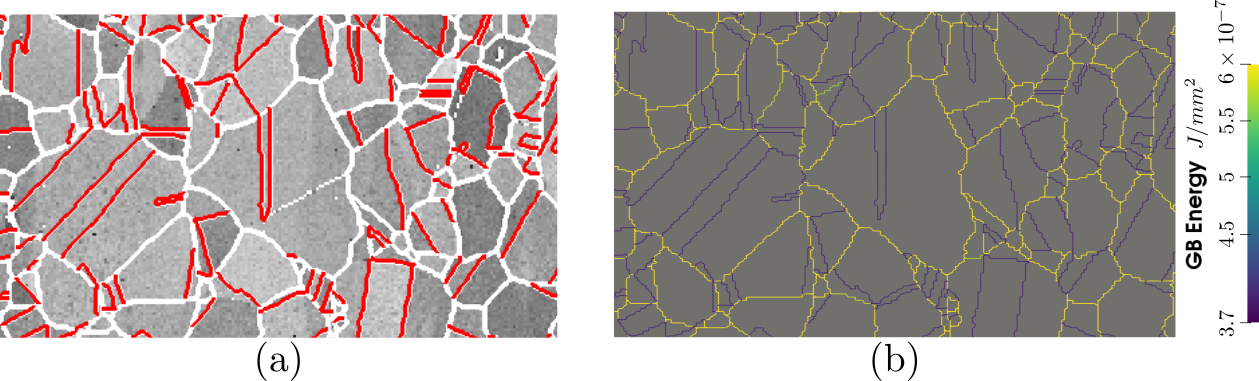}
  \caption{Detail of the (a) EBSD band contrast map and (b) its numerical twin showing the GB Energy field. Twin boundaries depicted in red on the left image have low energy on the right image. }
  \label{fig:ImmPX_Det}
\end{figure}

GB energy and mobility are defined using Equations~\ref{eqn:GammaP} and \ref{eqn:MobP}, respectively. The maximal value of GB energy is set to $\gamma_{max}=6 \times 10^{-7} \ J \cdot mm^{-2} $ and the maximal value of GB mobility is set to fit the mean grain size evolution. The maximal value of GB mobility for the Aniso($\mu$:Iso) and Aniso($\mu$:S) formulations are $\mu_{max}=0.146$ $mm^4\cdot J^{-1}\cdot s^{-1} $ and $\mu_{max}=0.272$ $mm^4\cdot J^{-1}\cdot s^{-1} $ respectively. Regarding the isotropic formulation, the value of GB reduced mobility remains constant $\mu \gamma = 0.414 \times 10^{-7} $ $mm^2 \cdot s^{-1} $. The changes in $\mu_{max}$ are due to the more complex geometry and the higher quantity of special boundaries that produce additional gradients of GB energy and GB mobility (see Equation~\ref{eqn:WeakFormAnisoSimp}). As stated before, the additional jump at $\theta_{\Sigma 3}$ is set to define special boundaries similar to $\Sigma 3$ TBs. Figure~\ref{fig:ImmPX_Det} confirms the good match between the TBs colored in red in the EBSD band contrast map (left side) and TBs colored in blue corresponding to a low GB energy in the numerical twin microstructure (right side).     

\begin{figure}[H]
  
  \centering
  \begin{subfigure}{0.495\textwidth}
    \centering
    \includegraphics[scale=0.295]{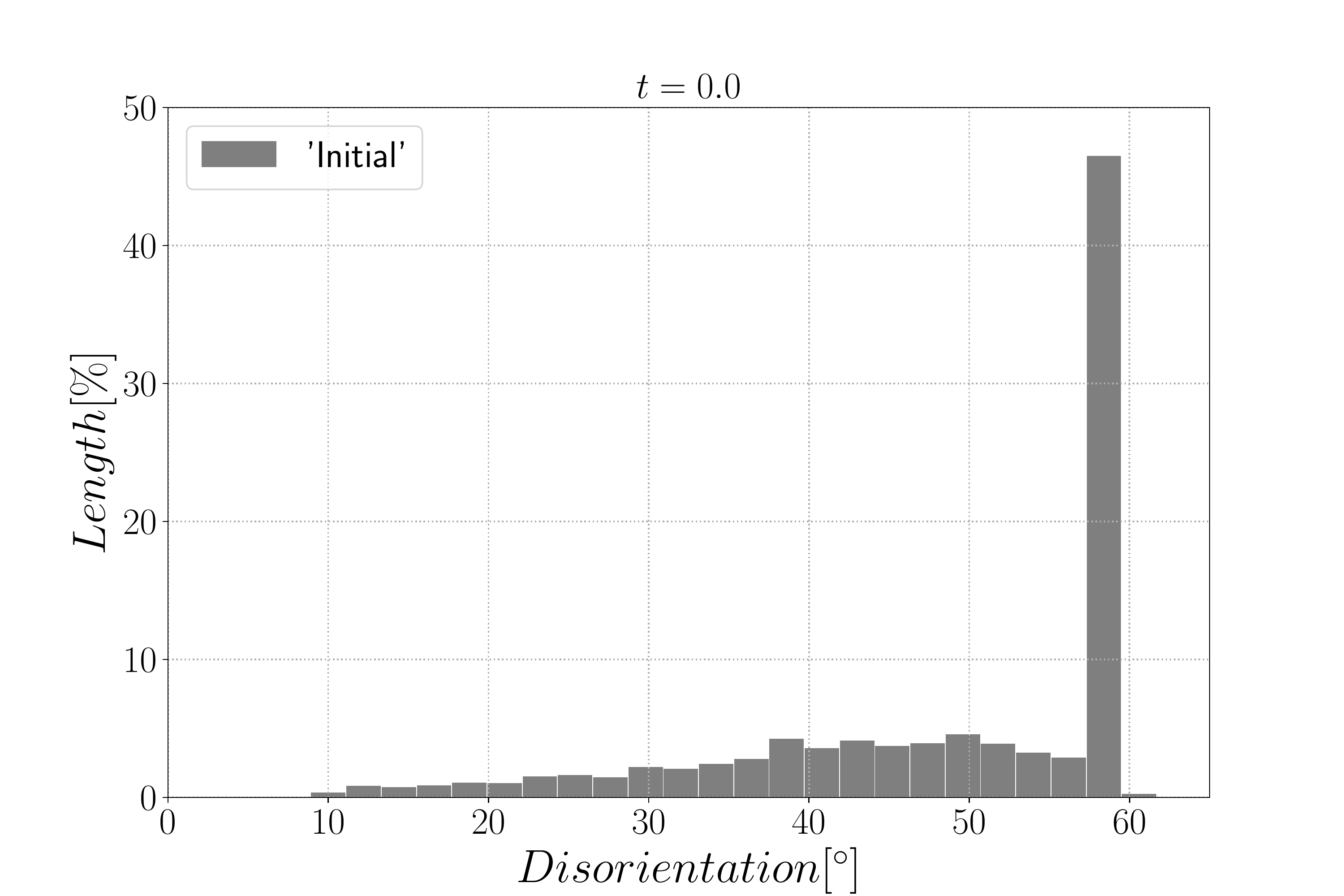}
    \caption{DDF}
  \end{subfigure}
  \begin{subfigure}{0.495\textwidth}
    \centering
    \includegraphics[scale=0.295]{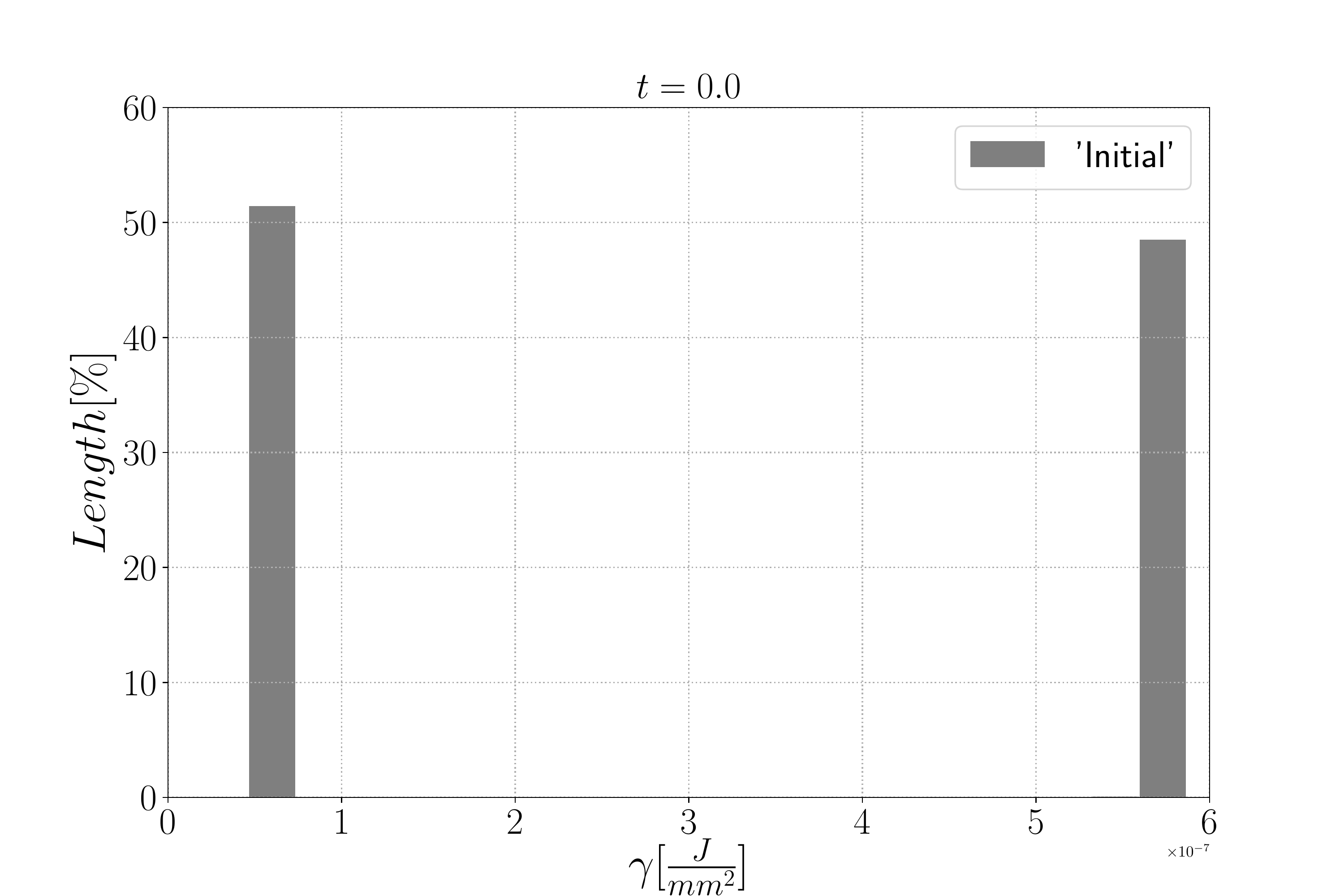}
    \caption{GBED}
    \label{fig:PXImmRSp}
  \end{subfigure}
  \caption{Initial DDF and GBED of the initial immersed microstructure produced by the modified Read-Shockley equation.}
  \label{fig:ImmPXDis}
\end{figure}

With the immersed data, one can obtain a close digital twin of the real microstructure with the initial GB distributions presented in Figure~\ref{fig:ImmPXDis}. The initial GB energy distribution (GBED) is shown in Figure~\ref{fig:PXImmRSp}. With this particular distribution, several junctions will have a high anisotropy level, and as stated in \cite{Elsey2013, ma14143883}, one can expect different behaviours using the different formulations.

\begin{figure}[H]
  
  \centering
  \begin{subfigure}{0.495\textwidth}
    \centering
    \includegraphics[scale=0.295]{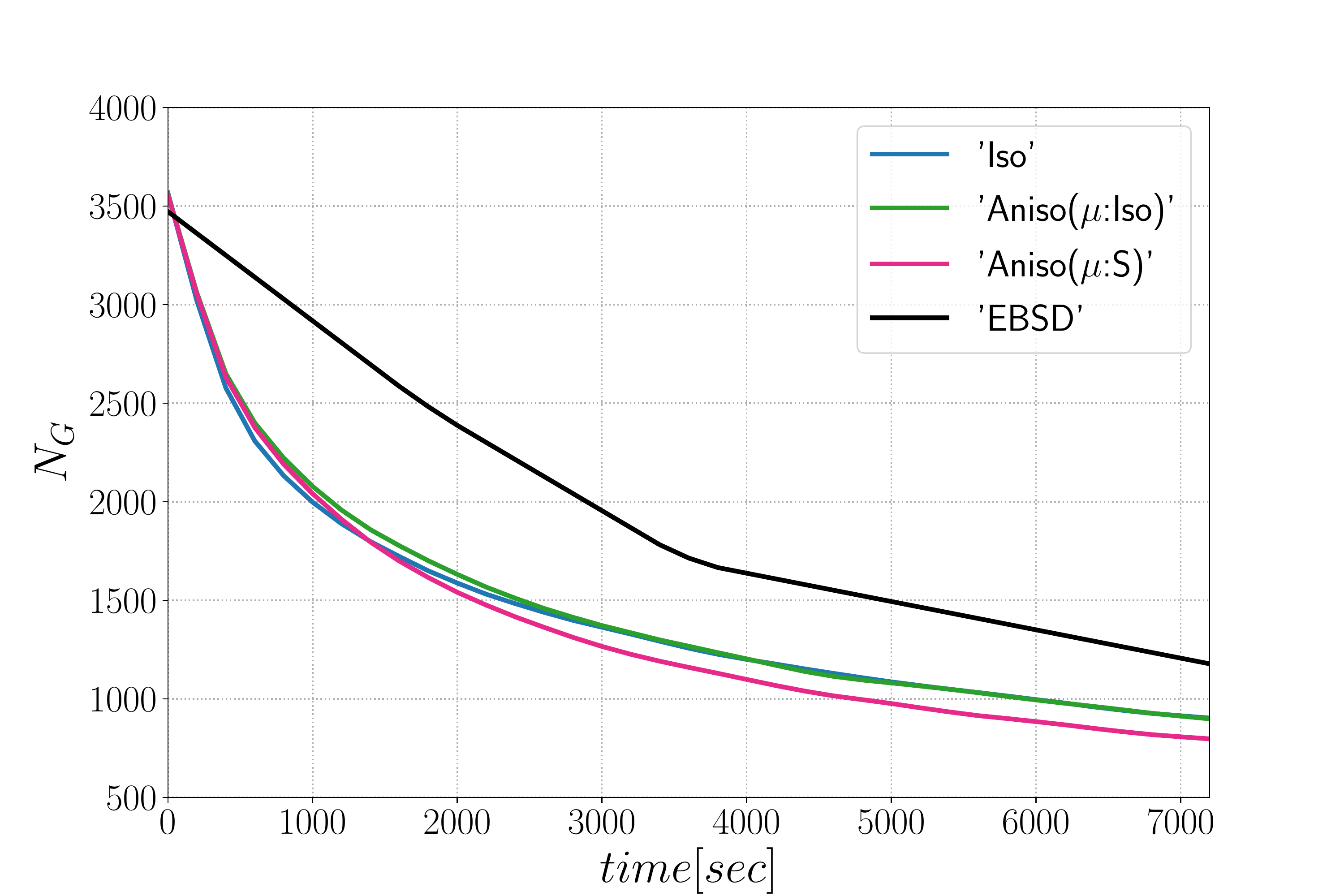}
    \caption{$N_g=f(t)$} 
  \end{subfigure}
  \begin{subfigure}{0.495\textwidth}
    \centering
    \includegraphics[scale=0.295]{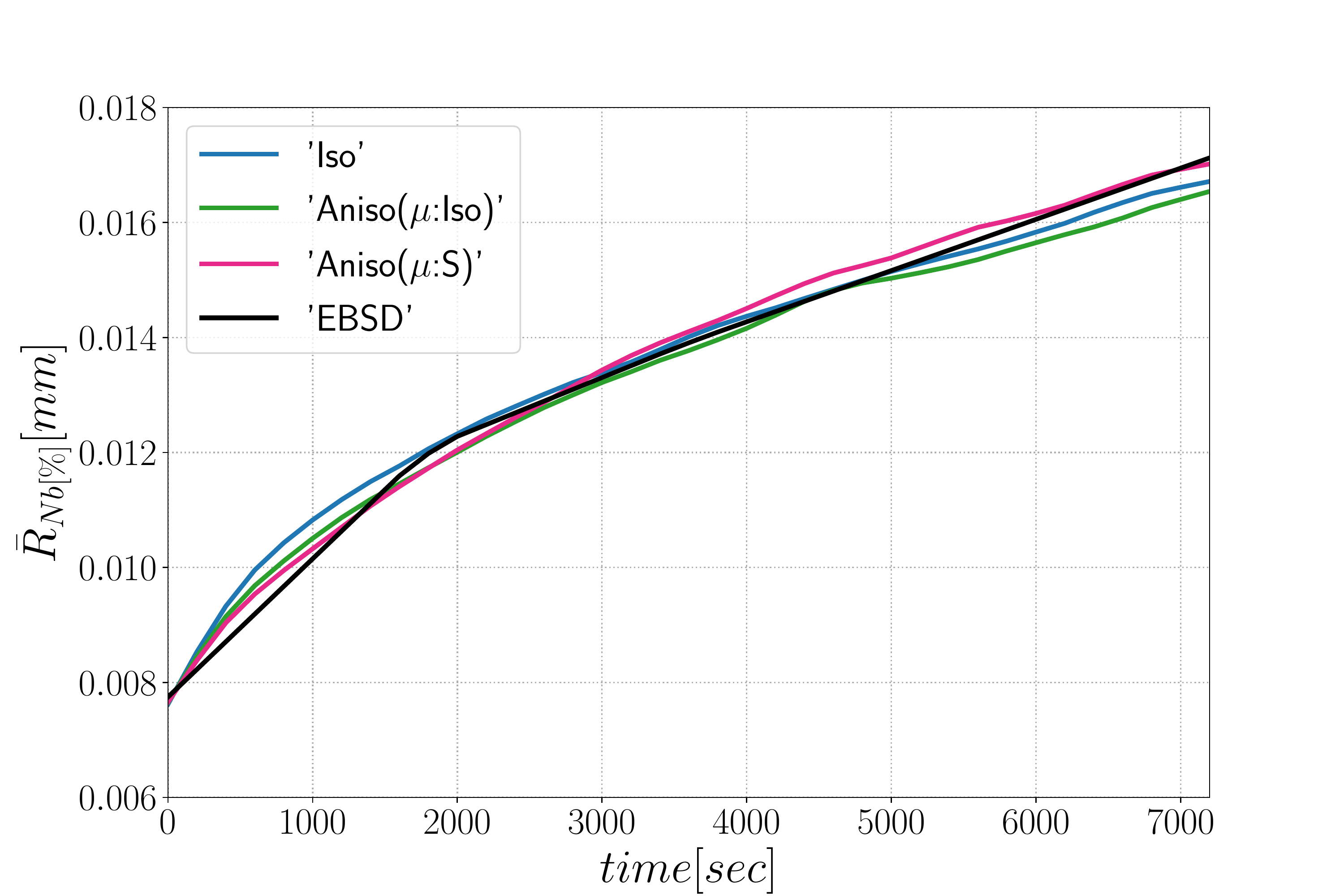}
    \caption{$\bar{R}_{Nb[\%]}=f(t)$}
  \end{subfigure}
 \\
  \begin{subfigure}{0.495\textwidth}
    \centering
    \includegraphics[scale=0.295]{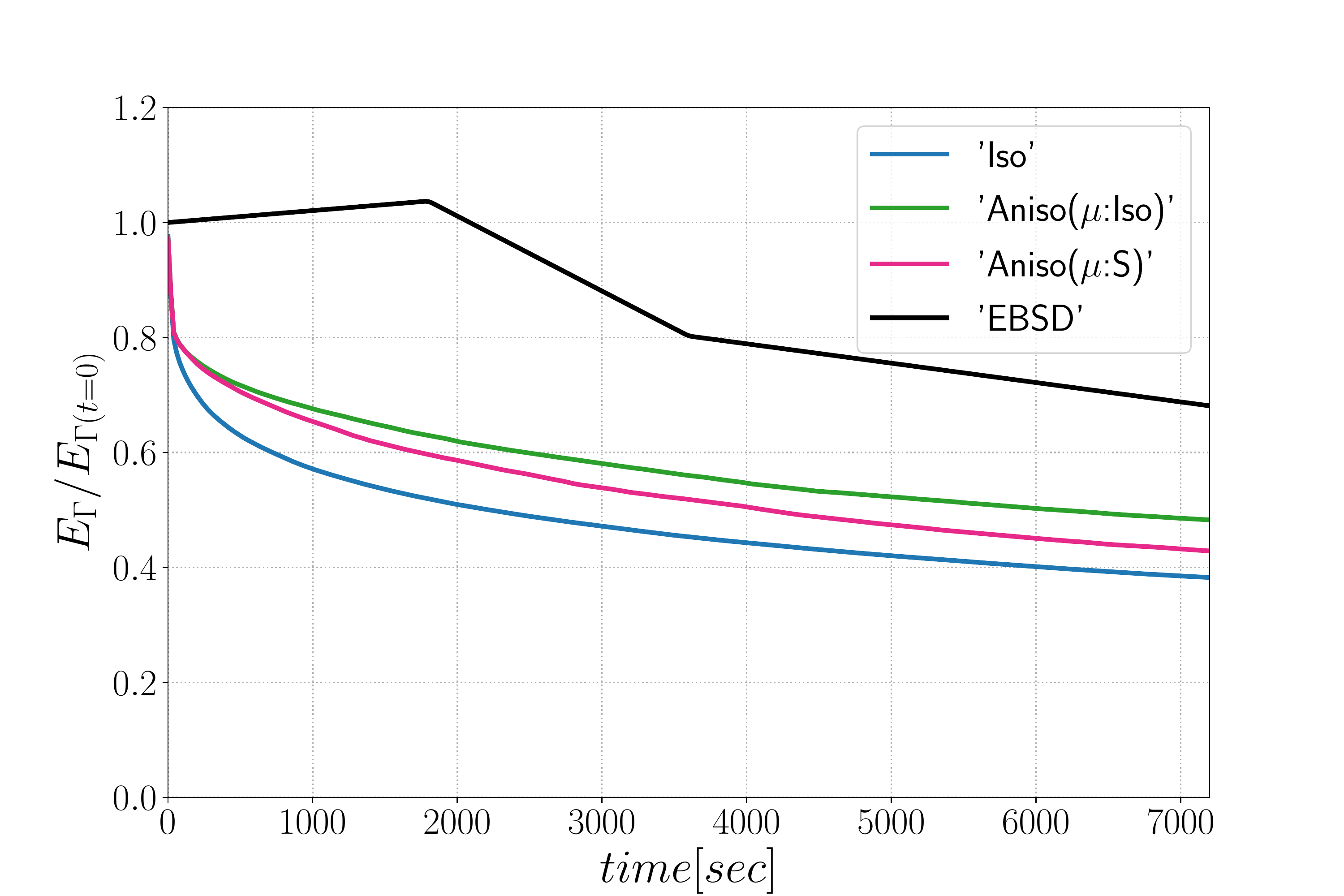}
    \caption{$E_{\Gamma}=f(t)$}   \vspace{3mm}
  \end{subfigure} 
  \begin{subfigure}{0.495\textwidth}
    \centering
    \includegraphics[scale=0.295]{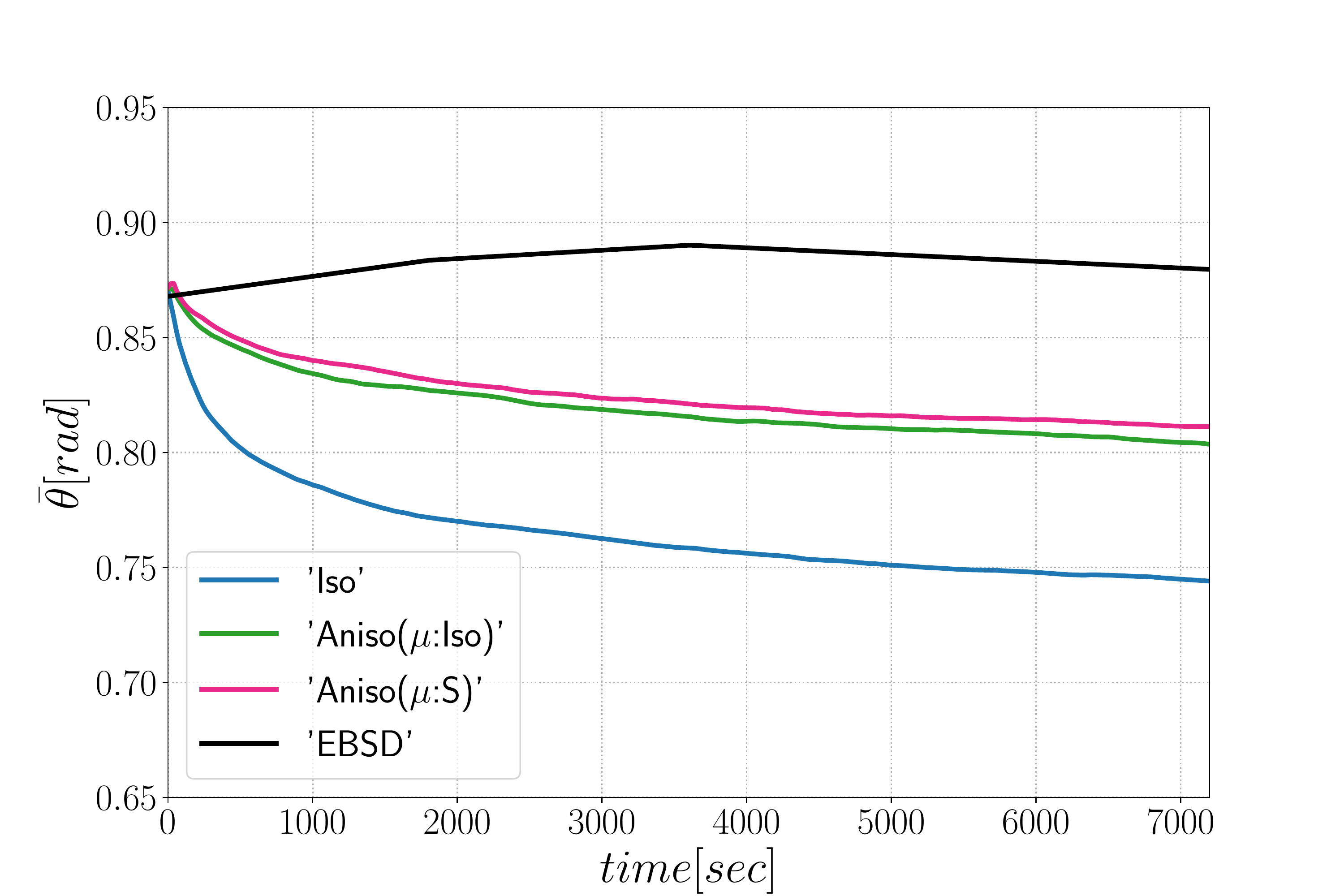}
    \caption{$\bar{\theta}=f(t)$}
  \end{subfigure} \\
  \begin{subfigure}{0.495\textwidth}
    \centering
    \includegraphics[scale=0.295]{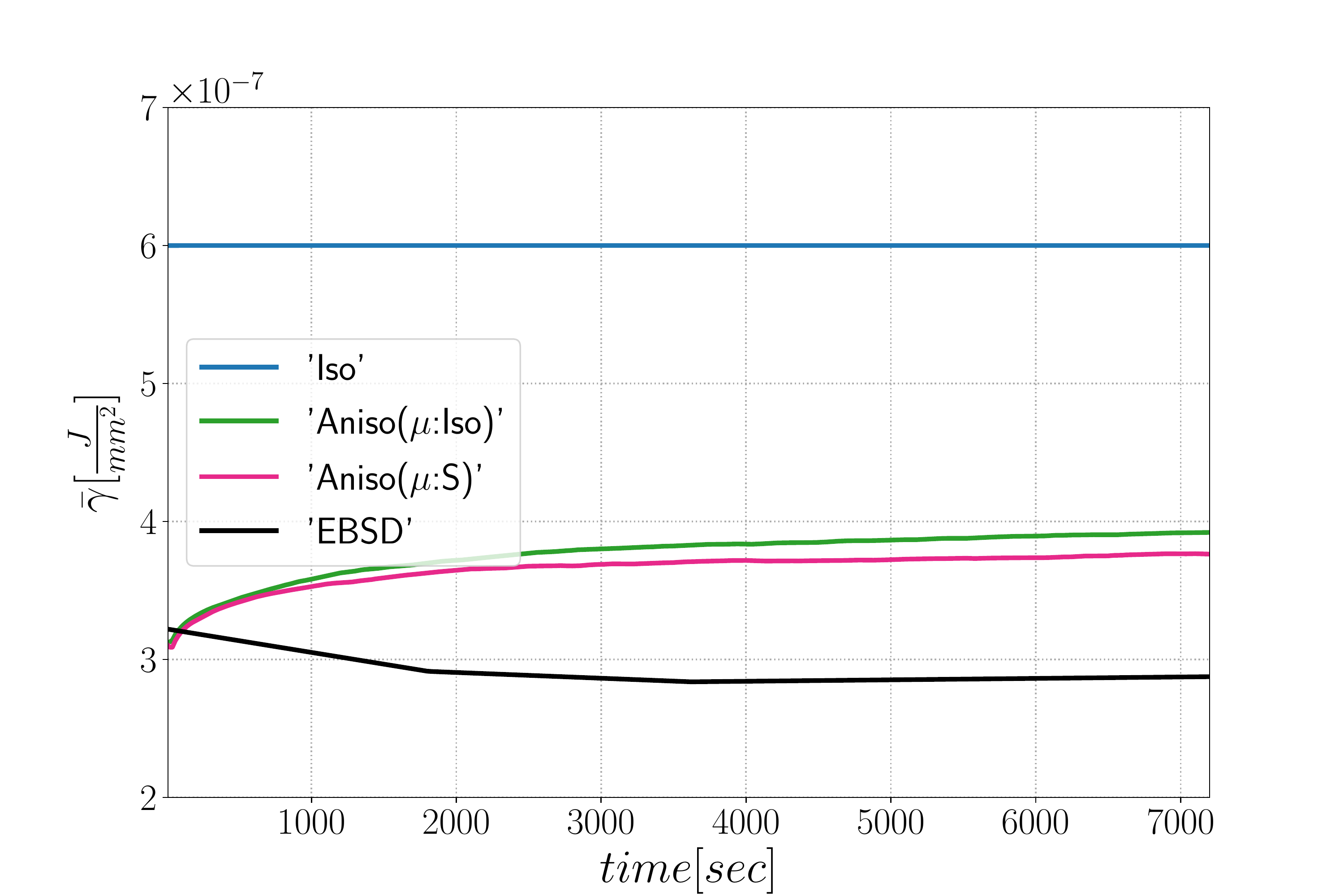}
    \caption{$\bar{\gamma}=f(t)$}
  \end{subfigure}
  \vspace{5pt}
  \caption{Mean values time evolution for the isotropic (Iso) formulation, anisotropic formulations with isotropic GB mobility (Aniso($\mu$:Iso)) and heterogeneous GB mobility (Aniso($\mu$:S)) and the experimental data (EBSD). Numerical results obtained from the initial microstructure shown in Figure~\ref{fig:ImmPX}.}\label{fig:ImmPXMuIsoMeanV}
\end{figure}

As illustrated in Figure~\ref{fig:ImmPXMuIsoMeanV}, the three simulations predict similar trends concerning the mean grain size and the grain number evolution. Comparatively to experimental EBSD data, these predictions are very good concerning the mean grain size prediction but all of them tend to predict, at the beginning, a faster disappearance of the small grains. Concerning the total energy, mean GB disorientation and mean GB energy evolutions, the Anisotropic formulation is closer to the EBSD data. This means that the Aniso formulation is more physical and promotes GBs with low GB energy.   

\clearpage

\begin{figure}[H]
  \centering
  \includegraphics[width=1.0\textwidth]{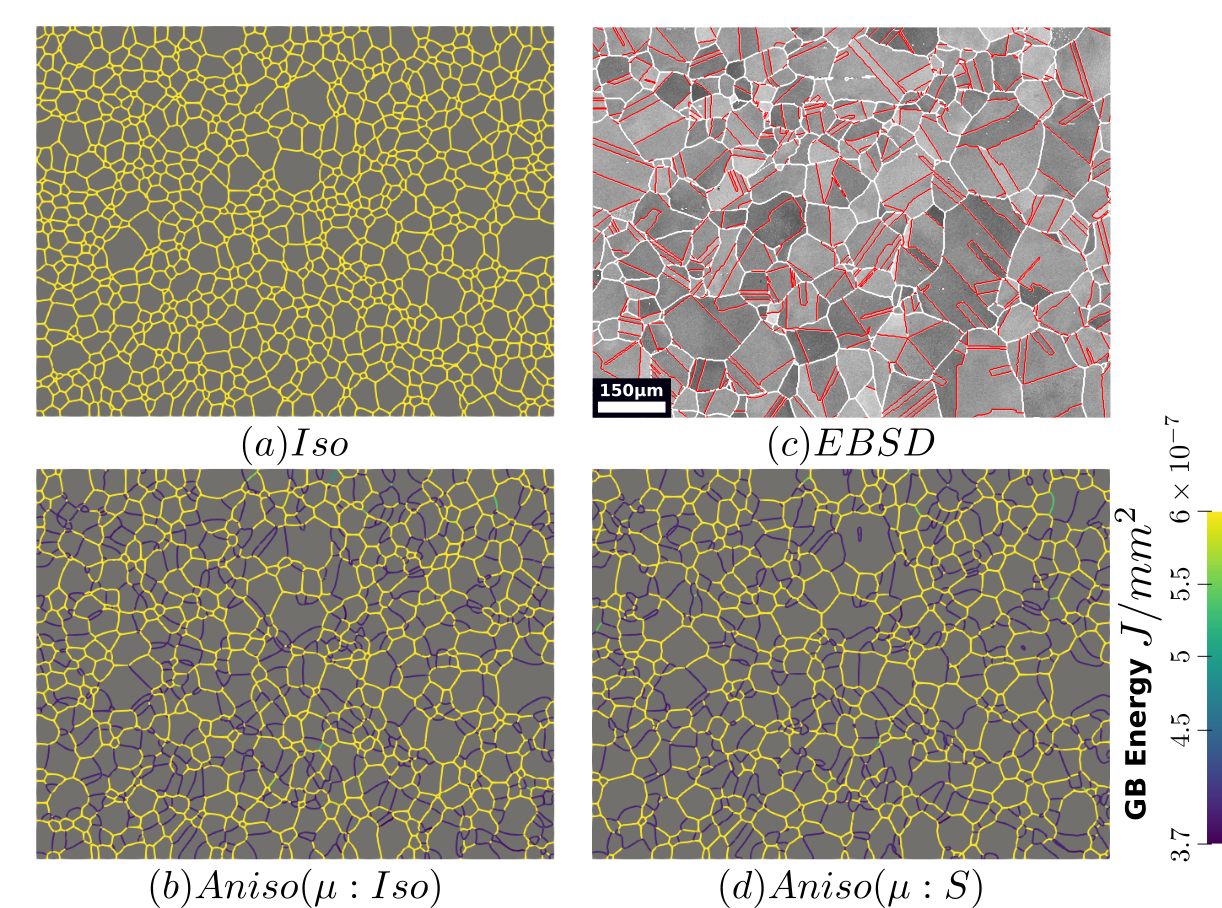}
  \caption{GB energy of the microstructures obtained numerically using the (a) isotropic formulation and the anisotropic formulations with (b) isotropic and (d) heterogeneous GB mobility and the experimental band contrast map at t=2h.}
  \label{fig:ImmPXMicroMuIso}
\end{figure}

\begin{figure}[H]
  \centering
  \begin{subfigure}{0.48\textwidth}
    \centering
    \includegraphics[scale=0.25]{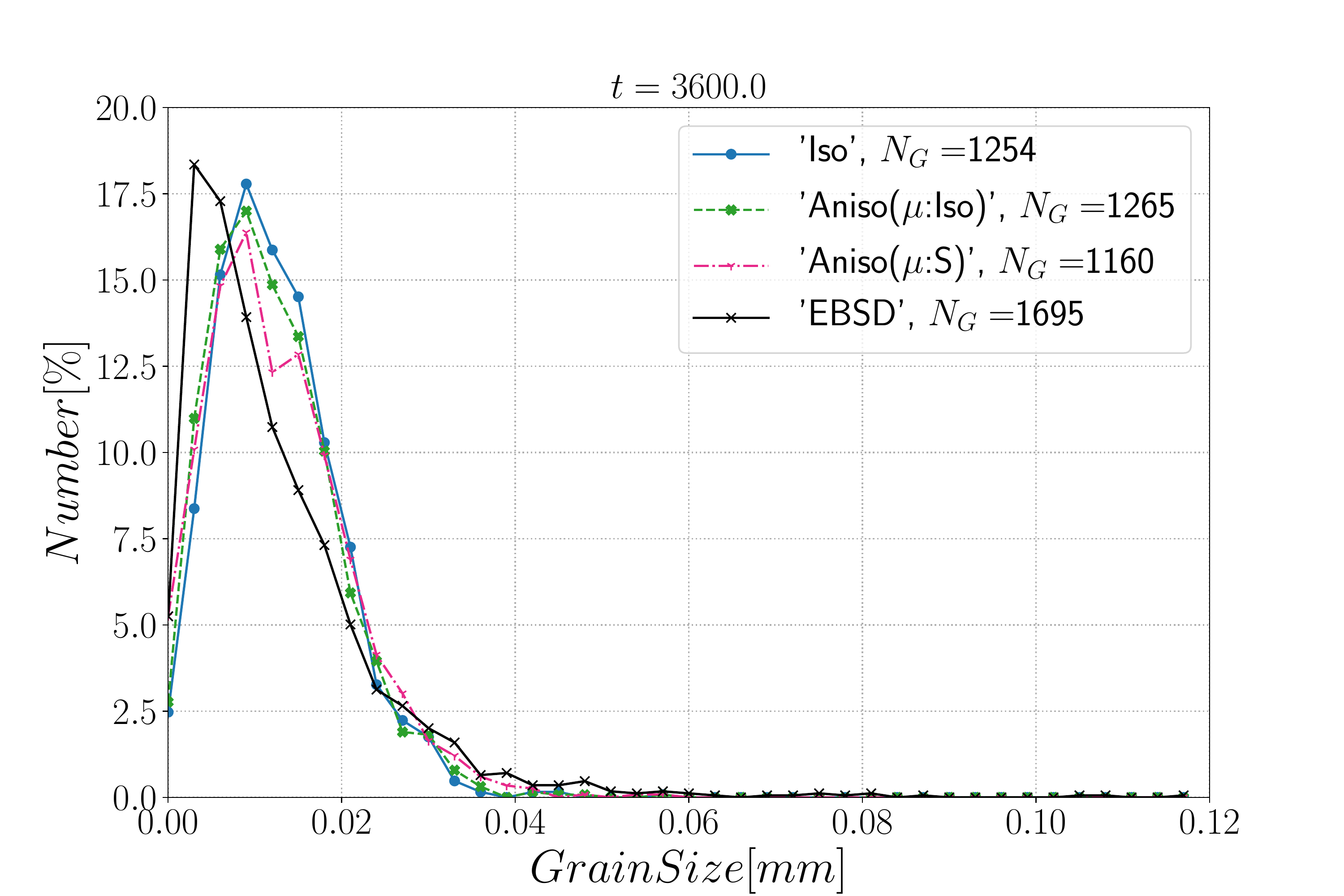}
    \caption{$ t=1h $}
  \end{subfigure}
  \begin{subfigure}{0.48\textwidth}
    \centering
    \includegraphics[scale=0.25]{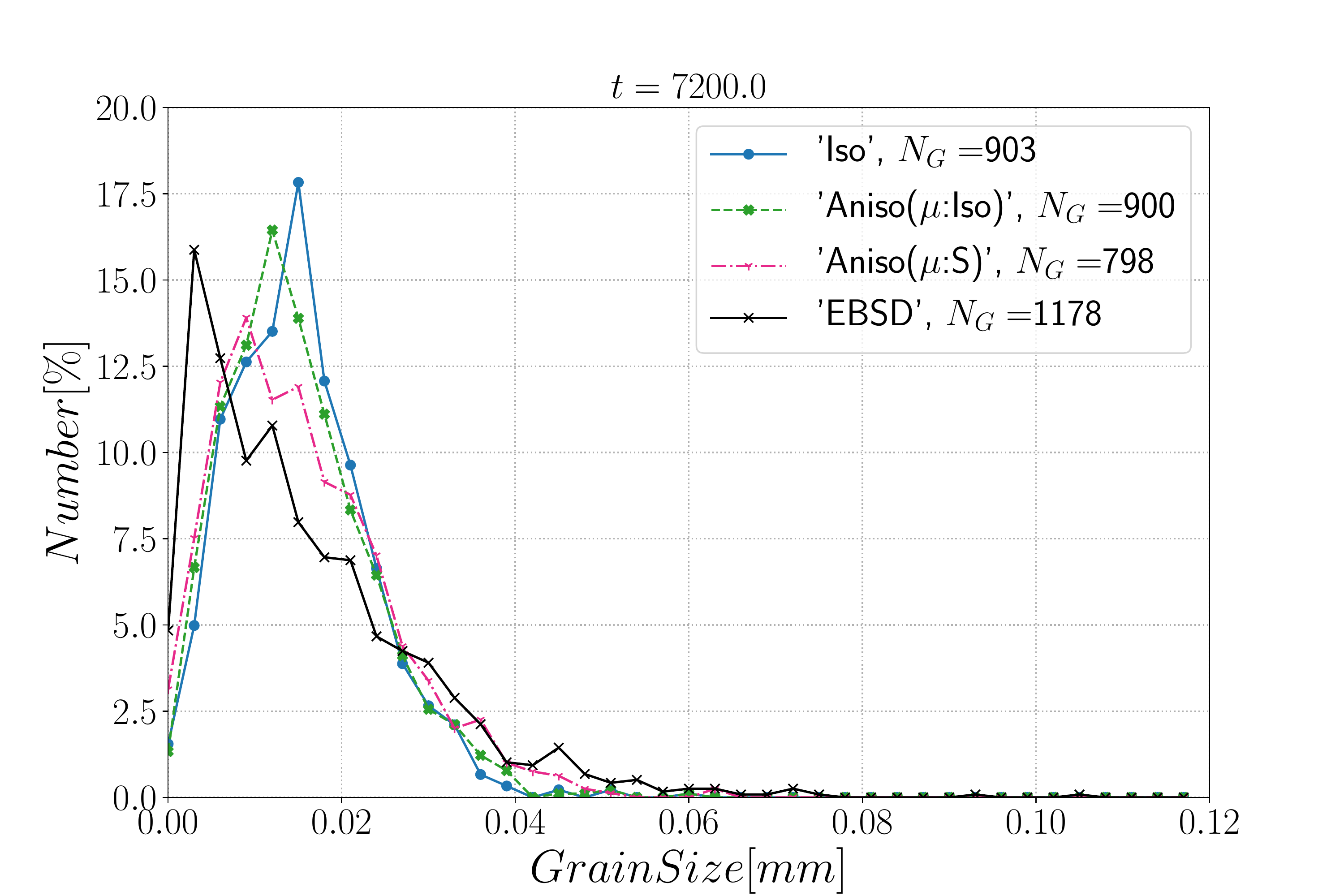}
    \caption{$ t=2h $}
  \end{subfigure}
  \caption{Grain Size Distributions obtained at (a) t=1h and (b) t=2h for the isotropic (Iso) formulation, anisotropic formulations with isotropic GB mobility (Aniso($\mu$:Iso)) and heterogeneous GB mobility (Aniso($\mu$:S)) and the experimental data (EBSD), $N_G$ refers to the number. Numerical results obtained from the initial immersed microstructure shown in Figure~\ref{fig:ImmPX} and the RS and Sigmoidal model to define GB energy and mobility.}
  \label{fig:ImmPXGSDMuIso}
\end{figure}

\begin{figure}[H]
  \centering
  \begin{subfigure}{0.48\textwidth}
    \centering
    \includegraphics[scale=0.25]{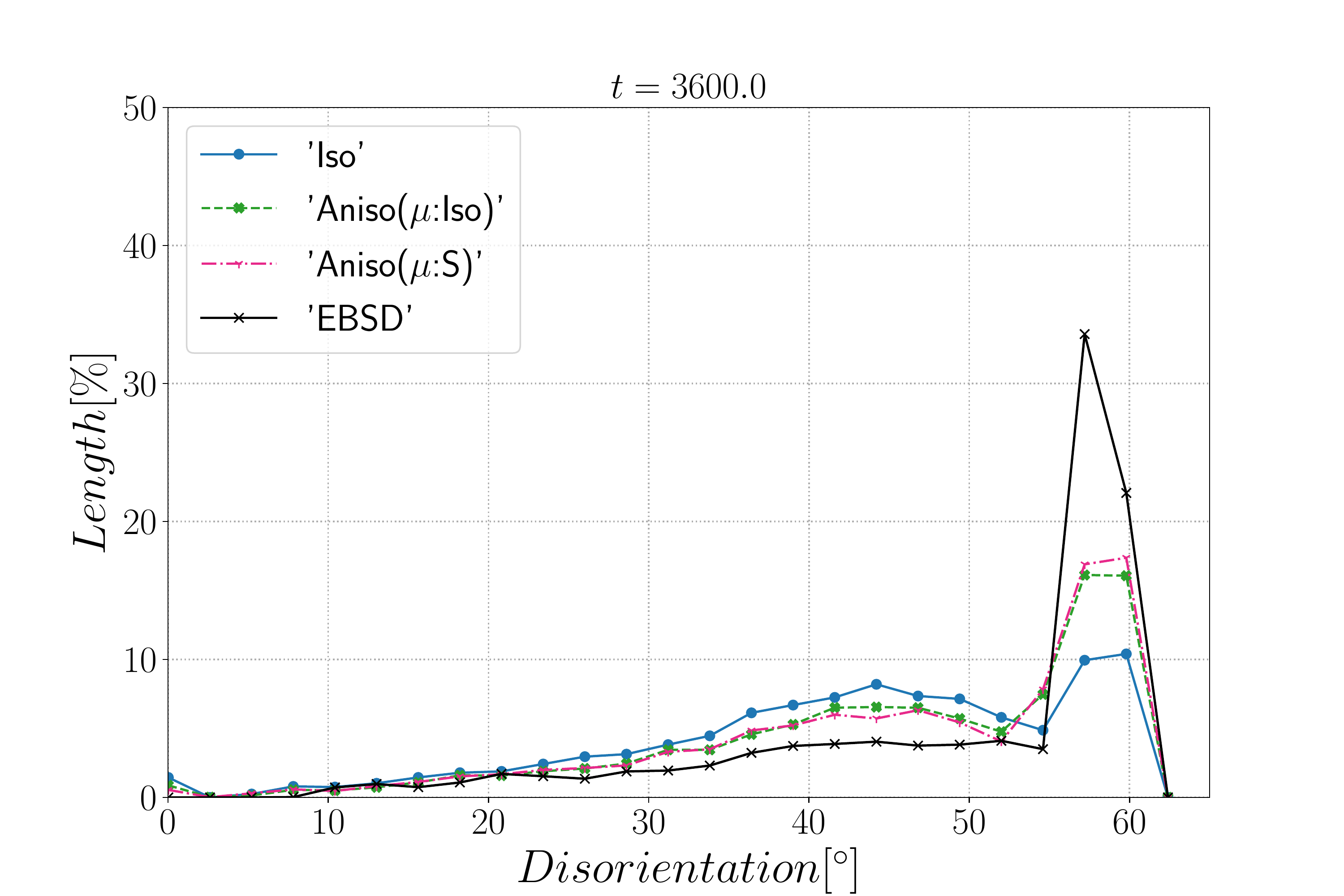}
    \caption{$ t=1h $}
  \end{subfigure}
  \begin{subfigure}{0.48\textwidth}
    \centering
    \includegraphics[scale=0.25]{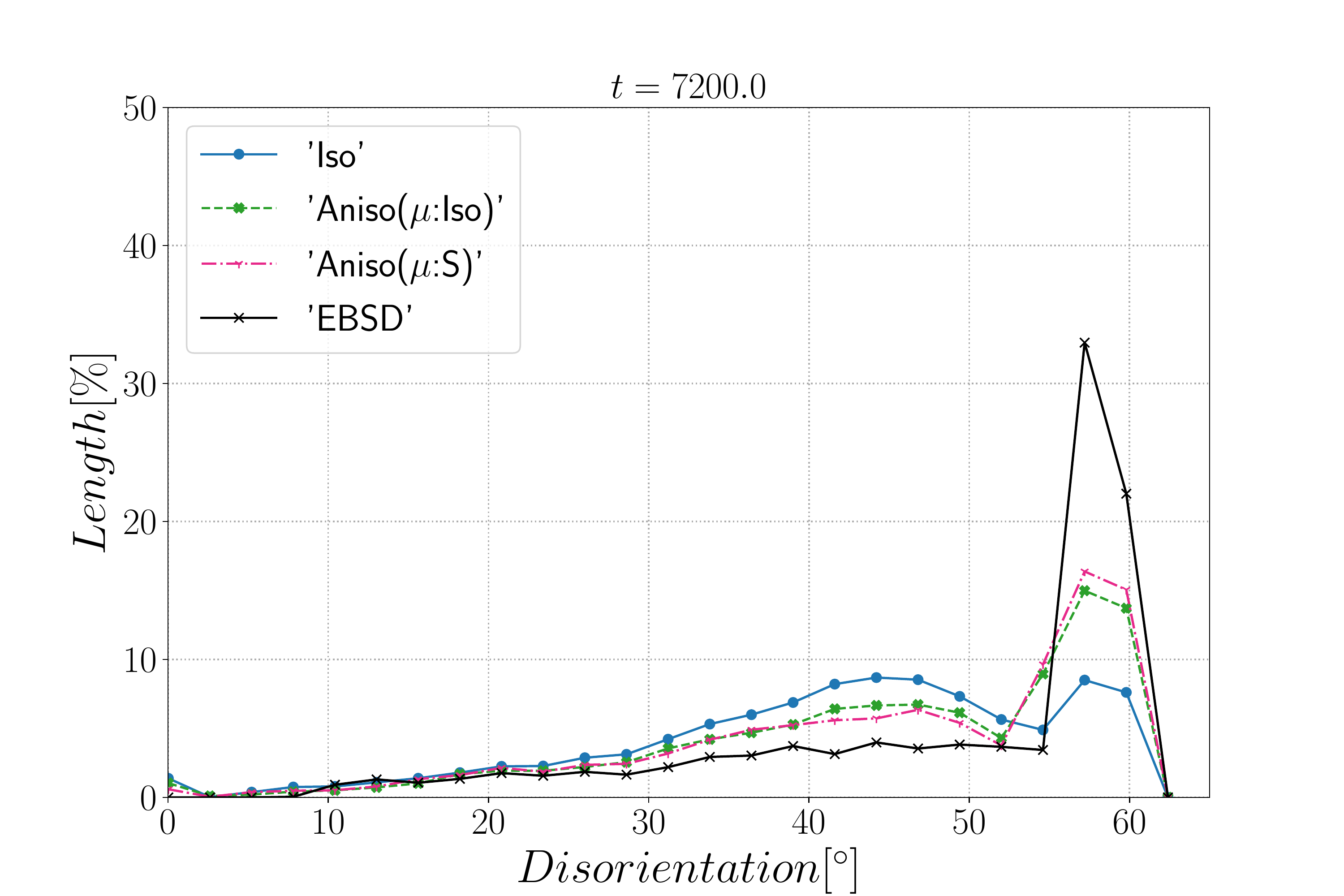}
    \caption{$ t=2h $}
  \end{subfigure}
  \caption{Disorientation Distribution obtained at (a) t=1h and (b) t=2h for the isotropic (Iso) formulation, anisotropic formulations with isotropic GB mobility (Aniso($\mu$:Iso)) and heterogeneous GB mobility (Aniso($\mu$:S)) and the experimental data (EBSD). The y-axis represents the GB length percentage. Numerical results obtained from the initial immersed microstructure shown in Figure~\ref{fig:ImmPX} and the RS and Sigmoidal model to define GB energy and mobility.}
  \label{fig:ImmPXDDFMuIso}
\end{figure}

Figure~ \ref{fig:ImmPXMicroMuIso} illustrates the topology of grains at $t=2h$. One can notice the higher quantity of GBs with low GB energy using the Anisotropic formulation and its similarity to the EBSD band contrast map even if one can notice from the EBSD data that the real microstructure contains more TBs that creates small grains as reflected in the GSD in Figure~\ref{fig:ImmPXGSDMuIso}. Another advantage of the Anisotropic formulation is the better reproduction of the DDF compared to the Iso formulation which contains a lower percentage of GBs with $\theta \approx \theta_{\Sigma 3}$ and tends to promote a Mackenzie-like DDF (see Figure~\ref{fig:ImmPXDDFMuIso}). One can also see in Figures~\ref{fig:ImmPXMicroMuIso}, \ref{fig:ImmPXGSDMuIso} and \ref{fig:ImmPXDDFMuIso} that the heterogeneous GB mobility improves the morphology of grain, the GSD and the DDF.  

In this section it has been shown that the immersed data gives a better insight of the real microstructure evolution. In terms of mean values, there is small differences between the results from the Laguerre-Voronoï tesselation and the immersed microstructure. However, the GSD and DDF distributions are better reproduced for the immersed case when the anisotropic formalism is adopted. The heterogeneous GB mobility affects the grain topology, the GSD and the DDF due to its additional retarding effect. Regarding the GSD, the Aniso($\mu$:S) formulation can reproduce the peak at low values of grain size and the peak around $\theta_{\Sigma 3}$ for the DDF which are due to the TBs. Nevertheless, the behavior of TBs is still not perfectly reproduced by the proposed simulations, being the anisotropic formulation the one that seems more physical. In the next section this issue is addressed using the GB5DOF code which allows to define the GB energy in terms of misorientation and GB inclination (in 2D) in order to better characterize the evolution of TBs. 

\section{Using anisotropic GB energy and heterogeneous GB mobility}
\label{sec:PXGB5DOF}
\subsection{Simulation results} 
\begin{figure}[H]

\centering
\begin{subfigure}{0.48\textwidth}
  \centering
  \includegraphics[scale=0.4]{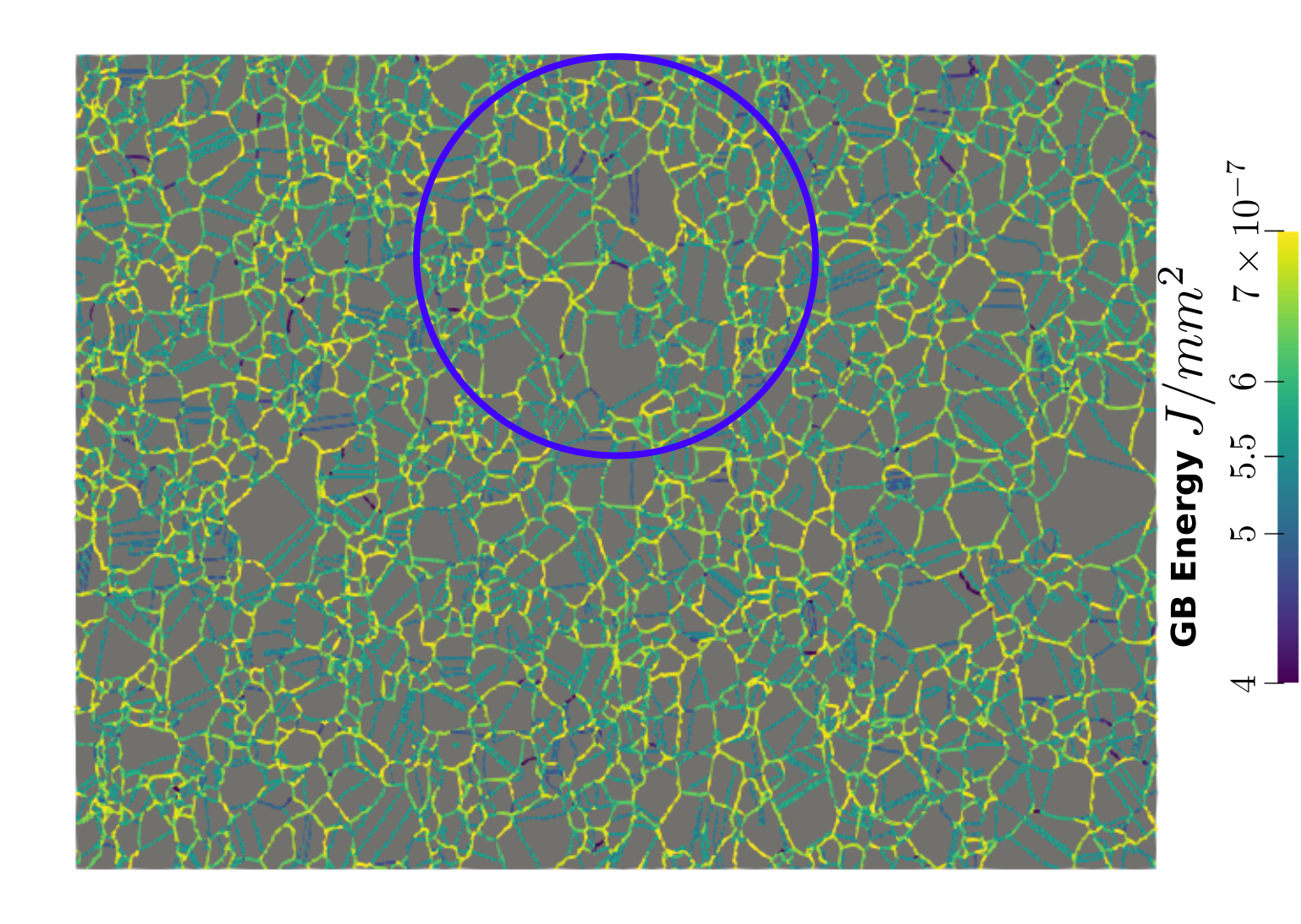}
  \caption{GB energy of the initial microstructure}
  \label{fig:GB5DOFImmPX}
\end{subfigure}
\begin{subfigure}{0.48\textwidth}
  \centering
  \includegraphics[scale=0.25]{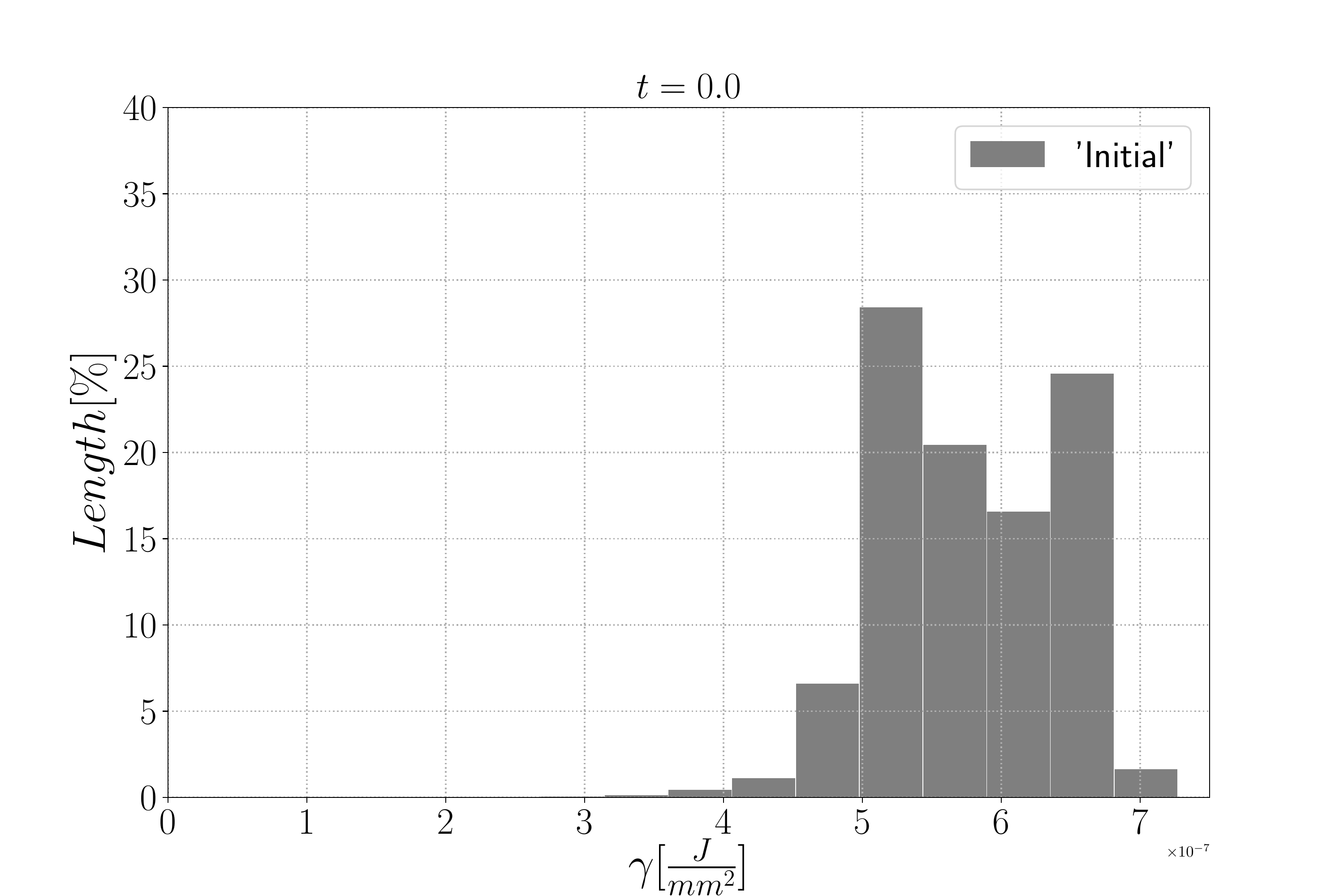}
  \caption{Initial GBED}
  \label{fig:GB5DOFImmPXInit}
\end{subfigure}
  \caption{(a) GB energy field and (b) GBED of the initial immersed microstructure obtained using the GB5DOF code with the parameters $\epsilon_{RGB} = 0.763 \ J mm^{-2}$ and AlCu-parameter$=0$. In (a) the blue circle shows a zone of interest with a twin boundary composed of a coherent and incoherent part, similar to the one shown in \cite{ma14143883}. }
\label{fig:GB5DOFImmPXInitChar}
\end{figure}

In this section, the immersed polycrystalline microstructure and the FE mesh presented in section~\ref{sec:PX} are used. Anisotropic GB energy values are defined using the GB5DOF code \cite{bulatov2014grain} and heterogeneous GB mobility are described using Equation~\ref{eqn:MobP}. It means that the GB energy can vary with the GB misorientation and inclination even if the torque terms are neglected. Note that the GB inclination is measured in 2D and not 3D, in other words, the GB is supposed to be perpendicular to the EBSD map. The GB Energy of the microstructure and its GBED are shown in Figure~\ref{fig:GB5DOFImmPXInitChar}. The initial microstructure is shown in Figure~\ref{fig:GB5DOFImmPX}, one can see that the maximum value of GB energy is set around $\gamma_{max} \approx 7 \times 10^{-7} J \cdot mm^{-2}$. As discussed in \cite{bulatov2014grain}, incoherent $\Sigma 3$ TBs have a GB energy defined as $\gamma_{\Sigma 3} \approx 0.6*\gamma_{max}$ meaning that the modified Read-Shockley model described by Equation~\ref{eqn:GammaP} seems exaggerated. In Figure~\ref{fig:GB5DOFImmPXInit} the GBED is concentrated within the values $4 \times 10^{-7} J\cdot mm^{-2} \leq \gamma \leq 7 \times 10^{-7} J\cdot mm^{-2}$ which means that the level of heterogeneity is low and the different formulations are expected to promote similar trends as stated in \cite{ma14143883}. The GB mobility was set to fit the evolution of the mean grain size, the maximal GB mobility for the Aniso($\mu$:Iso) and Aniso($\mu$:S) formulations are respectively set to $\mu_{max} = 0.0767\ mm^4\cdot J^{-1}\cdot s^{-1}$ and $\mu_{max} = 0.1423\ mm^4\cdot J^{-1}\cdot s^{-1}$. The difference between the $\mu_{max}$ values are generated by the higher values of GB energy produced by the GB5DOF code, note that with the RS model all the TBs are defined as coherent while the GB5DOF code can distinguish between a coherent twin boundaries and  incoherent twin boundaries as pointed out in \cite{hallberg2019modeling, ma14143883}.

\vspace{-0.7cm}

\begin{figure}[H]
  \centering
  \begin{subfigure}{0.48\textwidth}
    \centering
    \includegraphics[scale=0.25]{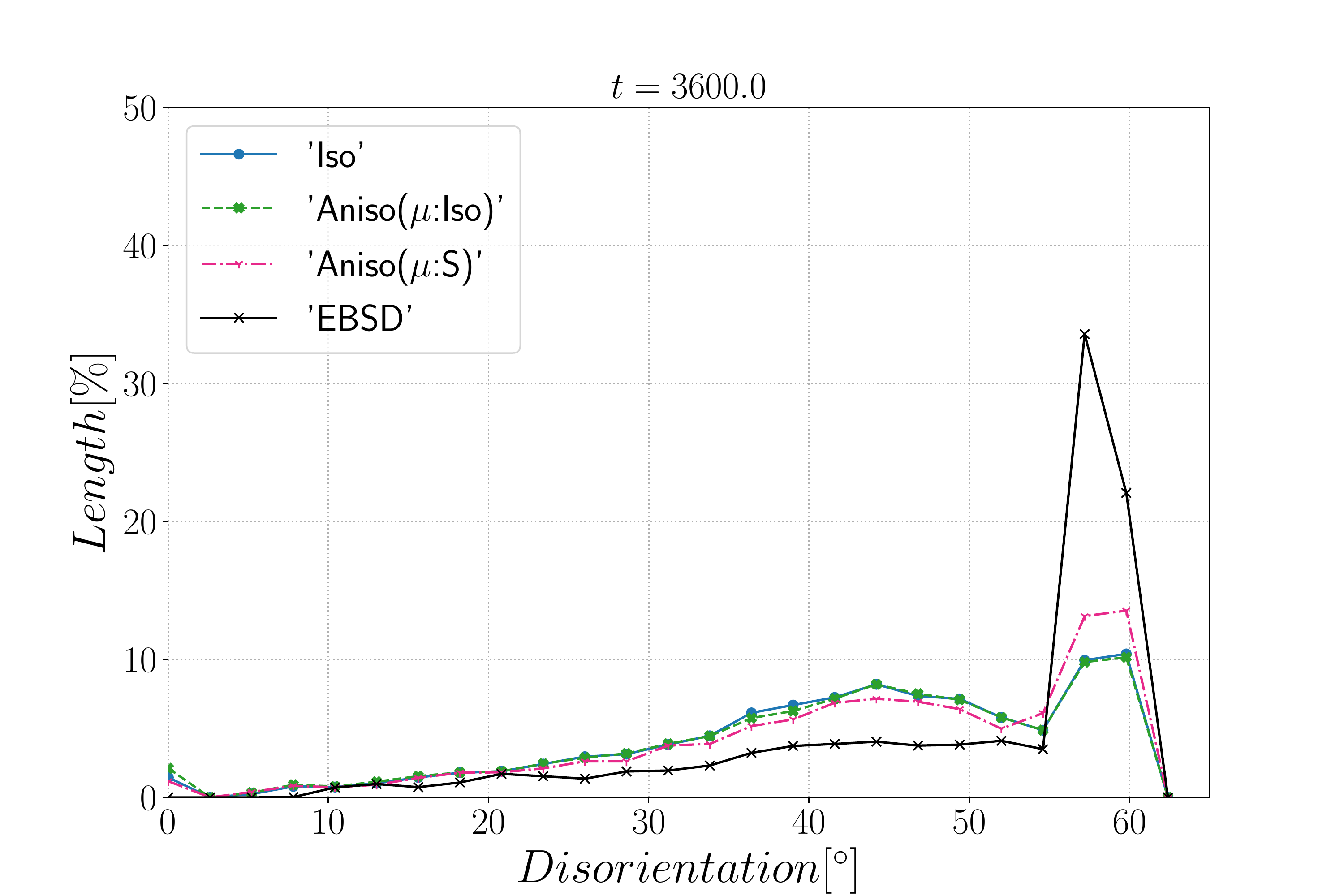}
    \caption{$ t=1\ h $}
  \end{subfigure}
  \begin{subfigure}{0.48\textwidth}
    \centering
    \includegraphics[scale=0.25]{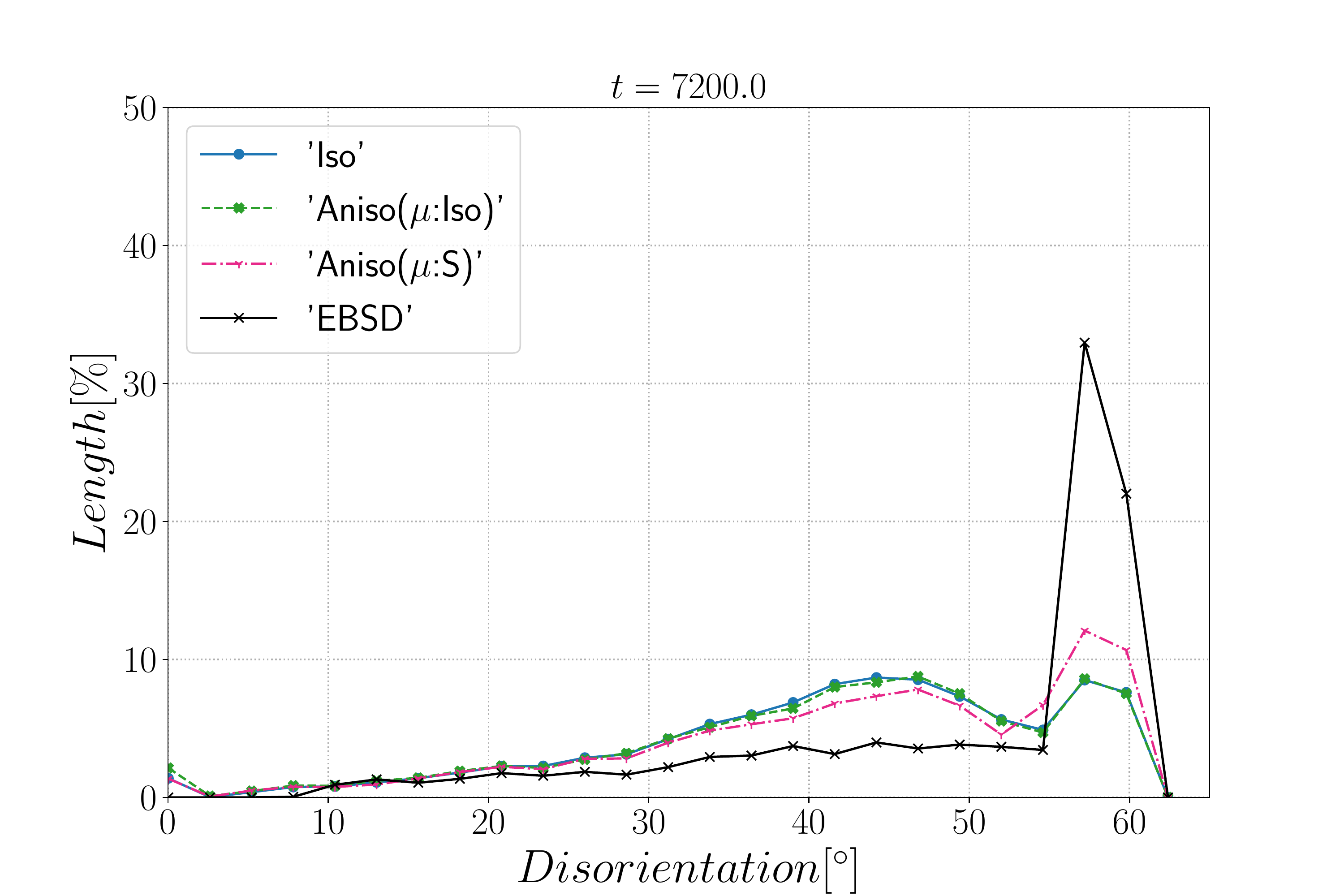}
    \caption{$ t=2\ h $}
  \end{subfigure}
  \caption{Disorientation Distribution obtained at (a) t=1h and (b) t=2h for the isotropic (Iso) formulation, anisotropic formulations with isotropic GB mobility (Aniso($\mu$:Iso)) and heterogeneous GB mobility (Aniso($\mu$:S)) and the experimental data (EBSD). The y-axis represents the GB length percentage. Numerical results obtained from the initial immersed microstructure shown in Figure~\ref{fig:GB5DOFImmPX} and the GB5DOF code to define the GB energy.}
  \label{fig:GB5DOFImmPXDDFMuIso}
\end{figure}

\begin{figure}[H]
  \centering
  \begin{subfigure}{0.48\textwidth}
    \centering
    \includegraphics[scale=0.25]{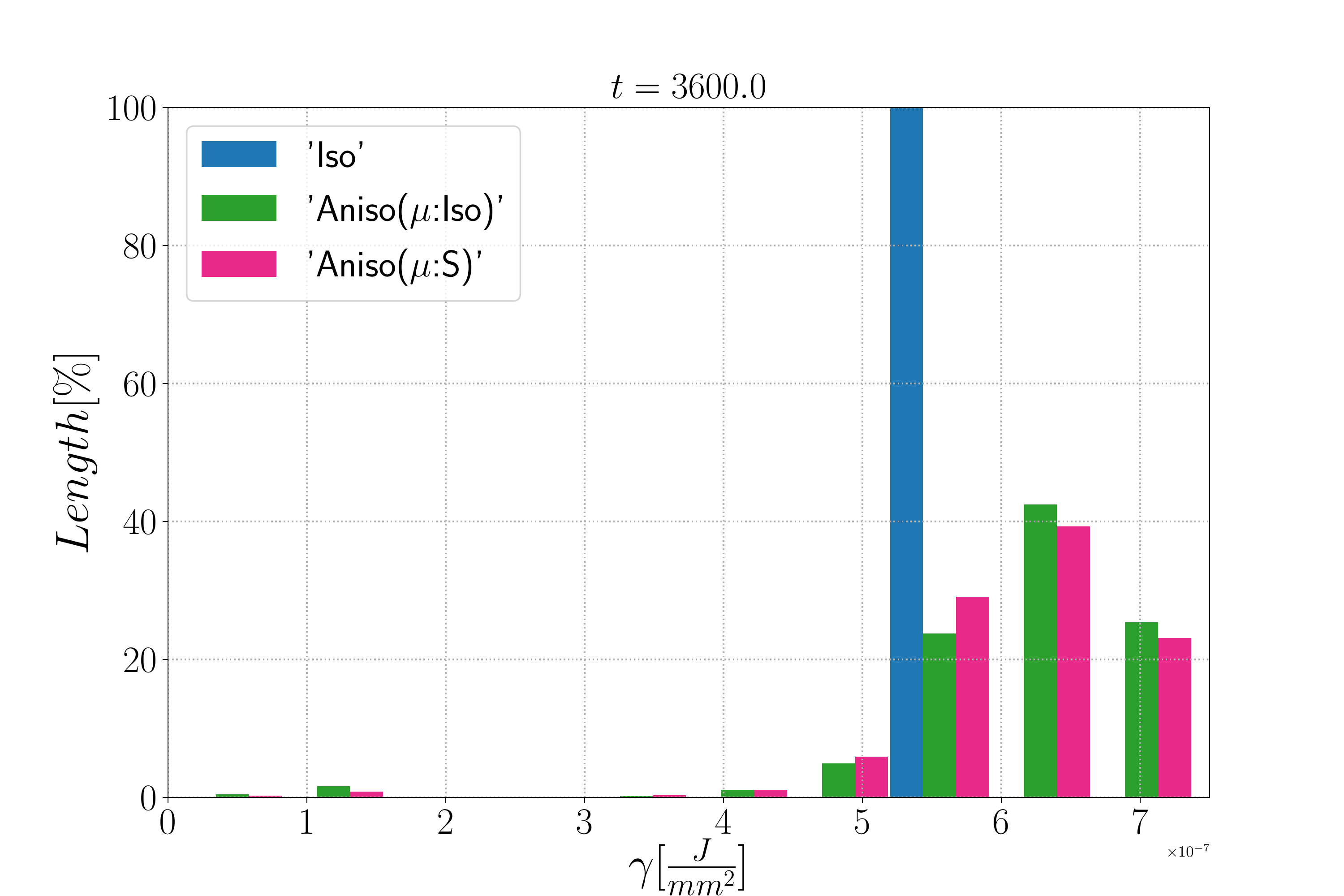}
    \caption{$ t=1\ h $}
  \end{subfigure}
  \begin{subfigure}{0.48\textwidth}
    \centering
    \includegraphics[scale=0.25]{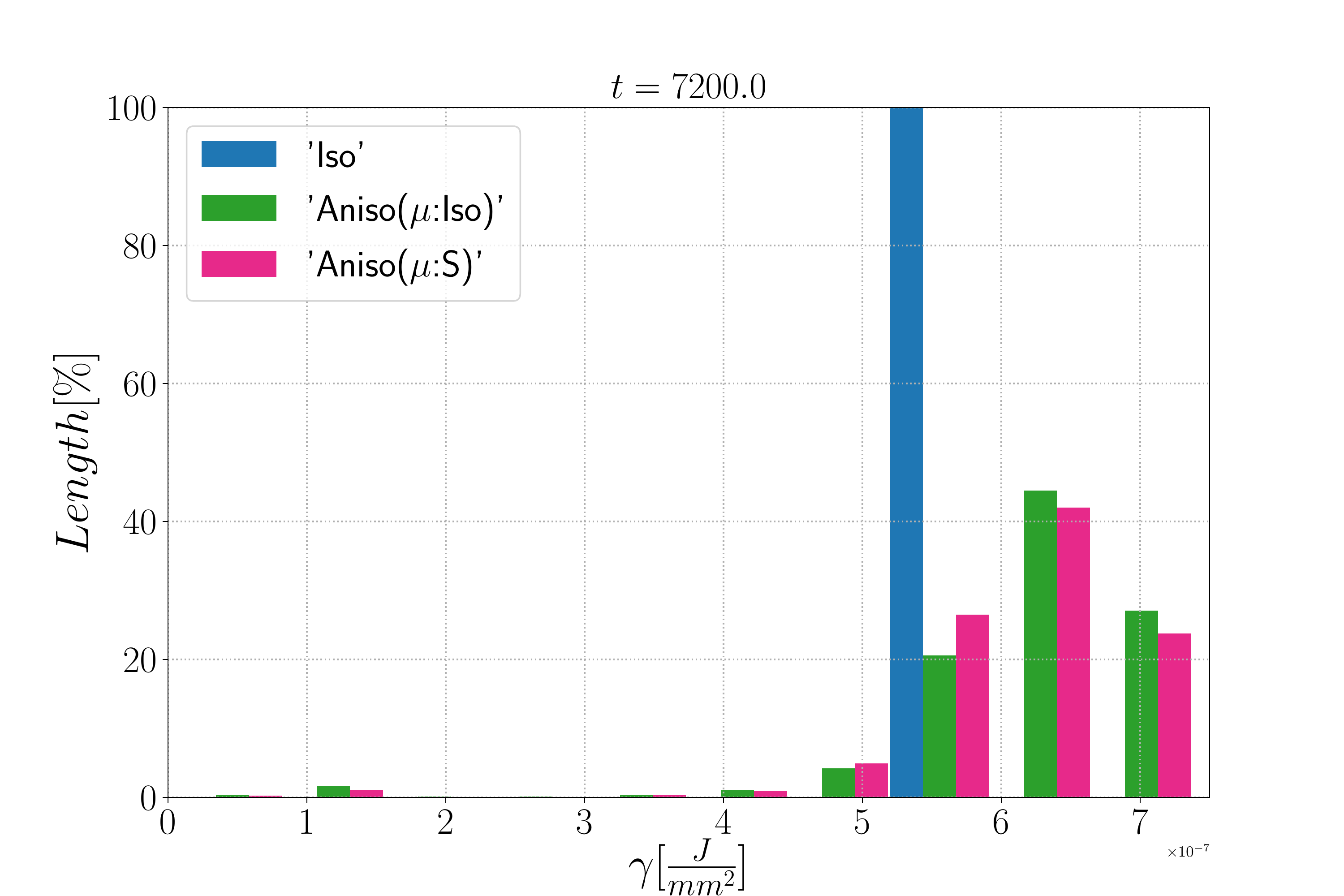}
    \caption{$ t=2\ h $}
  \end{subfigure}
  \caption{GBED obtained at (a) t=1h and (b) t=2h for the isotropic (Iso) formulation, anisotropic formulations with isotropic GB mobility (Aniso($\mu$:Iso)) and heterogeneous GB mobility (Aniso($\mu$:S)). The y-axis represents the GB length percentage. Numerical results obtained from the initial immersed microstructure shown in Figure~\ref{fig:GB5DOFImmPX} and the GB5DOF code to define the GB energy.}
  \label{fig:GB5DOFImmPXGBEDMuIso}
\end{figure}

\begin{figure}[H]
  \centering
  \includegraphics[width=1.0\textwidth]{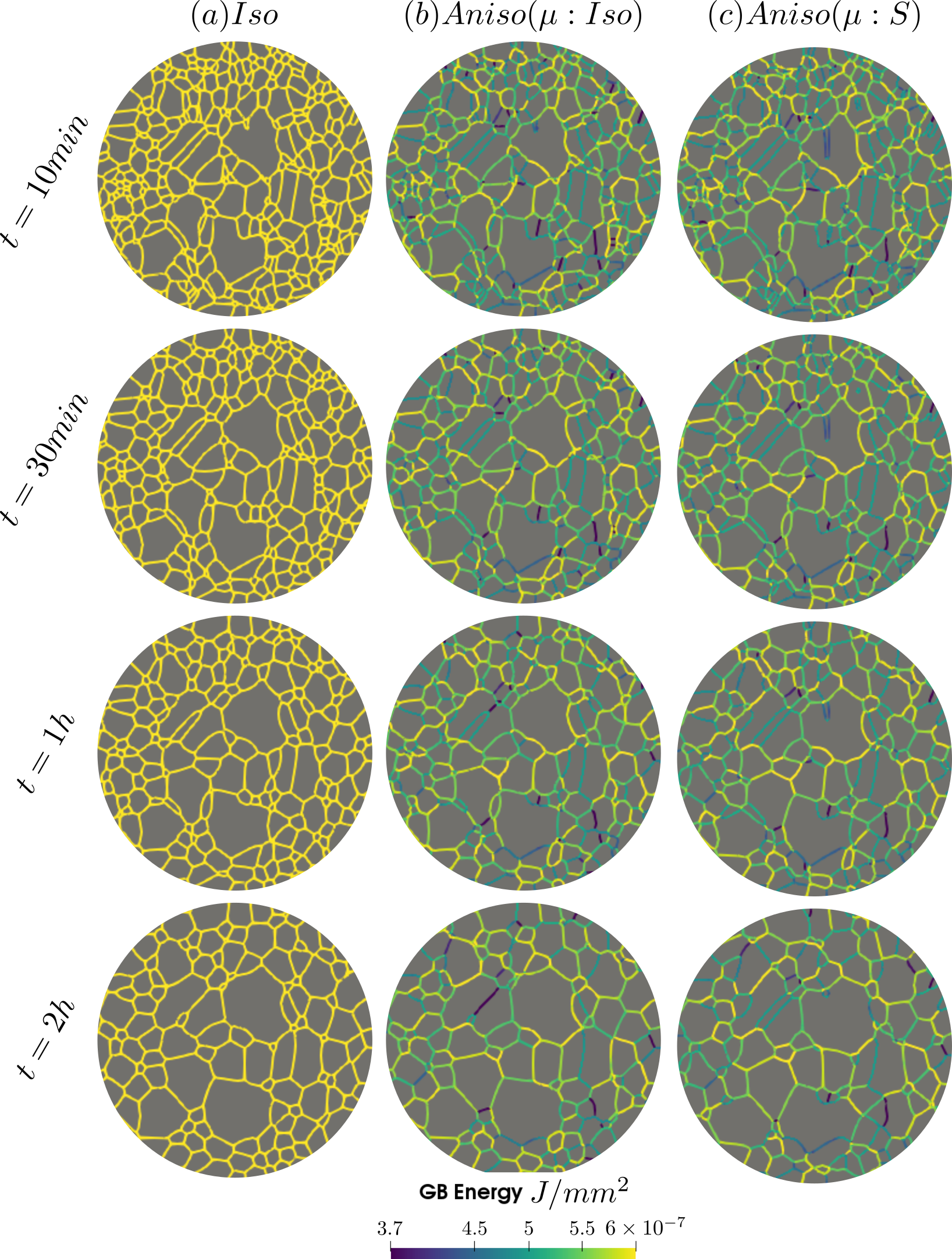}
  \caption{Microstructure evolution using the Isotropic formulation and Anisotropic formulation with isotropic and heterogeneous GB mobility at $t=30\ min$, $1\ h$ and $2\ h$. The zone shown here is encircled in blue in Figure~\ref{fig:GB5DOFImmPX}.}
  \label{fig:GB5DOFImmPXMicroMuIso}
\end{figure}

First the mean grain size evolution is well reproduced by the different simulations. The mean GB disorientation is not well represented by none of the formulations, the EBSD data show a stable value around $50 \degree$ while all numerical results exhibit a decreasing trend. This effect is due to the TBs and illustrates the inability for the numerical formulations to preserve or generate them. Additionally, the DDF from both formulations are similar and do not correspond to the experimental DDF (see Figure~\ref{fig:GB5DOFImmPXDDFMuIso}). The similarity between the isotropic and anisotropic simulations is due to the low anisotropy level, that may be produced by the lack of information of the GB inclination (see Figure~\ref{fig:GB5DOFImmPXGBEDMuIso}). As stated in \cite{ma14143883}, when the GBED is concentrated around a specific value, both formulations can present a similar trend. This is confirmed with Figure~\ref{fig:GB5DOFImmPXMicroMuIso} where a zoom on the GB network is shown at four different times and one cannot see any obvious difference among the obtained microstructures with the three different simulations. The main difference of these results lay in the ability of the anisotropic formulation to keep more $\Sigma 3$ TBs when the sigmoid description of $\mu$ is used.  

\vspace{-0.7cm}

\begin{figure}[H]
  \centering
  \begin{subfigure}{0.48\textwidth}
    \centering
    \includegraphics[scale=0.25]{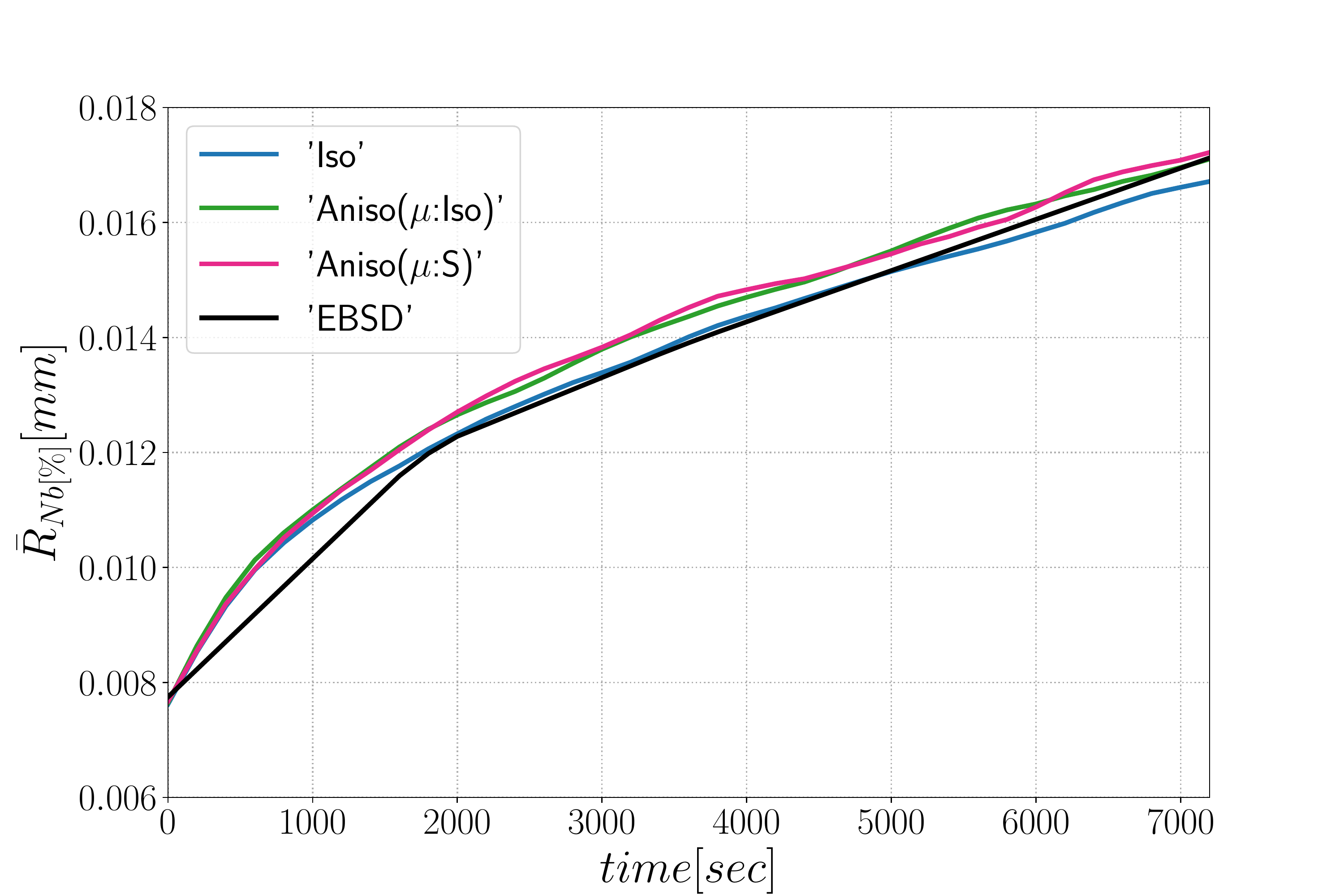}
    \caption{$N_g=f(t)$}
  \end{subfigure}
  \begin{subfigure}{0.48\textwidth}
    \centering
    \includegraphics[scale=0.25]{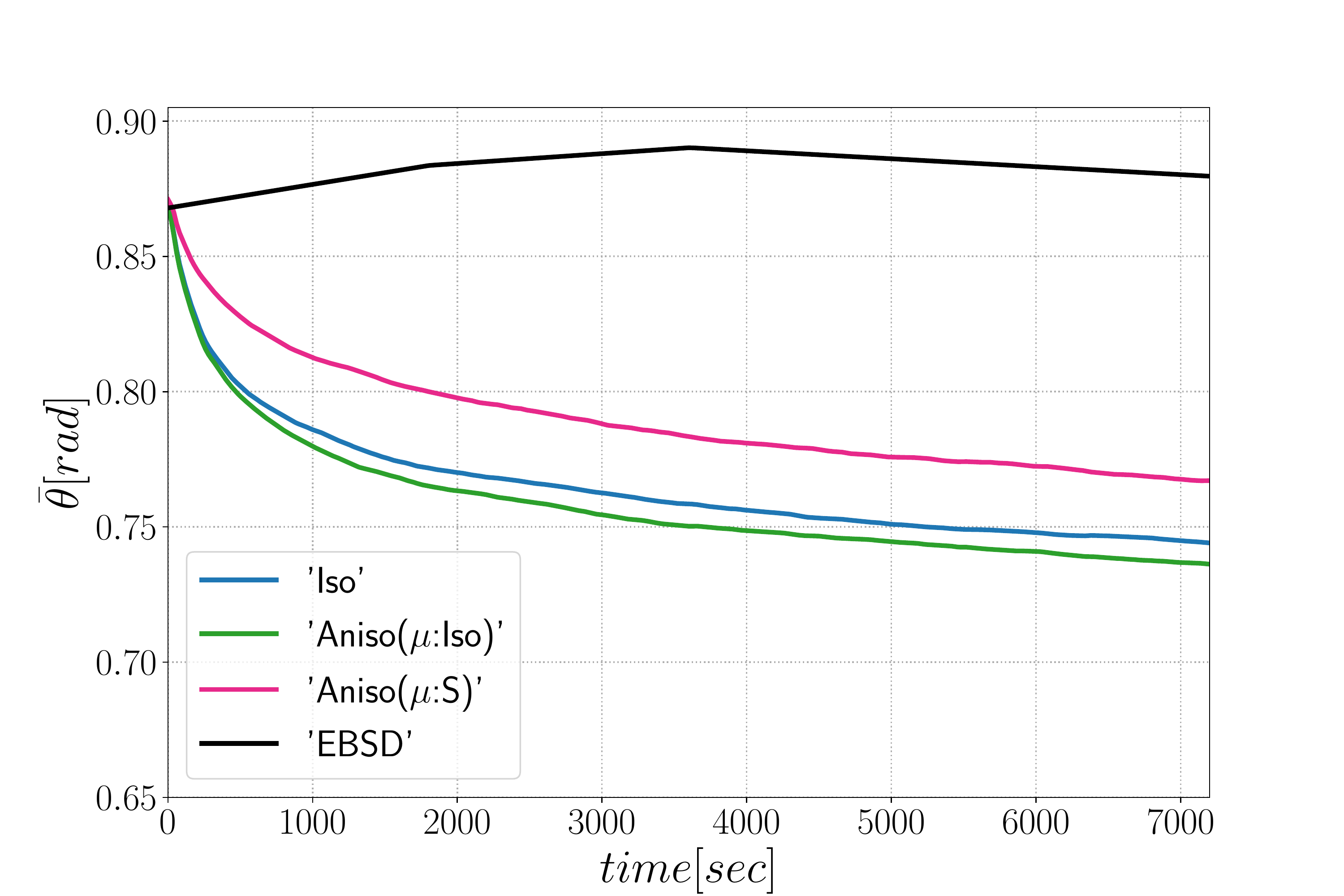}
    \caption{$\bar{R}_{Nb[\%]}=f(t)$}
  \end{subfigure}
  \caption{Mean values time evolution for the isotropic (Iso) formulation, anisotropic formulations with isotropic GB mobility (Aniso($\mu$:Iso)) and heterogeneous GB mobility (Aniso($\mu$:S)) and the experimental data (EBSD): (a) number of grains, (b) average grain radius. Numerical results obtained from the initial immersed microstructure shown in Figure~\ref{fig:GB5DOFImmPX} and the GB5DOF code to define the GB energy.}\label{fig:GB5DOFImmPXMuIsoMeanV}
\end{figure} 

\vspace{-1cm}

\begin{figure}[H]
  \centering
  \begin{subfigure}{0.48\textwidth}
    \centering
    \includegraphics[scale=0.25]{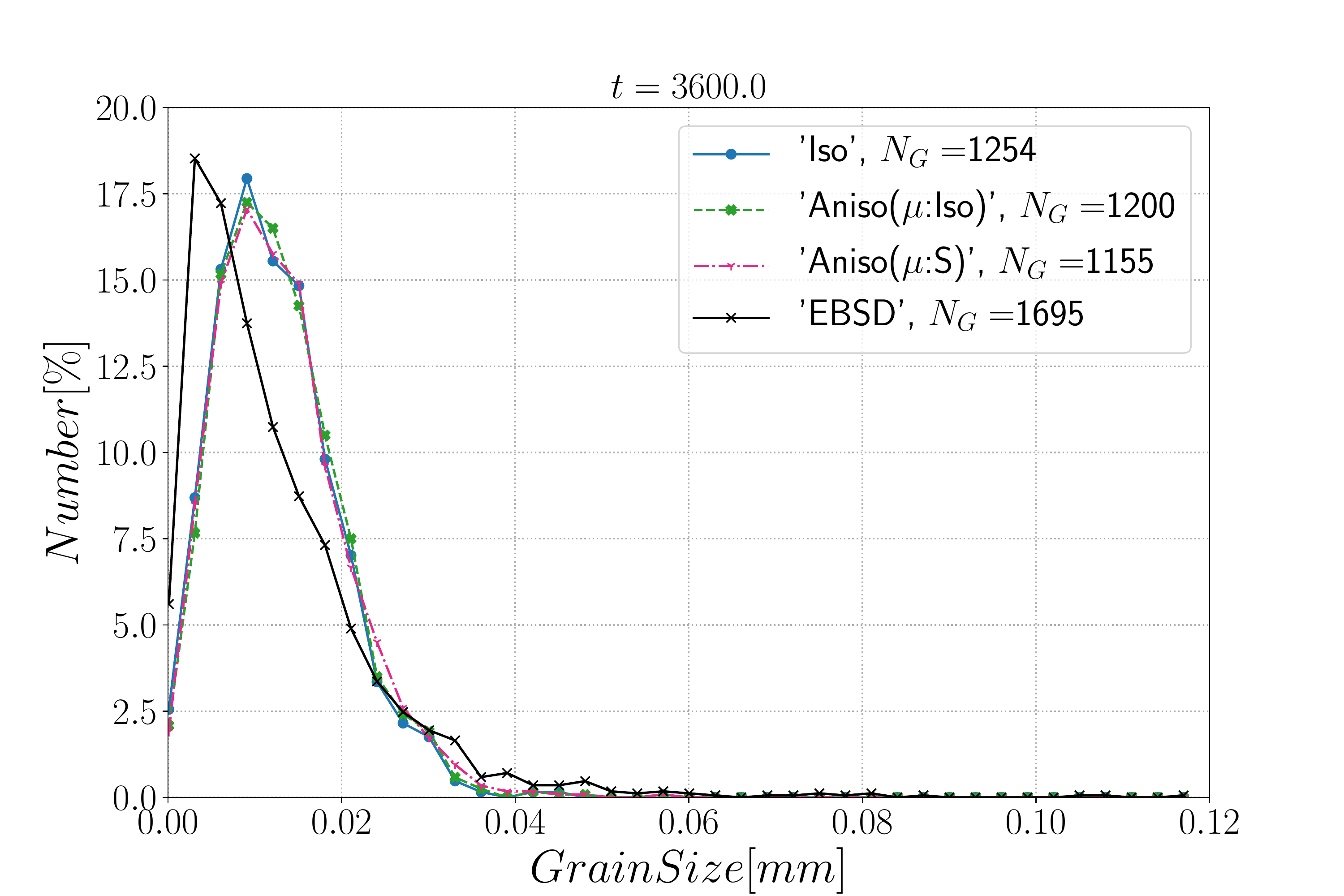}
    \caption{$ t=1\ h $}
  \end{subfigure}
  \begin{subfigure}{0.48\textwidth}
    \centering
    \includegraphics[scale=0.25]{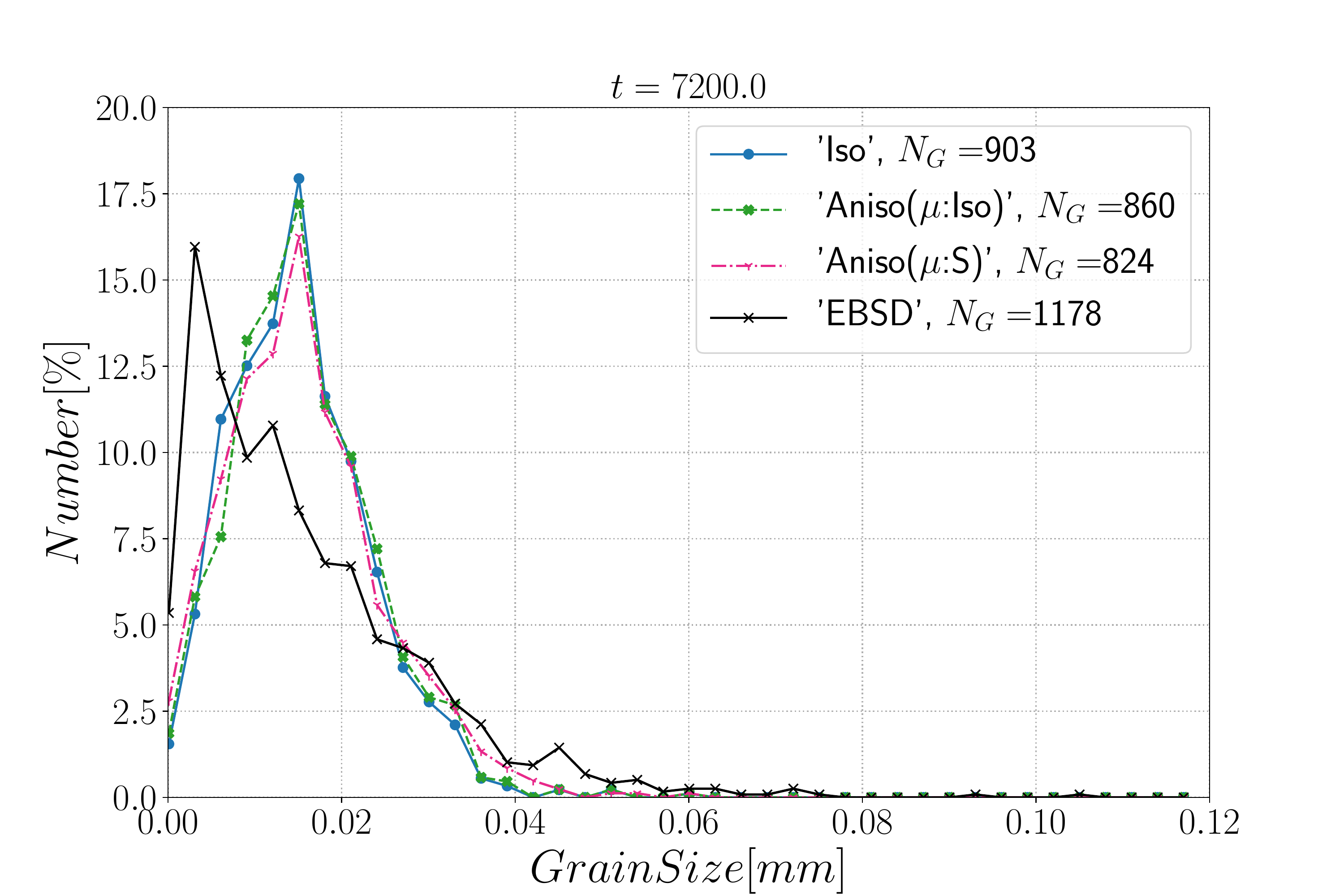}
    \caption{$ t=2\ h $}
  \end{subfigure}
  \caption{Grain Size Distributions obtained at (a) t=1h and (b) t=2h for the isotropic (Iso) formulation, anisotropic formulations with isotropic GB mobility (Aniso($\mu$:Iso)) and heterogeneous GB mobility (Aniso($\mu$:S)) and the experimental data (EBSD), $N_G$ refers to the number. Numerical results obtained from the initial immersed microstructure shown in Figure~\ref{fig:GB5DOFImmPX} and the GB5DOF code to define the GB energy.}
  \label{fig:GB5DOFImmPXGSDMuIso}
\end{figure}

\vspace{-1cm}

\vspace{-1cm}

\subsection{Current state of the modeling of 3D anisotropic grain growth}
\label{ssec:3DAnisoGG}

A final question regarding the anisotropy of GB properties is still open: do the 3D description of GB properties can affect the microstructure evolution? Until now, most of the studies of GG in 3D have presented simulations of polycrystalline microstructures using different textures and a mathematical description of GB properties \cite{FJELDBERG2010267, CHANG20191262, song2020effect, MIYOSHI2021109992} or using data bases of GB energy \cite{KIM20111152, kim2014phase}. The following conclusions are pointed out:
\begin{itemize}
  \item The effect of the heterogeneity is stronger when the material is textured or the disorientation transition between LAGBs and HAGBs, $\theta_0$, is high \cite{kim2014phase, song2020effect, MIYOSHI2021109992},
  \item The individual effect of GB energy and mobility is small on the GG \cite{MIYOSHI2021109992}.
\end{itemize}
Note that similar conclusions were presented in the first part of this work \cite{ma14143883}. In \cite{FJELDBERG2010267, CHANG20191262, song2020effect, MIYOSHI2021109992}, GB properties are defined as heterogeneous and not as anisotropic. The inclination dependency can have an important impact, hence, a complete description of the GB properties is necessary, i.e. $\mu(M(\theta, \vec{a}),\vec{n})$ and $\gamma(M(\theta, \vec{a}),\vec{n})$, as well as 3D non-desctructive in-situ characterization \cite{ZHANG2017229, JUULJENSEN2020100821, Fang:fc5052, wang2022reverse} in order to obtain more realistic values of GB mobility, must be considered as a key perspectives concerning full-field modeling of GG. 

\begin{figure}[H]
  
  \centering
  \begin{subfigure}[c]{0.495\textwidth}
    \centering
    \includegraphics[width=1.0\textwidth]{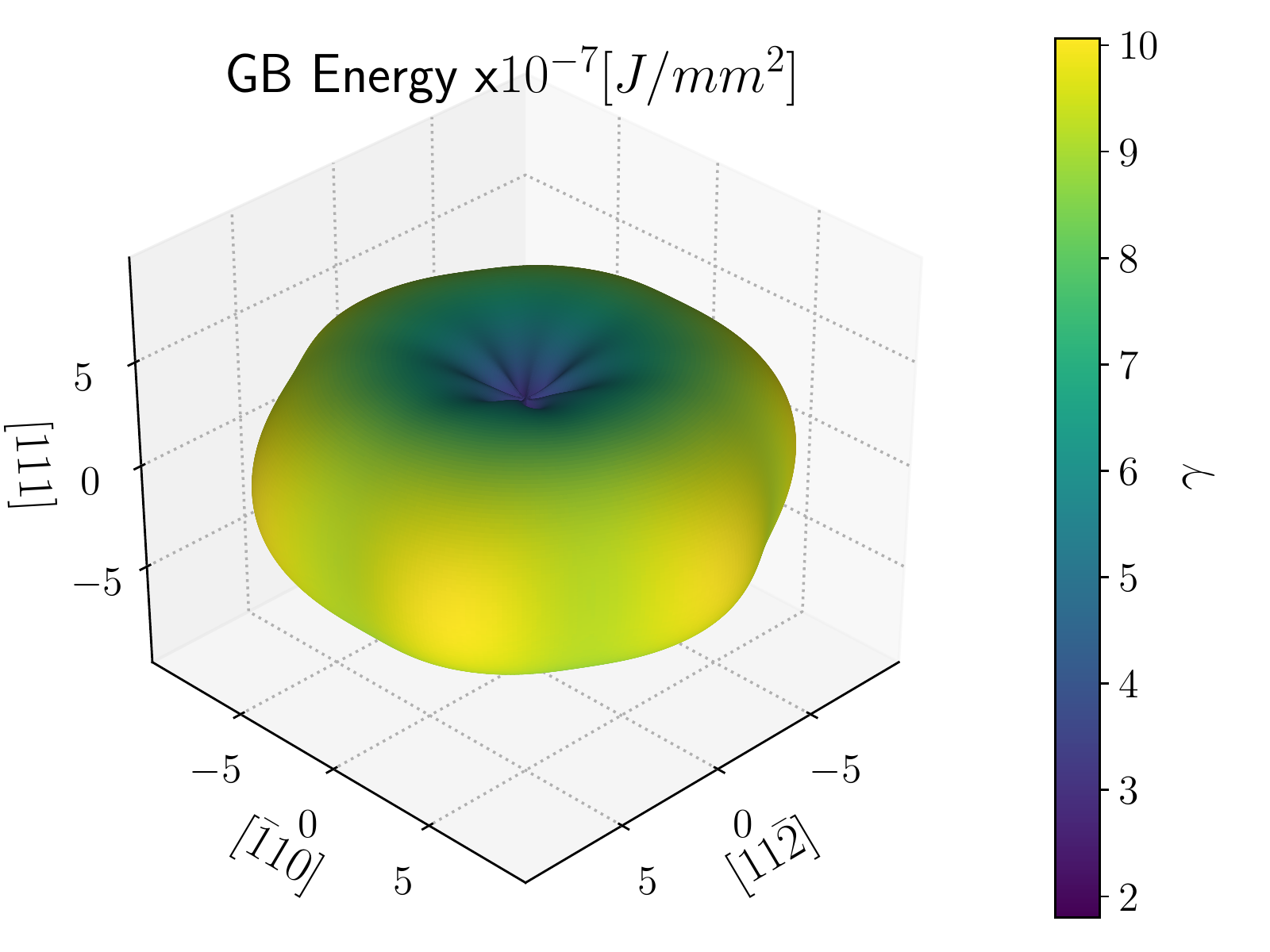}
    \caption{ $\gamma_{\Sigma 3}$ }
  \end{subfigure} 
  \begin{subfigure}[c]{0.495\textwidth}
    \centering
    \includegraphics[width=1.0\textwidth]{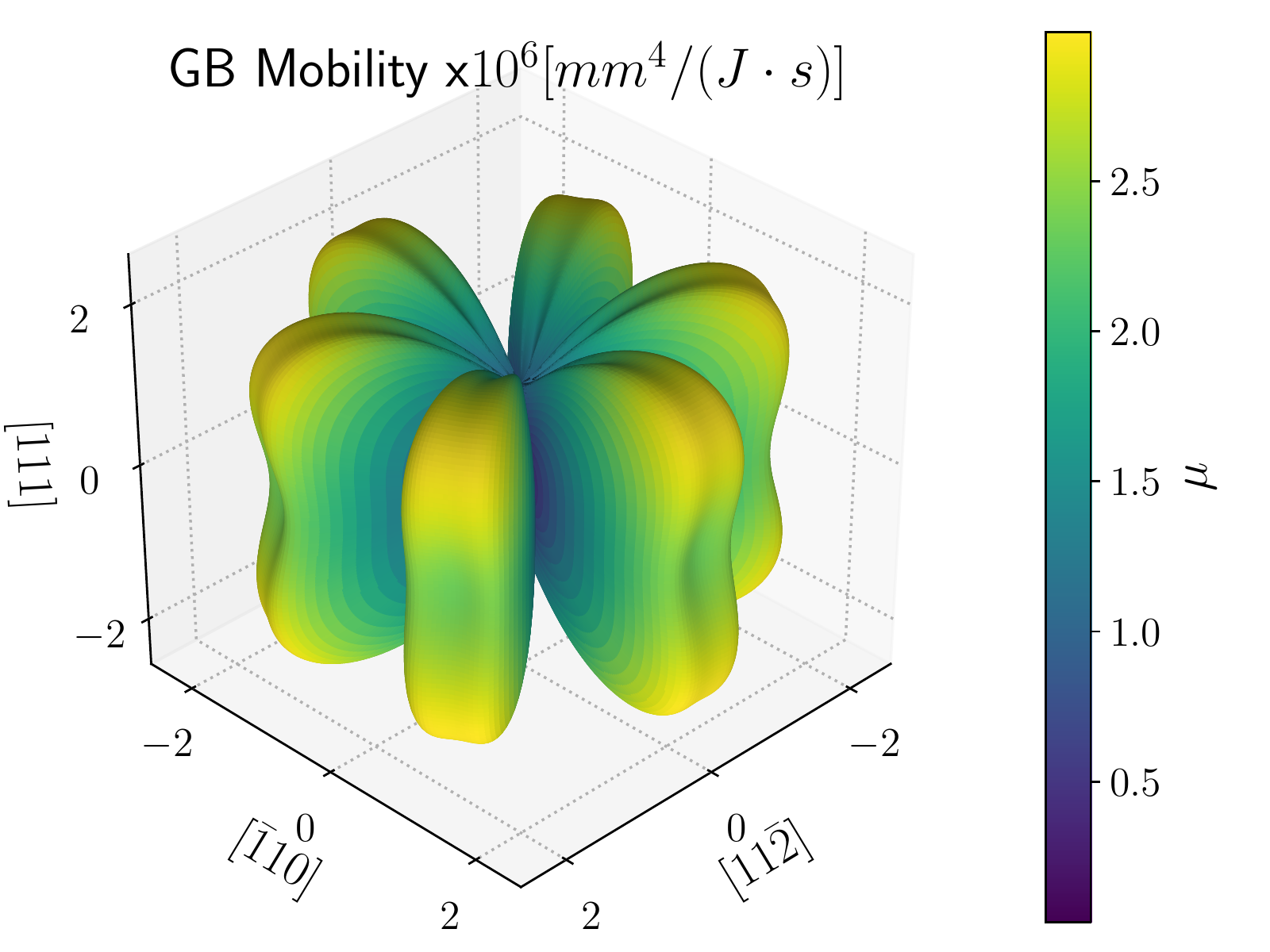}
    \caption{ $\mu_{\Sigma 3}$ }
  \end{subfigure}
  \caption{(a) GB Energy and (b) Mobility of a $\Sigma 3$ TB in Ni computed using the fits proposed in \cite{ABDELJAWAD2018440} of the atomistic simulation data in the study by Olmsted et al. \cite{olmsted2009survey, olmsted2009surveyii}. The minimum, maximum and average values of $\gamma$ and $\mu$ are $\{ 1.803, \ 6.793, \ 10.064 \}$ $\times 10^{-7} \ J \cdot mm^{-2}$ and $\{ 0.032, \ 1.518, \ 2.995 \}$ $\times 10^{6} \ mm^4 \cdot J^{-1} \cdot s^{-1} $. }
  \label{fig:GBEMS3C4}
\end{figure}

In the simulations presented in this section, the GB inclination is simplified as it is projected in the observation plane. In other words, the description of the GB properties is simplified and only a slice of the GB energy is considered. For instance, Figure~\ref{fig:GBEMS3C4} shows the GB energy and mobility of a $\Sigma 3$ TB. One can see that the 3D surface of the TB properties have a complex geometry. On the other hand, the anisotropy of the GB mobility is simplified to a Sigmoidal model with a cusp at $\theta_{\Sigma 3}$. Unluckily, the GB mobility data is not available for the complete GB space and at different temperatures.

\section{Summary and conclusions}

Different FE-LS formulations to study GG of 316L stainless steel were compared. The isotropic formulation is able to reproduce, for statistically generated or immersed polycrystals, the average grain size and grain size distribution for a wide range of anisotropy levels. \\

The results obtained using representative Laguerre-Voronoï polycrystals show that the heterogeneous GB mobility values do not affect the response of the different formulations and that the anisotropic formulation is more physical being the only formulation that enables to promote GBs with low energy. However, the anisotropy level was largely underestimated because of the initial Mackenzie-like DDF which could not be controlled during the polycrystal generation.

Two additional cases were presented with a twin numerical microstructure obtained directly from EBSD data. The main advantage is that the initial DDF and topology are accurately defined. First, GB energy and mobility were defined using the modified Read-Shockley and sigmoidal models already tested for the virtual polycrystals. Then, the model was coupled with an anisotropic model of GB energy that takes into account the GB misorientation and inclination \cite{bulatov2014grain}. However, it is highlighted again that the GB inclination is not well defined as the GB is supposed to be perpendicular to the observation plane. The proposed RS model seemed to exaggerate the anisotropy level comparatively to the GB5DOF code. However, the predictions are clearly better with the proposed RS and sigmoid models associated to the anisotropic formulation while not allowing to be predictive concerning the DDF whatever the method choosen.

These results illustrate that the prediction of grain growth at the polycrystal scale can be ambiguous depending the aimed attributes and the available data. First of all, 3D simulations should be considered. Of course, this aspect is, firstly, essential to improve the representativy of the considered polycrystals but is also essential to describe correctly the $\gamma$ dependence to the inclination. Indeed, the proposed 2D model/data context limits the actual use of the inclination as this parameter is described here with one degree of freedom and not in a 3D framework with 3D experimental data. This aspect can explain the low anisotropy level obtained using the GB5DOF code. Finally, this objective must also be correlated to the fact to integrate the torque effects, and so the GB stiffness tensor, in the simulations and analysis.  It should be highlighted that this conclusion is common to all existing works of the state of the art involving anisotropic 2D GG simulations and 3D simulations were the inclination dependence, or torque terms, or both are not taken into account.  

\section*{Acknowledgements}
The authors thank the ArcelorMittal, ASCOMETAL, AUBERT \& DUVAL, CEA, SAFRAN, FRAMATOME, TIMET, Constellium and TRANSVALOR companies and the ANR for their financial support through the DIGIMU consortium and ANR industrial Chair (Grant No. ANR-16-CHIN-0001).

\section*{Data availability}
The raw data required to reproduce these findings cannot be shared at this time as the data also forms part of an ongoing study. The processed data required to reproduce these findings cannot be shared at this time as the data also forms part of an ongoing study.


\bibliographystyle{unsrtnat}
\bibliography{ManuscriptEBSD}

\begin{thebibliography}{10}
\expandafter\ifx\csname url\endcsname\relax
  \def\url#1{\texttt{#1}}\fi
\expandafter\ifx\csname urlprefix\endcsname\relax\def\urlprefix{URL }\fi
\expandafter\ifx\csname href\endcsname\relax
  \def\href#1#2{#2} \def\path#1{#1}\fi

\bibitem{rollett2017recrystallization}
A.~Rollett, G.~S. Rohrer, J.~Humphreys, Recrystallization and Related Annealing
  Phenomena, Elsevier, 2017.

\bibitem{doi:10.1126/science.abj3210}
A.~Bhattacharya, Y.-F. Shen, C.~M. Hefferan, S.~F. Li, J.~Lind, R.~M. Suter,
  C.~E. Krill, G.~S. Rohrer, Grain boundary velocity and curvature are not
  correlated in ni polycrystals, Science 374~(6564) (2021) 189--193.

\bibitem{FLOREZ2022117459}
S.~Florez, K.~Alvarado, B.~Murgas, N.~Bozzolo, D.~Chatain, C.~E. Krill,
  M.~Wang, G.~S. Rohrer, M.~Bernacki, Statistical behaviour of interfaces
  subjected to curvature flow and torque effects applied to microstructural
  evolutions, Acta Materialia 222 (2022) 117459.

\bibitem{OlmstedI2009}
D.~L. Olmsted, S.~M. Foiles, E.~A. Holm, Survey of computed grain boundary
  properties in face-centered cubic metals: I. grain boundary energy, Acta
  Materialia 57 (2009) 3694--3703.

\bibitem{OlmstedII2009}
D.~L. Olmsted, S.~M. Foiles, E.~A. Holm, Survey of computed grain boundary
  properties in face-centered cubic metals: Ii. grain boundary mobility, Acta
  Materialia 57 (2009) 3704--3713.

\bibitem{bulatov2014grain}
V.~V. Bulatov, B.~W. Reed, M.~Kumar, Grain boundary energy function for fcc
  metals, Acta Materialia 65 (2014) 161--175.

\bibitem{RUNNELS2016174}
B.~Runnels, I.~J. Beyerlein, S.~Conti, M.~Ortiz, An analytical model of
  interfacial energy based on a lattice-matching interatomic energy, Journal of
  the Mechanics and Physics of Solids 89 (2016) 174--193.

\bibitem{garcke1999multiphase}
H.~Garcke, B.~Nestler, B.~Stoth, A multiphase field concept: numerical
  simulations of moving phase boundaries and multiple junctions, SIAM Journal
  on Applied Mathematics 60~(1) (1999) 295--315.

\bibitem{miyoshi2017multi}
E.~Miyoshi, T.~Takaki, Multi-phase-field study of the effects of anisotropic
  grain-boundary properties on polycrystalline grain growth, Journal of Crystal
  Growth 474 (2017) 160--165.

\bibitem{moelans2009comparative}
N.~Moelans, F.~Wendler, B.~Nestler, Comparative study of two phase-field models
  for grain growth, Computational Materials Science 46~(2) (2009) 479--490.

\bibitem{gao1996real}
J.~Gao, R.~Thompson, Real time-temperature models for monte carlo simulations
  of normal grain growth, Acta materialia 44~(11) (1996) 4565--4570.

\bibitem{upmanyu2002boundary}
M.~Upmanyu, G.~N. Hassold, A.~Kazaryan, E.~A. Holm, Y.~Wang, B.~Patton, D.~J.
  Srolovitz, Boundary mobility and energy anisotropy effects on microstructural
  evolution during grain growth, Interface Science 10~(2-3) (2002) 201--216.

\bibitem{hoffrogge2017grain}
P.~W. Hoffrogge, L.~A. Barrales-Mora, Grain-resolved kinetics and rotation
  during grain growth of nanocrystalline aluminium by molecular dynamics,
  Computational Materials Science 128 (2017) 207--222.

\bibitem{SAKOUT2020261}
S.~Sakout, D.~Weisz-Patrault, A.~Ehrlacher, Energetic upscaling strategy for
  grain growth. i: Fast mesoscopic model based on dissipation, Acta Materialia
  196 (2020) 261--279.

\bibitem{BarralesMora2010}
L.~A. {Barrales Mora}, {2D vertex modeling for the simulation of grain growth
  and related phenomena}, Mathematics and Computers in Simulation 80~(7) (2010)
  1411--1427.

\bibitem{wakai2000}
F.~Wakai, N.~Enomoto, H.~Ogawa, Three-dimensional microstructural evolution in
  ideal grain growth—general statistics, Acta Materialia 48~(6) (2000)
  1297--1311.

\bibitem{Florez2020}
S.~Florez, M.~Shakoor, T.~Toulorge, M.~Bernacki, A new finite element strategy
  to simulate microstructural evolutions, Computational Materials Science 172
  (2020) 109335.

\bibitem{Florez2020b}
S.~Florez, K.~Alvarado, D.~P. Muñoz, M.~Bernacki, A novel highly efficient
  lagrangian model for massively multidomain simulation applied to
  microstructural evolutions, Computer Methods in Applied Mechanics and
  Engineering 367 (2020) 113107.

\bibitem{bernacki2011level}
M.~Bernacki, R.~E. Log{\'e}, T.~Coupez, Level set framework for the
  finite-element modeling of recrystallization and grain growth in
  polycrystalline materials, Scripta Materialia 64~(6) (2011) 525--528.

\bibitem{miessen2015advanced}
C.~Mie{\ss}en, M.~Liesenjohann, L.~Barrales-Mora, L.~Shvindlerman,
  G.~Gottstein, An advanced level set approach to grain growth--accounting for
  grain boundary anisotropy and finite triple junction mobility, Acta
  Materialia 99 (2015) 39--48.

\bibitem{Fausty2020}
J.~Fausty, N.~Bozzolo, M.~Bernacki, A 2d level set finite element grain
  coarsening study with heterogeneous grain boundary energies, Applied
  Mathematical Modelling 78 (2020) 505--518.

\bibitem{ma14143883}
B.~Murgas, S.~Florez, N.~Bozzolo, J.~Fausty, M.~Bernacki, Comparative study and
  limits of different level-set formulations for the modeling of anisotropic
  grain growth, Materials 14~(14) (2021).

\bibitem{kim2021crystal}
J.~Kim, M.~Jacobs, S.~Osher, N.~C. Admal, A crystal symmetry-invariant
  kobayashi--warren--carter grain boundary model and its implementation using a
  thresholding algorithm, arXiv preprint arXiv:2102.02773 (2021).

\bibitem{Smith1948introduction}
C.~S. Smith, Introduction to grains, phases, and interfaces—an interpretation
  of microstructure, Transactions of the American Institute of Mining and
  Metallurgical Engineers 175~(2) (1948) 15--51.

\bibitem{kohara1958anisotropy}
S.~Kohara, M.~N. Parthasarathi, P.~A. Beck, \textbf{Anisotropy of boundary
  mobility}, Journal of Applied Physics 29~(7) (1958) 1125--1126.

\bibitem{anderson1984computer}
M.~Anderson, D.~Srolovitz, G.~Grest, P.~Sahni, Computer simulation of grain
  growth—i. kinetics, Acta metallurgica 32~(5) (1984) 783--791.

\bibitem{lazar2011more}
E.~A. Lazar, J.~K. Mason, R.~D. MacPherson, D.~J. Srolovitz, A more accurate
  three-dimensional grain growth algorithm, Acta Materialia 59~(17) (2011)
  6837--6847.

\bibitem{rollett1989simulation}
A.~Rollett, D.~J. Srolovitz, M.~Anderson, Simulation and theory of abnormal
  grain growth—anisotropic grain boundary energies and mobilities, Acta
  metallurgica 37~(4) (1989) 1227--1240.

\bibitem{hwang1998simulation}
N.~M. Hwang, Simulation of the effect of anisotropic grain boundary mobility
  and energy on abnormal grain growth, Journal of materials science 33~(23)
  (1998) 5625--5629.

\bibitem{Fausty2018}
J.~Fausty, N.~Bozzolo, D.~Pino~Mu{\~n}oz, M.~Bernacki, A novel level-set finite
  element formulation for grain growth with heterogeneous grain boundary
  energies, Materials \& Design 160 (2018) 578--590.

\bibitem{zollner2019texture}
D.~Z{\"o}llner, I.~Zlotnikov, Texture controlled grain growth in thin films
  studied by 3d potts model, Advanced Theory and Simulations 2~(8) (2019)
  1900064.

\bibitem{miyoshi2016validation}
E.~Miyoshi, T.~Takaki, Validation of a novel higher-order multi-phase-field
  model for grain-growth simulations using anisotropic grain-boundary
  properties, Computational Materials Science 112 (2016) 44--51.

\bibitem{chang2019effect}
K.~Chang, H.~Chang, Effect of grain boundary energy anisotropy in 2d and 3d
  grain growth process, Results in Physics 12 (2019) 1262--1268.

\bibitem{miyoshi2019accuracy}
E.~Miyoshi, T.~Takaki, M.~Ohno, Y.~Shibuta, Accuracy evaluation of phase-field
  models for grain growth simulation with anisotropic grain boundary
  properties, ISIJ International (2019) ISIJINT--2019.

\bibitem{holm2001misorientation}
E.~A. Holm, G.~N. Hassold, M.~A. Miodownik, On misorientation distribution
  evolution during anisotropic grain growth, Acta Materialia 49~(15) (2001)
  2981--2991.

\bibitem{kazaryan2002grain}
A.~Kazaryan, Y.~Wang, S.~Dregia, B.~Patton, Grain growth in anisotropic
  systems: comparison of effects of energy and mobility, Acta Materialia
  50~(10) (2002) 2491--2502.

\bibitem{FAUSTY202128}
J.~Fausty, B.~Murgas, S.~Florez, N.~Bozzolo, M.~Bernacki, A new analytical test
  case for anisotropic grain growth problems, Applied Mathematical Modelling 93
  (2021) 28--52.

\bibitem{hallberg2019modeling}
H.~Hallberg, V.~V. Bulatov, Modeling of grain growth under fully anisotropic
  grain boundary energy, Modeling and Simulation in Materials Science and
  Engineering 27~(4) (2019) 045002.

\bibitem{viswanathan1973kinetics}
R.~Viswanathan, C.~L. Bauer, Kinetics of grain boundary migration in copper
  bicrystals with [001] rotation axes, Acta Metallurgica 21~(8) (1973)
  1099--1109.

\bibitem{demianczuk1975effect}
D.~W. Demianczuk, K.~T. Aust, Effect of solute and orientation on the mobility
  of near-coincidence tilt boundaries in high-purity aluminum, Acta
  Metallurgica 23~(10) (1975) 1149--1162.

\bibitem{maksimova1988transformation}
E.~L. Maksimova, L.~S. Shvindlerman, B.~B. Straumal, Transformation of
  $\sigma$17 special tilt boundaries to general boundaries in tin, Acta
  Metallurgica 36~(6) (1988) 1573--1583.

\bibitem{gottstein1992true}
G.~Gottstein, L.~S. Shvindlerman, On the true dependence of grain boundary
  migration rate on driving force, Scripta metallurgica et materialia 27~(11)
  (1992) 1521--1526.

\bibitem{winning2002mechanisms}
M.~Winning, G.~Gottstein, L.~S. Shvindlerman, On the mechanisms of grain
  boundary migration, Acta Materialia 50~(2) (2002) 353--363.

\bibitem{ivanov2006kinetics}
V.~A. Ivanov, On kinetics and thermodynamics of high angle grain boundaries in
  aluminum: Experimental study on grain boundary properties in bi-and
  tricrystals, Tech. rep., Fakult{\"a}t f{\"u}r Georessourcen und
  Materialtechnik (2006).

\bibitem{ZHANG2017229}
J.~Zhang, S.~O. Poulsen, J.~W. Gibbs, P.~W. Voorhees, H.~F. Poulsen,
  Determining material parameters using phase-field simulations and
  experiments, Acta Materialia 129 (2017) 229--238.

\bibitem{zhang2018determination}
J.~Zhang, Determination of material parameters by comparison of 3d simulations
  and 3d experiments, Ph.D. thesis (2018).

\bibitem{ZHANG2020211}
J.~Zhang, W.~Ludwig, Y.~Zhang, H.~H.~B. Sørensen, D.~J. Rowenhorst,
  A.~Yamanaka, P.~W. Voorhees, H.~F. Poulsen, Grain boundary mobilities in
  polycrystals, Acta Materialia 191 (2020) 211--220.

\bibitem{JUULJENSEN2020100821}
D.~{Juul Jensen}, Y.~Zhang, Impact of 3d/4d methods on the understanding of
  recrystallization, Current Opinion in Solid State and Materials Science
  24~(2) (2020) 100821.

\bibitem{Fang:fc5052}
H.~Fang, D.~Juul~Jensen, Y.~Zhang, {Improved grain mapping by laboratory X-ray
  diffraction contrast tomography}, IUCrJ 8~(4) (2021) 559--573.

\bibitem{janssens2006computing}
K.~G. Janssens, D.~Olmsted, E.~A. Holm, S.~M. Foiles, S.~J. Plimpton, P.~M.
  Derlet, Computing the mobility of grain boundaries, Nature materials 5~(2)
  (2006) 124--127.

\bibitem{olmsted2009survey}
D.~L. Olmsted, S.~M. Foiles, E.~A. Holm, Survey of computed grain boundary
  properties in face-centered cubic metals: I. grain boundary energy, Acta
  Materialia 57~(13) (2009) 3694--3703.

\bibitem{olmsted2009surveyii}
D.~L. Olmsted, E.~A. Holm, S.~M. Foiles, Survey of computed grain boundary
  properties in face-centered cubic metals—ii: Grain boundary mobility, Acta
  materialia 57~(13) (2009) 3704--3713.

\bibitem{FJELDBERG2010267}
E.~Fjeldberg, K.~Marthinsen, A 3d monte carlo study of the effect of grain
  boundary anisotropy and particles on the size distribution of grains after
  recrystallisation and grain growth, Computational Materials Science 48~(2)
  (2010) 267--281.

\bibitem{CHANG20191262}
K.~Chang, H.~Chang, Effect of grain boundary energy anisotropy in 2d and 3d
  grain growth process, Results in Physics 12 (2019) 1262--1268.

\bibitem{song2020effect}
Y.-H. Song, M.-T. Wang, J.~Ni, J.-F. Jin, Y.-P. Zong, Effect of grain boundary
  energy anisotropy on grain growth in zk60 alloy using a 3d phase-field
  modeling, Chinese Physics B 29~(12) (2020) 128201.

\bibitem{MIYOSHI2021109992}
E.~Miyoshi, T.~Takaki, S.~Sakane, M.~Ohno, Y.~Shibuta, T.~Aoki, Large-scale
  phase-field study of anisotropic grain growth: Effects of
  misorientation-dependent grain boundary energy and mobility, Computational
  Materials Science 186 (2021) 109992.

\bibitem{KIM20111152}
H.-K. Kim, W.-S. Ko, H.-J. Lee, S.~G. Kim, B.-J. Lee, An identification scheme
  of grain boundaries and construction of a grain boundary energy database,
  Scripta Materialia 64~(12) (2011) 1152--1155.

\bibitem{kim2014phase}
H.-K. Kim, S.~G. Kim, W.~Dong, I.~Steinbach, B.-J. Lee, Phase-field modeling
  for 3d grain growth based on a grain boundary energy database, Modelling and
  Simulation in Materials Science and Engineering 22~(3) (2014) 034004.

\bibitem{ReadShockley}
W.~T. Read, W.~Shockley, Dislocation models of crystal grain boundaries,
  Physical Review 78~(3) (1950) 275--289.

\bibitem{humphreys1997unified}
F.~J. Humphreys, A unified theory of recovery, recrystallization and grain
  growth, based on the stability and growth of cellular microstructures—i.
  the basic model, Acta Materialia 45~(10) (1997) 4231--4240.

\bibitem{Bernacki2008}
M.~Bernacki, Y.~Chastel, T.~Coupez, R.~E. Log{\'{e}}, {Level set framework for
  the numerical modelling of primary recrystallization in polycrystalline
  materials}, Scripta Materialia 58~(12) (2008) 1129--1132.

\bibitem{Bernacki2009}
M.~Bernacki, H.~Resk, T.~Coupez, R.~E. Log{\'{e}}, {Finite element model of
  primary recrystallization in polycrystalline aggregates using a level set
  framework}, Modelling and Simulation in Materials Science and Engineering
  17~(6) (2009) 64006.

\bibitem{scholtes2015new}
B.~Scholtes, M.~Shakoor, A.~Settefrati, P.-O. Bouchard, N.~Bozzolo,
  M.~Bernacki, New finite element developments for the full field modeling of
  microstructural evolutions using the level-set method, Computational
  Materials Science 109 (2015) 388--398.

\bibitem{Maire2017}
L.~Maire, B.~Scholtes, C.~Moussa, N.~Bozzolo, D.~P. Mu{\~{n}}oz, A.~Settefrati,
  M.~Bernacki, {Modeling of dynamic and post-dynamic recrystallization by
  coupling a full field approach to phenomenological laws}, Materials and
  Design 133 (2017) 498--519.

\bibitem{Hitti2012}
K.~Hitti, P.~Laure, T.~Coupez, L.~Silva, M.~Bernacki, {Precise generation of
  complex statistical Representative Volume Elements (RVEs) in a finite element
  context}, Computational Materials Science 61 (2012) 224--238.

\bibitem{Osher1988}
S.~Osher, J.~A. Sethian, {Fronts propagating with curvature-dependent speed:
  Algorithms based on Hamilton-Jacobi formulations}, Journal of Computational
  Physics 79~(1) (1988) 12--49.

\bibitem{Merriman1994}
B.~Merriman, J.~K. Bence, S.~J. Osher, {Motion of multiple junctions: A level
  set approach} (1994).

\bibitem{Zhao1996}
H.~Zhao, T.~Chan, B.~Merriman, S.~Osher, A variational level set approach to
  multiphase motion, Journal of Computational Physics 127 (1996) 179--195.

\bibitem{Bernacki2011}
M.~Bernacki, R.~E. Log{\'{e}}, T.~Coupez, {Level set framework for the
  finite-element modelling of recrystallization and grain growth in
  polycrystalline materials}, Scripta Materialia 64~(6) (2011) 525--528.

\bibitem{Shakoor2015amm}
M.~Shakoor, B.~Scholtes, P.-O. Bouchard, M.~Bernacki, {An efficient and
  parallel level set reinitialization method – Application to micromechanics
  and microstructural evolutions}, Applied Mathematical Modelling 39~(23-24)
  (2015) 7291--7302.

\bibitem{morawiec2003orientations}
A.~Morawiec, Orientations and rotations, Springer, 2003.

\bibitem{ABDELJAWAD2018440}
F.~Abdeljawad, S.~M. Foiles, A.~P. Moore, A.~R. Hinkle, C.~M. Barr, N.~M.
  Heckman, K.~Hattar, B.~L. Boyce, The role of the interface stiffness tensor
  on grain boundary dynamics, Acta Materialia 158 (2018) 440--453.

\bibitem{DU2007467}
D.~Du, H.~Zhang, D.~J. Srolovitz, Properties and determination of the interface
  stiffness, Acta Materialia 55~(2) (2007) 467--471.

\bibitem{MOORE2021117220}
R.~D. Moore, T.~Beecroft, G.~S. Rohrer, C.~M. Barr, E.~R. Homer, K.~Hattar,
  B.~L. Boyce, F.~Abdeljawad, The grain boundary stiffness and its impact on
  equilibrium shapes and boundary migration: Analysis of the $\sigma$5, 7, 9,
  and 11 boundaries in ni, Acta Materialia (2021) 117220.

\bibitem{BURKE1952220}
J.~Burke, D.~Turnbull, Recrystallization and grain growth, Progress in Metal
  Physics 3 (1952) 220--292.

\bibitem{CRUZFABIANO2014305}
A.~Cruz-Fabiano, R.~Logé, M.~Bernacki, Assessment of simplified 2d grain
  growth models from numerical experiments based on a level set framework,
  Computational Materials Science 92 (2014) 305 -- 312.

\bibitem{alvarado2021dissolution}
K.~Alvarado, I.~Janeiro, S.~Florez, B.~Flipon, J.-M. Franchet, D.~Locq,
  C.~Dumont, N.~Bozzolo, M.~Bernacki, Dissolution of the primary $\gamma$'
  precipitates and grain growth during solution treatment of three nickel base
  superalloys, Metals 11~(12) (2021) 1921.

\bibitem{agnoli2014development}
A.~Agnoli, N.~Bozzolo, R.~Log{\'e}, J.-M. Franchet, J.~Laigo, M.~Bernacki,
  Development of a level set methodology to simulate grain growth in the
  presence of real secondary phase particles and stored energy--application to
  a nickel-base superalloy, Computational Materials Science 89 (2014) 233--241.

\bibitem{maire2016}
L.~Maire, B.~Scholtes, C.~Moussa, D.~Pino~Mu{\~n}oz, N.~Bozzolo, M.~Bernacki,
  Improvement of {3-D} mean field models for pure grain growth based on full
  field simulations, Journal of Materials Science 51~(24) (2016) 10970--10981.

\bibitem{alvarado2021level}
K.~Alvarado, S.~Florez, B.~Flipon, N.~Bozzolo, M.~Bernacki, A level set
  approach to simulate grain growth with an evolving population of second phase
  particles, Modelling and Simulation in Materials Science and Engineering
  29~(3) (2021) 035009.

\bibitem{hitti2013optimized}
K.~Hitti, M.~Bernacki, Optimized dropping and rolling (odr) method for packing
  of poly-disperse spheres, Applied Mathematical Modelling 37~(8) (2013)
  5715--5722.

\bibitem{ROUX201332}
E.~Roux, M.~Bernacki, P.~Bouchard, A level-set and anisotropic adaptive
  remeshing strategy for the modeling of void growth under large plastic
  strain, Computational Materials Science 68 (2013) 32--46.

\bibitem{ratanaphan2019atomistic}
S.~Ratanaphan, R.~Sarochawikasit, N.~Kumanuvong, S.~Hayakawa, H.~Beladi, G.~S.
  Rohrer, T.~Okita, Atomistic simulations of grain boundary energies in
  austenitic steel, Journal of Materials Science 54~(7) (2019) 5570--5583.

\bibitem{chang2014effect}
K.~Chang, N.~Moelans, Effect of grain boundary energy anisotropy on highly
  textured grain structures studied by phase-field simulations, Acta materialia
  64 (2014) 443--454.

\bibitem{gruber2009misorientation}
J.~Gruber, H.~Miller, T.~Hoffmann, G.~Rohrer, A.~Rollett, Misorientation
  texture development during grain growth. part i: Simulation and experiment,
  Acta Materialia 57~(20) (2009) 6102--6112.

\bibitem{elsey2013simulations}
M.~Elsey, S.~Esedog, P.~Smereka, et~al., Simulations of anisotropic grain
  growth: Efficient algorithms and misorientation distributions, Acta
  materialia 61~(6) (2013) 2033--2043.

\bibitem{TU2019268}
X.~Tu, A.~Shahba, J.~Shen, S.~Ghosh, Microstructure and property based
  statistically equivalent rves for polycrystalline-polyphase aluminum alloys,
  International Journal of Plasticity 115 (2019) 268--292.

\bibitem{Elsey2013}
M.~Elsey, S.~Esedog¯lu, P.~Smereka, {Simulations of anisotropic grain growth:
  Efficient algorithms and misorientation distributions}, Acta Materialia
  61~(6) (2013) 2033--2043.

\bibitem{wang2022reverse}
M.~Wang, Reverse engineering the kinetics of grain growth in al-based
  polycrystals by microstructural mapping in 4d, Ph.D. thesis, Universit{\"a}t
  Ulm (2022).

\end{thebibliography}

\end{document}